\documentclass[12pt,titlepage,letterpaper]{utarticle}

\usepackage{amsmath}
\usepackage{mathrsfs}
\usepackage{amsfonts}
\usepackage{amssymb}
\usepackage{amsthm}
\usepackage{mathtools}
\usepackage{mathbbol}
\usepackage{graphicx}
\usepackage{color}
\usepackage{xparse}
\usepackage{tikz}
\usepackage{tocloft}
\usepackage[titletoc,title]{appendix}
\usepackage[nosort]{cite}
\usepackage{hyperref}
\usepackage{longtable}

\usepackage{caption}

\definecolor{lightmauve}{RGB}{255,187,255}
\definecolor{lightblue}{RGB}{238,238,255}
\definecolor{lightred}{RGB}{255,238,238}
\definecolor{midgreen}{RGB}{15,170,15}

\definecolor{aqua}{rgb}{0, 1.0, 1.0}
\definecolor{fuschia}{rgb}{1.0, 0, 1.0}
\definecolor{gray}{rgb}{0.502, 0.502, 0.502}
\definecolor{lime}{rgb}{0, 1.0, 0}
\definecolor{maroon}{rgb}{0.502, 0, 0}
\definecolor{navy}{rgb}{0, 0, 0.502}
\definecolor{olive}{rgb}{0.502, 0.502, 0}
\definecolor{purple}{rgb}{0.502, 0, 0.502}
\definecolor{silver}{rgb}{0.753, 0.753, 0.753}
\definecolor{teal}{rgb}{0, 0.502, 0.502}

\makeatletter
\newdimen\itex@wd%
\newdimen\itex@dp%
\newdimen\itex@thd%
\def\itexspace#1#2#3{\itex@wd=#3em%
\itex@wd=0.1\itex@wd%
\itex@dp=#2ex%
\itex@dp=0.1\itex@dp%
\itex@thd=#1ex%
\itex@thd=0.1\itex@thd%
\advance\itex@thd\the\itex@dp%
\makebox[\the\itex@wd]{\rule[-\the\itex@dp]{0cm}{\the\itex@thd}}}
\makeatother

\makeatletter
\newif\if@sup
\newtoks\@sups
\def\append@sup#1{\edef\act{\noexpand\@sups={\the\@sups #1}}\act}%
\def\reset@sup{\@supfalse\@sups={}}%
\def\mk@scripts#1#2{\if #2/ \if@sup ^{\the\@sups}\fi \else%
  \ifx #1_ \if@sup ^{\the\@sups}\reset@sup \fi {}_{#2}%
  \else \append@sup#2 \@suptrue \fi%
  \expandafter\mk@scripts\fi}
\def\tensor#1#2{\reset@sup#1\mk@scripts#2_/}
\def\multiscripts#1#2#3{\reset@sup{}\mk@scripts#1_/#2%
  \reset@sup\mk@scripts#3_/}
\makeatother

\makeatletter
\newbox\slashbox \setbox\slashbox=\hbox{$/$}
\def\itex@pslash#1{\setbox\@tempboxa=\hbox{$#1$}
  \@tempdima=0.5\wd\slashbox \advance\@tempdima 0.5\wd\@tempboxa
  \copy\slashbox \kern-\@tempdima \box\@tempboxa}
\def\slash{\protect\itex@pslash}
\makeatother

\def\clap#1{\hbox to 0pt{\hss#1\hss}}

\let\oldroot\root
\def\root#1#2{\oldroot #1 \of{#2}}
\renewcommand{\sqrt}[2][]{\oldroot #1 \of{#2}}

\DeclareSymbolFont{symbolsC}{U}{txsyc}{m}{n}
\SetSymbolFont{symbolsC}{bold}{U}{txsyc}{bx}{n}
\DeclareFontSubstitution{U}{txsyc}{m}{n}

\DeclareSymbolFont{stmry}{U}{stmry}{m}{n}
\SetSymbolFont{stmry}{bold}{U}{stmry}{b}{n}

\DeclareFontFamily{OMX}{MnSymbolE}{}
\DeclareSymbolFont{mnomx}{OMX}{MnSymbolE}{m}{n}
\SetSymbolFont{mnomx}{bold}{OMX}{MnSymbolE}{b}{n}
\DeclareFontShape{OMX}{MnSymbolE}{m}{n}{
    <-6>  MnSymbolE5
   <6-7>  MnSymbolE6
   <7-8>  MnSymbolE7
   <8-9>  MnSymbolE8
   <9-10> MnSymbolE9
  <10-12> MnSymbolE10
  <12->   MnSymbolE12}{}

\makeatletter
\def\re@DeclareMathSymbol#1#2#3#4{%
    \let#1=\undefined
    \DeclareMathSymbol{#1}{#2}{#3}{#4}}
\re@DeclareMathSymbol{\neArrow}{\mathrel}{symbolsC}{116}
\re@DeclareMathSymbol{\neArr}{\mathrel}{symbolsC}{116}
\re@DeclareMathSymbol{\seArrow}{\mathrel}{symbolsC}{117}
\re@DeclareMathSymbol{\seArr}{\mathrel}{symbolsC}{117}
\re@DeclareMathSymbol{\nwArrow}{\mathrel}{symbolsC}{118}
\re@DeclareMathSymbol{\nwArr}{\mathrel}{symbolsC}{118}
\re@DeclareMathSymbol{\swArrow}{\mathrel}{symbolsC}{119}
\re@DeclareMathSymbol{\swArr}{\mathrel}{symbolsC}{119}
\re@DeclareMathSymbol{\nequiv}{\mathrel}{symbolsC}{46}
\re@DeclareMathSymbol{\Perp}{\mathrel}{symbolsC}{121}
\re@DeclareMathSymbol{\Vbar}{\mathrel}{symbolsC}{121}
\re@DeclareMathSymbol{\sslash}{\mathrel}{stmry}{12}
\re@DeclareMathSymbol{\boxslash}{\mathrel}{stmry}{27}
\re@DeclareMathSymbol{\boxbslash}{\mathrel}{stmry}{28}
\re@DeclareMathSymbol{\boxast}{\mathrel}{stmry}{24}
\re@DeclareMathSymbol{\boxcircle}{\mathrel}{stmry}{29}
\re@DeclareMathSymbol{\boxbox}{\mathrel}{stmry}{30}
\re@DeclareMathSymbol{\obslash}{\mathrel}{stmry}{20}
\re@DeclareMathSymbol{\obar}{\mathrel}{stmry}{58}
\re@DeclareMathSymbol{\olessthan}{\mathrel}{stmry}{60}
\re@DeclareMathSymbol{\ogreaterthan}{\mathrel}{stmry}{61}
\re@DeclareMathSymbol{\bigsqcap}{\mathop}{stmry}{"64}
\re@DeclareMathSymbol{\biginterleave}{\mathop}{stmry}{"6}
\re@DeclareMathSymbol{\invamp}{\mathrel}{symbolsC}{77}
\re@DeclareMathSymbol{\parr}{\mathrel}{symbolsC}{77}
\makeatother

\makeatletter
\def\Decl@Mn@Delim#1#2#3#4{%
  \if\relax\noexpand#1%
    \let#1\undefined
  \fi
  \DeclareMathDelimiter{#1}{#2}{#3}{#4}{#3}{#4}}
\def\Decl@Mn@Open#1#2#3{\Decl@Mn@Delim{#1}{\mathopen}{#2}{#3}}
\def\Decl@Mn@Close#1#2#3{\Decl@Mn@Delim{#1}{\mathclose}{#2}{#3}}
\Decl@Mn@Open{\llangle}{mnomx}{'164}
\Decl@Mn@Close{\rrangle}{mnomx}{'171}
\Decl@Mn@Open{\lmoustache}{mnomx}{'245}
\Decl@Mn@Close{\rmoustache}{mnomx}{'244}
\Decl@Mn@Open{\llbracket}{stmry}{'112}
\Decl@Mn@Close{\rrbracket}{stmry}{'113}
\makeatother

\makeatletter
\DeclareRobustCommand\widecheck[1]{{\mathpalette\@widecheck{#1}}}
\def\@widecheck#1#2{%
    \setbox\z@\hbox{\m@th$#1#2$}%
    \setbox\tw@\hbox{\m@th$#1%
       \widehat{%
          \vrule\@width\z@\@height\ht\z@
          \vrule\@height\z@\@width\wd\z@}$}%
    \dp\tw@-\ht\z@
    \@tempdima\ht\z@ \advance\@tempdima2\ht\tw@ \divide\@tempdima\thr@@
    \setbox\tw@\hbox{%
       \raise\@tempdima\hbox{\scalebox{1}[-1]{\lower\@tempdima\box
\tw@}}}%
    {\ooalign{\box\tw@ \cr \box\z@}}}
\makeatother

\makeatletter
\NewDocumentCommand\mathraisebox{moom}{%
\IfNoValueTF{#2}{\def\@temp##1##2{\raisebox{#1}{$\m@th##1##2$}}}{%
\IfNoValueTF{#3}{\def\@temp##1##2{\raisebox{#1}[#2]{$\m@th##1##2$}}%
}{\def\@temp##1##2{\raisebox{#1}[#2][#3]{$\m@th##1##2$}}}}%
\mathpalette\@temp{#4}}
\makeatletter

\makeatletter
\def\udots{\mathinner{\mkern2mu\raise\p@\hbox{.}
\mkern2mu\raise4\p@\hbox{.}\mkern1mu
\raise7\p@\vbox{\kern7\p@\hbox{.}}\mkern1mu}}
\makeatother

\theoremstyle{plain}

\theoremstyle{definition}

\theoremstyle{remark}

\usetikzlibrary{positioning,positioning,arrows,decorations.markings}

\begin{document}

%-------------------------------------------------------------------
\preprint{
UTTG--02--2022\\
}

\title{Two Applications of Nilpotent Higgsing in Class-S}

\author{Jacques Distler, Grant Elliot
     \oneaddress{
      Theory Group\\
      Department of Physics,\\
      University of Texas at Austin,\\
      Austin, TX 78712, USA \\
      {~}\\
      \email{distler@golem.ph.utexas.edu}\\
      \email{gelliot123@utexas.edu}
      }
}
%\date{\today}
\date{March 10, 2022}

\Abstract{
We introduce a class of Higgs-branch RG flows in theories of class-S, which flow between $d=4$ $\mathcal{N}=2$ SCFTs of the same ADE type. We discuss two applications of this class of RG flows: 1) determining the current-algebra levels in SCFTs where they were previously unknown --- a program we carry out for the class-S theories of type $E_6$ and $E_7$ --- and 2) constructing a multitude of examples of pairs of $\mathcal{N}=2$ SCFTs whose ``conventional invariants" coincide. We disprove the conjecture of \cite{Distler:2020tub} that the global form of the flavour symmetry group is a reliable diagnostic for determining when two such theories are isomorphic.
}

\maketitle

\tocloftpagestyle{empty}
\tableofcontents
\vfill
\newpage
\setcounter{page}{1}

\section{Introduction}\label{introduction}

In some sense, ``most" quantum field theories are non-Lagrangian. This presents a challenge to understanding their properties. In the case of $d=4$, $\mathcal{N}=2$ superconformal field theories of class-S, considerable progress can be made. These theories are obtained as the compactification of a $6d$ (2,0) SCFT on a Riemann surface, $C_{g,n}$, with punctures labeled by nilpotent orbits in the appropriate Lie algebra. For each pants-decomposition of $C_{g,n}$, we have a presentation of the SCFT as a gauging (with vanishing (super-)Yang-Mills $\beta$-function) of a product of SCFTs associated to the compactification on 3-punctured spheres (``fixtures"). Different pants-decompositions yield different (``S-dual") presentations of the same SCFT. The central task, then, is to understand the properties of the SCFTs associated to fixtures.

Many of those properties are calculable, in algorithmic fashion, from the data of the (2,0) theory and the choice of a triple of nilpotent orbits. But others have proven elusive. One example concerns the current-algebra levels associated to the ``enhanced" flavour symmetries (enhanced relative to the manifest flavour symmetries associated to the nilpotent orbits at the punctures) of some of these SCFTs. In most cases, these levels can be determined from the current algebra levels of the manifest flavour symmetries. But the authors of \cite{Chacaltana:2014jba,Chacaltana:2015bna,Chacaltana:2017boe,Chacaltana:2018vhp} found a long list of fixtures in the class-S theories of type $E_{6,7,8}$ for which some of the levels could not be so-determined. In \S\ref{DeterminingLevels}, we rectify this by determining all of the missing levels in type $E_6$ and all but four of them in type $E_7$. Our results are tabulated in \S\ref{results} and the 
 \href{https://golem.ph.utexas.edu/class-S/E7/}{online application} for the $E_7$ theory has been updated accordingly.

The key tool which allows this progress is the Higgs-branch RG flows between SCFTs in class-S.

One well-known class of such RG flows is known as partial puncture closure, where one starts with the full puncture (which has a manifest $\mathfrak{g}$ flavour symmetry) and turns on a VEV for the highest root moment map of some $\mathfrak{su}(2)\subset \mathfrak{g}$. After decoupling some Nambu-Goldstone bosons, the resulting SCFT is the one associated to replacing the full puncture with the puncture associated to the nilpotent orbit determined (via Jacobson-Morozov) by that  $\mathfrak{su}(2)$ embedding (see \cite{Tachikawa:2015bga} for a review).

Unfortunately, that Higgsing does not yield any useful information for us. A more useful one, introduced in \cite{Beem:2019snk} and further explored in \cite{Distler-Martone} is one where we turn on a VEV for the highest root moment map for some simple factor $\mathfrak{f}_i$ in the flavour symmetry of the theory. As with partial puncture closure, the resulting RG flow is realized in the VOA as Drinfeld-Sokolov reduction \cite{Beem:2014rza, Beem:2019snk}. This has implications for the IR SCFT: certain quantities are RG-invariant and are the same in the UV and IR SCFTs; other quantities change, but in computable ways, in flowing from the UV to the IR.
 
As a second application of these ideas, we revisit a question posed in \cite{Distler:2020tub}: how can we determine whether two $\mathcal{N}=2$ SCFTs are isomorphic? That paper found examples of pairs of SCFTs whose easy-to-compute (``conventional") invariants are identical, but where the SCFTs themselves are in fact not isomorphic. That paper suggested that the global form of the flavour symmetry \emph{group}, $F$,  (supplementing the conventional invariants) might be an effective diagnostic for determining whether two theories are isomorphic.

In \S\ref{Distinguishing}, we examine this proposal further, by studying \emph{families} of pairs of theories (whose conventional invariants coincide) which are related by a sequence of Higgs branch RG flows. Generalizing the proposal of  \cite{Bhardwaj:2021ojs} for determining the global form of $F$ to include the $\mathbb{Z}_{2}$ symmetry $e^{2\pi i(R+j_1+j_2)}$ (which is the center of the $\mathcal{N}=2$ superconformal supergroup), we find that the conjecture of  \cite{Distler:2020tub} seems to hold in many examples. Unfortunately, the same technique also generates a family of counter-examples: pairs of theories whose conventional invariants \emph{and} global form of the flavour symmetry  coincide, but which are nonetheless non-isomorphic.

\section{Nilpotent Higgsings}\label{nilpotent_higgsings}

Let $\mathfrak{f}\subset \mathfrak{f}_\mathcal{T}$ be a simple subalgebra of the flavour symmetry of some SCFT, $\mathcal{T}$, of class-S. We will restrict ourselves to the case where $\mathfrak{f}$ is a simple factor in the manifest flavour symmetry associated to a puncture, $O$. Turning on a VEV for the moment map ($\hat{B}_1$ operator) which is the highest root of $\mathfrak{f}$ induces an RG flow which, in favourable circumstances \cite{Distler-Martone}, yields in the infrared another class-S SCFT, $\mathcal{T}'$, with the puncture $O$ replaced by the puncture $O'$, where the nilpotent orbit $O$ lies in the closure of the orbit $O'$. The key observation of \cite{Beem:2019snk} is that the flow from $\mathcal{T}$ to $\mathcal{T}'$ is implemented on the level of the chiral algebras as Drinfeld-Sokolov reduction.

This has several implications \cite{ Beem:2019snk,Distler-Martone}:

\begin{enumerate}%
\item The \emph{rest} of the current algebra of $\mathcal{T}$ is unmolested. In particular the flavour symmetry algebra of $\mathcal{T}'$, $\mathfrak{f}_{\mathcal{T}'}$ is some infrared enhancement of $\mathfrak{f}_\mathcal{T}/\mathfrak{f}$.

\item The rank of the theory is either preserved ($k_\mathfrak{f}$ odd) or decreases by 1 ($k_\mathfrak{f}$ even). More specifically,

\begin{itemize}%
\item When $k$ is even, the dimension of the Coulomb branch decreases by one. We lose a Coulomb branch parameter, $\varepsilon$, with $\Delta(\varepsilon)=k/2$.
\item When $k$ is odd, the dimension of the Coulomb branch is preserved, But a Coulomb branch parameter, $\varepsilon$, with $\Delta(\varepsilon)=k-1$, is replaced by $\varepsilon'$, where $\varepsilon = (\varepsilon')^2$ (i.e. $\Delta(\varepsilon')= (k-1)/2$). The nilpotent orbits in question lie in the same special piece (i.e., $d(O)=d(O')$), such that the non-special orbit is Higgsed to the special one.

\end{itemize}

\item $n_v$ decreases\footnote{This was the key observation in \cite{Beem:2019snk}, though there is was expressed in terms of $\delta c$ and $\delta\dim_\mathbb{H}(\text{Higgs})$.} by $k_\mathfrak{f}-1$.

\item The (quaternionic) dimension of the Higgs branch decreases by $h^\vee(\mathfrak{f})-1$.

\end{enumerate}
The diagram of nilpotent Higgsings for $E_6$ is

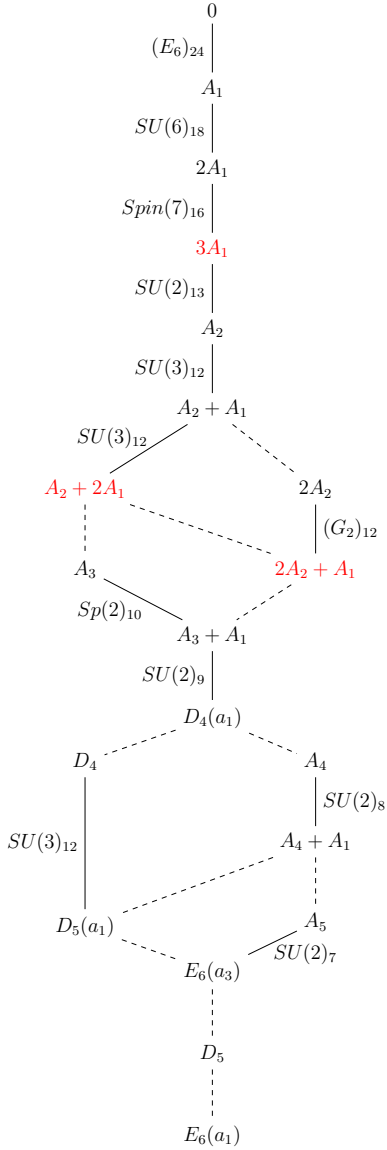
\begin{equation*}
\scalebox{.65}{
\begin{tikzpicture}\node (0) at (0,0) {$0$};
\node[below=1cm of   0] (A1) {$A_1$};
\node[below=1cm of   A1] (2A1) {$2A_1$};
\node[below=1cm of   2A1] (3A1) {${\color{red}3A_1}$};
\node[below=1cm of   3A1] (A2) {$A_2$};
\node[below=1cm of   A2] (A2A1) {$A_2+A_1$};
\node[below left=1cm and .75cm of A2A1] (A22A1)  {$\color{red}A_2+2A_1$};
\node[below right=1cm and .75cm of A2A1] (2A2)  {$2A_2$};
\node[below=1cm of   A22A1] (A3) {$A_3$};
\node[below=1cm of   2A2] (2A2A1) {$\color{red}2A_2+A_1$};
\node[below=4cm of   A2A1] (A3A1) {$A_3+A_1$};
\node[below=1cm of   A3A1] (D4a1) {$D_4(a_1)$};
\node[below=3.3cm of   A3] (D4)  {$D_4$};
\node[below=3.3cm of 2A2A1] (A4)  {$A_4$};
\node[below=1cm of   A4] (A4A1) {$A_4+A_1$};
\node[below=1cm of A4A1] (A5)  {$A_5$};
\node[below=2.7cmof D4] (D5a1)  {$D_5(a_1)$};
\node[below=4.5cm of   D4a1] (E6a3) {$E_6(a_3)$};
\node[below=1cm of   E6a3] (D5) {$D_5$};
\node[below=1cm of   D5] (E6a1) {$E_6(a_1)$};
\path (0) edge node[left] {$(E_6)_{24}$}  (A1)
(A1) edge node[left] {$SU(6)_{18}$}  (2A1)
(2A1) edge node[left] {$Spin(7)_{16}$}  (3A1)
(3A1) edge node[left] {$SU(2)_{13}$}  (A2)
(A2) edge node[left] {$SU(3)_{12}$}  (A2A1)
(A2A1) edge node[above left=-.125cm and -.125cm] {$SU(3)_{12}$} (A22A1)
(A2A1) edge[dashed] (2A2)
(2A2) edge node[right] {$(G_2)_{12}$}  (2A2A1)
%(A22A1) edge node[left] {${\color{blue}SU(2)_{34}}$}  (A3)
(A22A1) edge [dashed]  (A3)
(A22A1) edge [dashed]  (2A2A1)
(A3) edge node[below left=-.125cm and -.125cm] {$Sp(2)_{10}$} (A3A1)
%(2A2A1) edge node[below right=-.125cm and -.125cm] {${\color{blue}SU(2)_{17}}$} (A3A1)
(2A2A1) edge [dashed] (A3A1)
(A3A1) edge node[left] {$SU(2)_9$}  (D4a1)
(D4a1) edge [dashed]  (D4)
(D4a1) edge [dashed]  (A4)
(A4) edge node[right] {$SU(2)_8$} (A4A1)
(D4) edge node[left] {$SU(3)_{12}$} (D5a1)
(A4A1) edge [dashed]  (D5a1)
(A4A1) edge [dashed]  (A5)
(D5a1) edge [dashed]  (E6a3)
(A5) edge node[below right=-.125cm and -.125cm] {$SU(2)_7$} (E6a3)
(E6a3) edge [dashed]  (D5)
(D5) edge [dashed]  (E6a1)
%(3A1) edge[bend left,blue] node[right] {$\color{blue}SU(3)_{24}$} (A2A1)
;
\end{tikzpicture}
}
\end{equation*}

This is just a decorated version of the Hasse diagram for nilpotent orbit closure, flipped on its head and decorated by the $\mathfrak{f}_k$ which get Higgsed at each stage. A nilpotent Higgsing from $O$ to $O'$ is denoted by a downward-directed solid edge from $O$ to $O'$. A dashed edge indicates that the nilpotent orbit $O$ is contained in the closure of the orbit $O'$, though there's no corresponding nilpotent Higgsing. Some of the nilpotent orbits in the diagram are denoted in red. These have a simple factor in their flavour symmetry (e.g.~$SU(3)_{24}\subset SU(3)_{24}\times SU(2)_{13}$, in the case of $\color{red}3A_1$) which cannot be Higgsed in this fashion. More precisely, the Higgsing (whether or not it yields a nontrivial SCFT in the IR) does not keep us within this family of class-S theories.

The solid edges in the diagram are the ones which correspond to minimal singularities in the work of \cite{MR3570131}; the transverse slice to the singularity is the minimal nilpotent orbit for the (complexified) Lie algebra listed. In the classical Lie algebras \cite{MR604841,MR694606}, the transverse slices are either minimal nilpotent orbits or ADE surface singularities $\mathbb{C}^2/\Gamma$. For nilpotent orbit closures in the exceptional Lie algebras, the classification is more complicated \cite{MR3570131}.

While the dimension of the minimal nilpotent orbit in $\mathfrak{f}$ is also $h^\vee(\mathfrak{f})-1$, there isn't an a-priori connection between our Higgsings and the nilpotent orbit closures considered by Kraft-Procesi and Fu \emph{et al}. Here, when we Higgs $O\xrightarrow{\; (F)_k\;}O'$, the dimension of the Higgs branch \emph{decreases} by $h^\vee(\mathfrak{f})-1$. By contrast, the dimension of the nilpotent orbit $\dim(O')$ is \emph{greater} than $\dim(O)$. For these particular orbits, $\dim_\mathbb{H}(O') = \dim_\mathbb{H}(O) + (h^\vee(\mathfrak{f})-1)$ and the infinitesimal neighbourhood of a generic point in $O$ is isomorphic to the minimal nilpotent orbit of $\mathfrak{f}$.

In the twisted sector of $E_6$, punctures are labeled by nilpotent orbits in $F_4$. The associated diagram of nilpotent Higgsings is

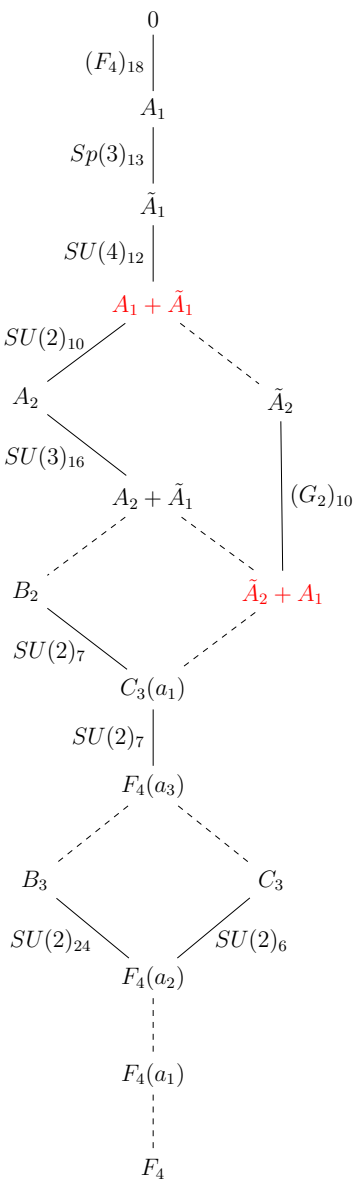
\begin{equation*}
\scalebox{.75}{
\begin{tikzpicture}
\node (0) at (0,0) {$0$};
\node[below=1cm of   0] (A1) {$A_1$};
\node[below=1cm of   A1] (tA1) {$\tilde{A}_1$};
\node[below=1cm of   tA1] (A1ptA1)  {${\color{red}A_1+\tilde{A}_1}$};
\node[below left=1cm and 1cm of A1ptA1] (A2)  {$A_2$};
\node[below right=1cm and 1cm of A1ptA1] (tA2)  {$\tilde{A}_2$};
\node[below=2.7cm of   A1ptA1] (A2tA1) {$A_2+\tilde{A}_1$};
\node[below left=1cm and 1cm of A2tA1] (B2)  {$B_2$};
\node[right=3.3cm  of B2] (tA2A1)  {${\color{red}\tilde{A}_2+A_1}$};
\node[below=2.7cm of   A2tA1] (C3a1) {$C_3(a_1)$};
\node[below=1cm of   C3a1] (F4a3) {$F_4(a_3)$};
\node[below left=1cm and 1cm of F4a3] (B3)  {$B_3$};
\node[below right=1cm and 1cm of F4a3] (C3)  {$C_3$};
\node[below=2.7cm of   F4a3] (F4a2) {$F_4(a_2)$};
\node[below=1cm of   F4a2] (F4a1) {$F_4(a_1)$};
\node[below=1cm of   F4a1] (F4) {$F_4$};
\path (0) edge node[left] {$(F_4)_{18}$}  (A1)
(A1) edge node[left] {$Sp(3)_{13}$}  (tA1)
(tA1) edge node[left] {$SU(4)_{12}$}  (A1ptA1)
(A1ptA1) edge node[above left=-.125cm and -.125cm] {$SU(2)_{10}$} (A2)
(A1ptA1) edge[dashed] (tA2)
(A2) edge node[below left=-.125cm and -.125cm] {$SU(3)_{16}$} (A2tA1)
(tA2) edge node[right] {$(G_2)_{10}$}  (tA2A1)
(A2tA1)edge[dashed](B2)
(A2tA1)edge[dashed] (tA2A1)
(tA2A1)edge [dashed] (C3a1)
(B2) edge node[below left=-.125cm and -.125cm] {$SU(2)_7$} (C3a1)
(C3a1) edge node[left] {$SU(2)_{7}$} (F4a3)
(F4a3) edge[dashed] (B3)
(F4a3) edge[dashed] (C3)
(B3) edge node[below left=-.125cm and -.125cm] {$SU(2)_{24}$} (F4a2)
(C3) edge node[below right=-.125cm and -.125cm] {$SU(2)_6$} (F4a2)
(F4a2) edge[dashed] (F4a1)
(F4a1) edge[dashed] (F4)
;
\end{tikzpicture}
}
\end{equation*}

The diagram of nilpotent Higgsings for $E_7$ is

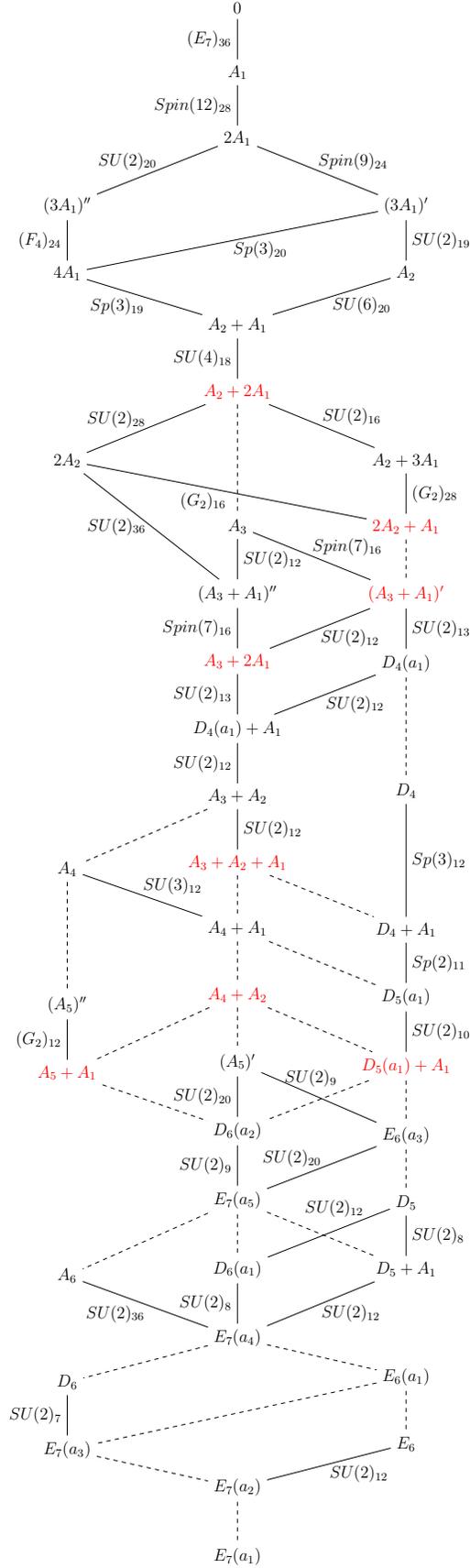
\begin{equation}\label{E7Hasse}
\begin{matrix}\scalebox{.585}{
\begin{tikzpicture}
\node (0) at (0,0) {$0$};
\node[below=1cm of 0] (A1) {$A_1$};
\node[below=1cm of A1] (A1A1) {$2A_1$};
\node[below left=1cm and 3cm of A1A1] (3A1pp) {$(3A_1)''$};
\node[below right=1cm and 3cm of A1A1] (3A1p) {$(3A_1)'$};
\node[below=1cm of 3A1pp] (A1A1A1A1) {$4A_1$};
\node[below=1cm of 3A1p] (A2) {$A_2$};
\node[below=4cm of A1A1] (A2A1) {$A_2+A_1$};
\node[below=1cm of A2A1] (A2A1A1) {${\color{red}A_2+2A_1}$};
\node[below=4cm of A1A1A1A1] (A2A2) {$2A_2$};
\node[below=4cm of A2] (A2A1A1A1) {$A_2+3A_1$};
\node[below=2.7cm of A2A1A1] (A3) {$A_3$};
\node[below=1cm of A2A1A1A1] (A2A2A1) {${\color{red}2A_2+A_1}$};
\node[below=1cm of A3] (A3A1pp) {$(A_3+A_1)''$};
\node[below=1cm of A2A2A1] (A3A1p) {${\color{red}(A_3+A_1)'}$};
\node[below=1cm of A3A1pp] (A3A1A1) {${\color{red}A_3+2A_1}$};
\node[below=1cm of A3A1p] (D4a1) {$D_4(a_1)$};
\node[below=1cm of A3A1A1] (D4a1A1) {$D_4(a_1)+A_1$};
\node[below=1cm of D4a1A1] (A3A2) {$A_3+A_2$};
\node[below=2.5cm of D4a1] (D4) {$D_4$};
\node[below=9.5cm of A2A2] (A4) {$A_4$};
\node[below=1cm of A3A2] (A3A2A1) {${\color{red}A_3+A_2+A_1}$};
\node[below=1cm of A3A2A1] (A4A1) {$A_4+A_1$};
\node[below=2.8cm of D4] (D4A1) {$D_4+A_1$};
\node[below=2.7cm of A4] (A5pp) {$(A_5)''$};
\node[below=1cm of A4A1] (A4A2) {${\color{red}A_4+A_2}$};
\node[below=1cm of D4A1] (D5a1) {$D_5(a_1)$};
\node[below=1cm of A5pp] (A5A1) {${\color{red}A_5+A_1}$};
\node[below=1cm of A4A2] (A5p) {$(A_5)'$};
\node[below=1cm of D5a1] (D5a1A1) {${\color{red}D_5(a_1)+A_1}$};
\node[below=1cm of A5p] (D6a2) {$D_6(a_2)$};
\node[below=1cm of D5a1A1] (E6a3) {$E_6(a_3)$};
\node[below=1cm of D6a2] (E7a5) {$E_7(a_5)$};
\node[below=1cm of E6a3] (D5) {$D_5$};
\node[below=4.4cm of A5A1] (A6) {$A_6$};
\node[below=1cm of E7a5] (D6a1) {$D_6(a_1)$};
\node[below=1cm of D5] (D5A1) {$D_5+A_1$};
\node[below=1cm of D6a1] (E7a4) {$E_7(a_4)$};
\node[below=2cm of A6] (D6) {$D_6$};
\node[below=2cm of D5A1] (E6a1) {$E_6(a_1)$};
\node[below=1cm of D6] (E7a3) {$E_7(a_3)$};
\node[below=1cm of E6a1] (E6) {$E_6$};
\node[below=3cm of E7a4] (E7a2) {$E_7(a_2)$};
\node[below=1cm of E7a2] (E7a1) {$E_7(a_1)$};
\path (0) edge node[left] {$(E_7)_{36}$} (A1)
(A1) edge node[left] {$Spin(12)_{28}$} (A1A1)
(A1A1) edge node[above left=-.125cm and -.125cm] {$SU(2)_{20}$} (3A1pp)
(A1A1) edge node[above right=-.125cm and -.125cm] {$Spin(9)_{24}$} (3A1p)
(3A1pp) edge node[left] {$(F_4)_{24}$} (A1A1A1A1)
(3A1p) edge node[below right=-.125cm and -.125cm] {$Sp(3)_{20}$} (A1A1A1A1)
(3A1p) edge node[right] {$SU(2)_{19}$} (A2)
(A1A1A1A1) edge node[below left=-.125cm and -.125cm] {$Sp(3)_{19}$} (A2A1)
(A2) edge node[below right=-.125cm and -.125cm] {$SU(6)_{20}$} (A2A1)
(A2A1) edge node[left] {$SU(4)_{18}$} (A2A1A1)
(A2A1A1) edge node[above left=-.125cm and -.125cm] {$SU(2)_{28}$} (A2A2)
(A2A1A1) edge node[above right=-.125cm and -.125cm] {$SU(2)_{16}$} (A2A1A1A1)
(A2A1A1) edge[dashed] (A3)
(A2A2) edge node[below left=-.125cm and -.125cm] {$(G_2)_{16}$} (A2A2A1)
(A2A2) edge node[left] {$SU(2)_{36}$} (A3A1pp)
(A2A1A1A1) edge node[right] {$(G_2)_{28}$} (A2A2A1)
(A3) edge node[right] {$SU(2)_{12}$} (A3A1pp)
(A3) edge node[above right=-.125cm and -.125cm] {$Spin(7)_{16}$} (A3A1p)
(A2A2A1) edge[dashed] (A3A1p)
(A3A1pp) edge node[left] {$Spin(7)_{16}$} (A3A1A1)
(A3A1p) edge node[below right=-.125cm and -.125cm] {$SU(2)_{12}$} (A3A1A1)
(A3A1p) edge node[right] {$SU(2)_{13}$} (D4a1)
(A3A1A1) edge node[left] {$SU(2)_{13}$} (D4a1A1)
(D4a1) edge node[below right=-.125cm and -.125cm] {$SU(2)_{12}$} (D4a1A1)
(D4a1) edge[dashed] (D4)
(D4a1A1) edge node[left] {$SU(2)_{12}$} (A3A2)
(A3A2) edge[dashed] (A4)
(A3A2) edge node[right] {$SU(2)_{12}$} (A3A2A1)
(D4) edge node[right] {$Sp(3)_{12}$} (D4A1)
(A4) edge[dashed] (A5pp)
(A4) edge node[above right=-.125cm and -.125cm] {$SU(3)_{12}$} (A4A1)
(A3A2A1) edge[dashed] (A4A1)
(A3A2A1) edge[dashed] (D4A1)
(A4A1) edge[dashed] (A4A2)
(A4A1) edge[dashed] (D5a1)
(D4A1) edge node[right] {$Sp(2)_{11}$} (D5a1)
(A5pp) edge node[left] {$(G_2)_{12}$} (A5A1)
(A4A2) edge[dashed] (A5A1)
(A4A2) edge[dashed] (A5p)
(A4A2) edge[dashed] (D5a1A1)
(D5a1) edge node[right] {$SU(2)_{10}$} (D5a1A1)
(A5A1) edge[dashed] (D6a2)
(A5p) edge node[left] {$SU(2)_{20}$} (D6a2)
(A5p) edge node[above right=.125cm and -1cm] {$SU(2)_9$} (E6a3)
(D5a1A1) edge[dashed] (D6a2)
(D5a1A1) edge[dashed] (E6a3)
(D6a2) edge node[left] {$SU(2)_{9}$} (E7a5)
(E6a3) edge node[above left=-.125cm and -.125cm] {$SU(2)_{20}$} (E7a5)
(E6a3) edge[dashed] (D5)
(E7a5) edge[dashed] (A6)
(E7a5) edge[dashed] (D6a1)
(E7a5) edge[dashed] (D5A1)
(D5) edge node[above left=.25cm and -1cm] {$SU(2)_{12}$} (D6a1)
(D5) edge node[right] {$SU(2)_{8}$} (D5A1)
(A6) edge node[below left=-.125cm and -.125cm] {$SU(2)_{36}$} (E7a4)
(D6a1) edge node[left] {$SU(2)_{8}$} (E7a4)
(D5A1) edge node[below right=-.125cm and -.125cm] {$SU(2)_{12}$} (E7a4)
(E7a4) edge[dashed] (D6)
(E7a4) edge[dashed] (E6a1)
(D6) edge node[left] {$SU(2)_{7}$} (E7a3)
(E6a1) edge[dashed] (E7a3)
(E6a1) edge[dashed] (E6)
(E7a3) edge[dashed] (E7a2)
(E6) edge node[below right=-.125cm and -.125cm] {$SU(2)_{12}$} (E7a2)
(E7a2) edge[dashed] (E7a1)
;
\end{tikzpicture}
}\end{matrix}
\end{equation}

The diagram for $E_8$ is

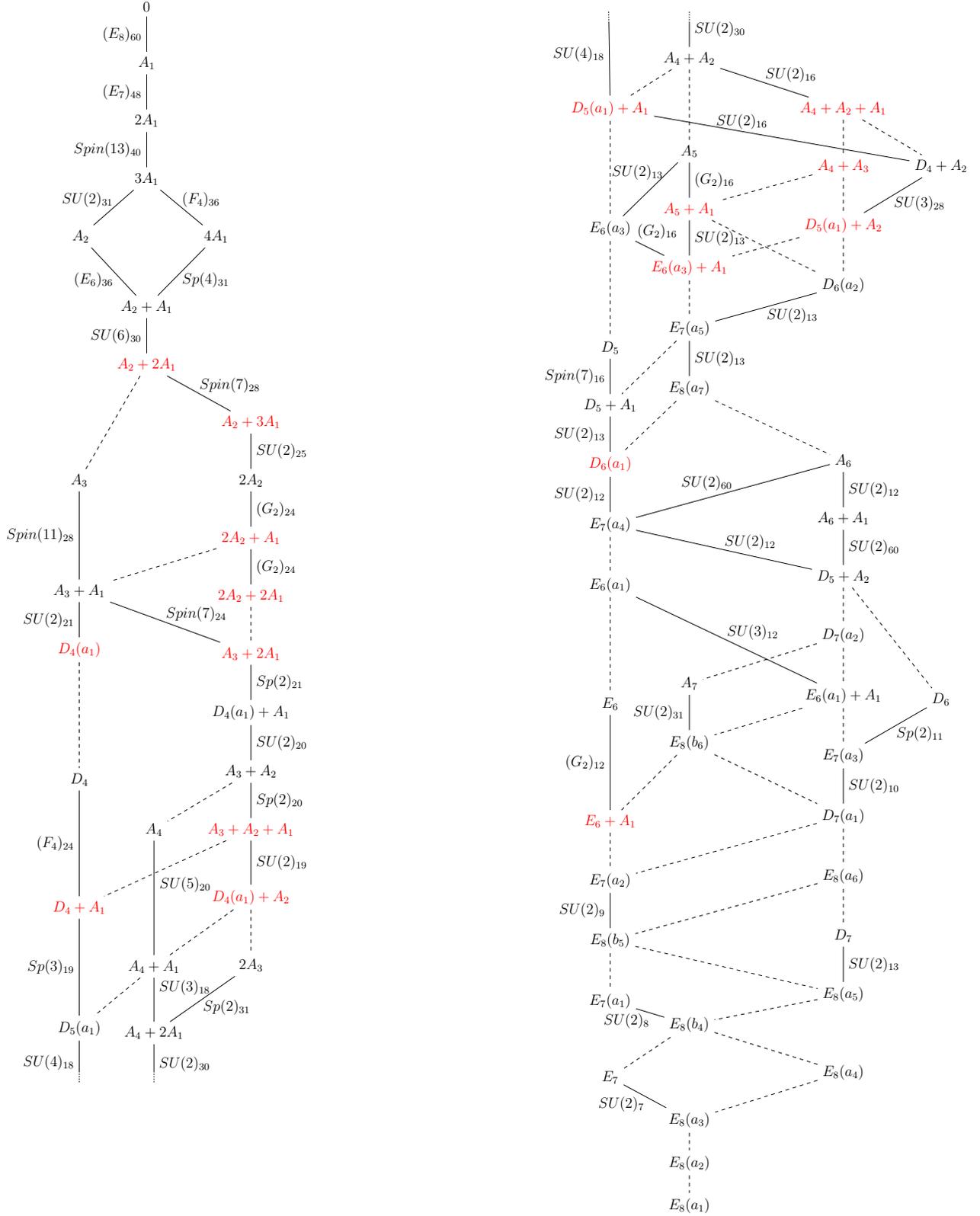
\begin{equation*}
\scalebox{.615}{
\begin{tikzpicture}
\node (0) at (0,0) {$0$};
\node[below=1cm of   0] (A1) {$A_1$};
\node[below=1cm of  A1] (2A1)  {$2A_1$};
\node[below=1cm of 2A1] (3A1)  {$3A_1$};
\node[below left=1cm and 1cm of 3A1] (A2)  {$A_2$};
\node[below right=1cm and 1cm of 3A1] (4A1)  {$4A_1$};
\node[below=3cm of 3A1] (A2A1)  {$A_2+A_1$};
\node[below=1cm of A2A1] (A2A1A1)  {${\color{red}A_2+2A_1}$};
\node[below right=1cm and 1cm of A2A1A1] (A2A1A1A1)  {${\color{red}A_2+3A_1}$};
\node[below=1cm of A2A1A1A1] (A2A2)  {$2A_2$};
\node[left=4cm of A2A2] (A3) {$A_3$};
\node[below=1cm of A2A2] (A2A2A1)  {${\color{red}2A_2+A_1}$};
\node[below=1cm of A2A2A1] (A2A2A1A1)  {${\color{red}2A_2+2A_1}$};
\node[below=2.5cm of A3] (A3A1) {$A_3+A_1$};
\node[below=1cm of A2A2A1A1] (A3A1A1)  {${\color{red}A_3+2A_1}$};
\node[below=1cm of A3A1A1] (D4a1A1)  {$D_4(a_1)+A_1$};
\node[below=1cm of D4a1A1] (A3A2)  {$A_3+A_2$};
\node[below=1cm of A3A2] (A3A2A1)  {${\color{red}A_3+A_2+A_1}$};
\node[below=1.25cm of A3A2A1] (D4a1A2) {${\color{red}D_4(a_1)+A_2}$};
\node[below=1.25cm of D4a1A2] (A3A3) {$2A_3$};
\node[left=1cm of A3A2A1] (A4)  {$A_4$};
\node[below=3.25cm of A4] (A4A1) {$A_4+A_1$};
\node[below=1.25cm of A4A1] (A4A1A1) {$A_4+2A_1$};
\node[below=1.1cm of A4A1A1] (A4A2dummy) {};
\node[right=15cm of 0] (A4A1A1dummy) {};
\node[below=1cm of A4A1A1dummy] (A4A2) {$A_4+A_2$};
\node[below=1cm of A3A1] (D4a1) {${\color{red}D_4(a_1)}$};
\node[below=3cm of D4a1] (D4) {$D_4$};
\node[below=3cm of D4] (D4A1) {${\color{red}D_4+A_1}$};
\node[below=2.75cm of D4A1] (D5a1) {$D_5(a_1)$};
\node[below=1.25cm of D5a1] (D5a1A1dummy) {};
\node[left=2cm of A4A1A1dummy] (D5a1dummy) {};
\node[below left=.75cm and .125cm of A4A2] (D5a1A1) {${\color{red}D_5(a_1)+A_1}$};
\node[right=4cm of D5a1A1] (A4A2A1) {${\color{red}A_4+A_2+A_1}$};
\node[below=2cm of A4A2] (A5) {$A_5$};
\node[below=1cm of A4A2A1] (A4A3) {${\color{red}A_4+A_3}$};
\node[right=1cm of A4A3] (D4A2) {$D_4+A_2$};
\node[below=2.7cm of D5a1A1] (E6a3) {$E_6(a_3)$};
\node[below=1cm of A5] (A5A1) {${\color{red}A_5+A_1}$};
\node[below=1cm of A4A3] (D5a1A2) {${\color{red}D_5(a_1)+A_2}$};
\node[below=1cm of A5A1] (E6a3A1) {${\color{red}E_6(a_3)+A_1}$};
\node[below=1cm of D5a1A2] (D6a2) {$D_6(a_2)$};
\node[below=2.7cm of E6a3] (D5) {$D_5$};
\node[below=1cm of E6a3A1] (E7a5) {$E_7(a_5)$};
\node[below=1cm of D5] (D5A1) {$D_5+A_1$};
\node[below=1cm of E7a5] (E8a7) {$E_8(a_7)$};
\node[below=1cm of D5A1] (D6a1) {${\color{red}D_6(a_1)}$};
\node[below=4.3cm of D6a2] (A6) {$A_6$};
\node[below=1cm of D6a1] (E7a4) {$E_7(a_4)$};
\node[below=1cm of A6] (A6A1) {$A_6+A_1$};
\node[below=1cm of E7a4] (E6a1) {$E_6(a_1)$};
\node[below=1cm of A6A1] (D5A2) {$D_5+A_2$};
\node[below=1cm of D5A2] (D7a2) {$D_7(a_2)$};
\node[below=2.7cm of E6a1] (E6) {$E_6$};
\node[below=7.7cm of E8a7] (A7) {$A_7$};
\node[below=1cm of D7a2] (E6a1A1) {$E_6(a_1)+A_1$};
\node[below=14.5cm of D4A2] (D6) {$D_6$};
\node[below=1cm of A7] (E8b6) {$E_8(b_6)$};
\node[below=1cm of E6a1A1] (E7a3) {$E_7(a_3)$};
\node[below=2.7cm of E6] (E6A1) {${\color{red}E_6+A_1}$};
\node[below=1cm of E7a3] (D7a1) {$D_7(a_1)$};
\node[below=1cm of E6A1] (E7a2) {$E_7(a_2)$};
\node[below=1cm of D7a1] (E8a6) {$E_8(a_6)$};
\node[below=1cm of E7a2] (E8b5) {$E_8(b_5)$};
\node[below=1cm of E8a6] (D7) {$D_7$};
\node[below=1cm of E8b5] (E7a1) {$E_7(a_1)$};
\node[below=1cm of D7] (E8a5) {$E_8(a_5)$};
\node[below=7.3cm of E8b6] (E8b4) {$E_8(b_4)$};
\node[below=1.5cm of E7a1] (E7) {$E_7$};
\node[below=1.5cm of E8a5] (E8a4) {$E_8(a_4)$};
\node[below=2cm of E8b4] (E8a3) {$E_8(a_3)$};
\node[below=.5cm of E8a3] (E8a2) {$E_8(a_2)$};
\node[below=.5cm of E8a2] (E8a1) {$E_8(a_1)$};
\path (0) edge node[left] {$ (E_8)_{60}$}  (A1)
     (A1) edge node[left] {$ (E_7)_{48}$} (2A1)
    (2A1) edge node[left] {$Spin(13)_{40}$} (3A1)
    (3A1) edge node[above left=-.125cm and -.125cm] {$ SU(2)_{31}$} (A2)
    (3A1) edge node[above right=-.125cm and -.125cm] {$ (F_4)_{36}$} (4A1)
     (A2) edge node[below left=-.125cm and -.125cm] {$ (E_6)_{36}$} (A2A1)
    (4A1) edge node[below right=-.125cm and -.125cm] {$ Sp(4)_{31}$} (A2A1)
   (A2A1) edge node[left] {$ SU(6)_{30}$} (A2A1A1)
 (A2A1A1) edge node[above right=-.125cm and -.125cm] {$ Spin(7)_{28}$} (A2A1A1A1)
(A2A1A1A1) edge node[right] {$ SU(2)_{25}$} (A2A2)
   (A2A2) edge node[right] {$(G_2)_{24}$} (A2A2A1)
 (A2A1A1) edge[dashed] (A3)
     (A3) edge node[left] {$Spin(11)_{28}$} (A3A1)
 (A2A2A1) edge[dashed] (A3A1)
 (A2A2A1) edge node[right] {$(G_2)_{24}$} (A2A2A1A1)
(A2A2A1A1) edge[dashed] (A3A1A1)
   (A3A1) edge node[above right=-.125cm and -.125cm] {$Spin(7)_{24}$} (A3A1A1)
   (A3A1) edge node[left] {$SU(2)_{21}$} (D4a1)
 (A3A1A1) edge node[right] {$Sp(2)_{21}$} (D4a1A1)
 (D4a1A1) edge node[right] {$SU(2)_{20}$} (A3A2)
   (D4a1) edge[dashed] (D4)
   (A3A2) edge node[right] {$Sp(2)_{20}$} (A3A2A1)
   (A3A2) edge[dashed] (A4)
     (D4) edge node[left] {$(F_4)_{24}$} (D4A1)
 (A3A2A1) edge[dashed] (D4A1)
 (A3A2A1) edge node[right] {$SU(2)_{19}$} (D4a1A2)
 (D4a1A2) edge[dashed] (A4A1)
     (A4) edge node[above right=.025cm] {$SU(5)_{20}$}(A4A1)
 (D4a1A2) edge[dashed] (A3A3)
   (A4A1) edge node[above right=0cm and 0cm] {$SU(3)_{18}$}(A4A1A1)
   (A4A1) edge[dashed] (D5a1)
  (D4A1) edge node[left] {$Sp(3)_{19}$}(D5a1)
  (A3A3) edge node[below right=-.125cm and -.125cm] {$Sp(2)_{31}$}(A4A1A1)
  (A4A1A1) edge[dash pattern=on .8cm off 1pt on 1pt  off 1pt on 1pt  off 1pt on 1pt  off 1pt on 1pt ] node[right] {$SU(2)_{30}$}(A4A2dummy)
  (A4A1A1dummy) edge[dash pattern=on 1pt  off 1pt on 1pt  off 1pt on 1pt  off 1pt on 1pt  off 1pt on 2cm] node[right] {$SU(2)_{30}$}(A4A2)
  (D5a1) edge[dash pattern=on .9cm off 1pt on 1pt  off 1pt on 1pt  off 1pt on 1pt  off 1pt on 1pt ] node[left] {$SU(4)_{18}$} (D5a1A1dummy)
    (D5a1dummy) edge[dash pattern=on 1pt  off 1pt on 1pt  off 1pt on 1pt  off 1pt on 1pt  off 1pt on 2.1cm] node[left] {$SU(4)_{18}$} (D5a1A1)
  (A4A2) edge[dashed] (D5a1A1)
  (A4A2) edge[dashed] (A5)
  (A4A2) edge node[above right=-.125cm and -.125cm] {$SU(2)_{16}$} (A4A2A1)
(D5a1A1) edge[dashed] (E6a3)
(D5a1A1) edge node[above right=.125cm and -2cm] {$SU(2)_{16}$} (D4A2)
(A4A2A1) edge[dashed] (A4A3)
(A4A2A1) edge[dashed] (D4A2)
(A5)edge node[above left=.125cm and -.5cm] {$SU(2)_{13}$} (E6a3)
(A5)edge node[right] {$(G_2)_{16}$} (A5A1)
(A4A3) edge[dashed] (A5A1)
(A4A3) edge[dashed] (D5a1A2)
(D4A2)edge node[below right=-.125cm and -.125cm] {$SU(3)_{28}$} (D5a1A2)
(E6a3) edge[dashed] (D5)
(E6a3)edge node[above right=.125cm and -.5cm] {$(G_2)_{16}$} (E6a3A1)
(A5A1)edge node[right] {$SU(2)_{13}$} (E6a3A1)
(A5A1) edge[dashed] (D6a2)
(D5a1A2) edge[dashed] (E6a3A1)
(D5a1A2) edge[dashed] (D6a2)
(E6a3A1) edge[dashed] (E7a5)
(D6a2)edge node[below right=-.125cm and -.125cm] {$SU(2)_{13}$} (E7a5)
(D5) edge node[left] {$Spin(7)_{16}$} (D5A1)
(E7a5) edge[dashed] (D5A1)
(E7a5)edge node[right] {$SU(2)_{13}$} (E8a7)
(D5A1)edge node[left] {$SU(2)_{13}$} (D6a1)
(E8a7)edge[dashed](D6a1)
(E8a7)edge[dashed](A6)
(D6a1)edge node[left] {$SU(2)_{12}$} (E7a4)
(A6)edge node[above left=-.125cm and -.125cm] {$SU(2)_{60}$} (E7a4)
(A6)edge node[right] {$SU(2)_{12}$} (A6A1)
(E7a4)edge[dashed] (E6a1)
(E7a4)edge node[above right=-.125cm and -.125cm] {$SU(2)_{12}$}  (D5A2)
(A6A1)edge node[right] {$SU(2)_{60}$}(D5A2)
(E6a1) edge[dashed](E6)
(E6a1) edge node[above right=-.125cm and -.125cm] {$SU(3)_{12}$}(E6a1A1)
(D5A2) edge[dashed] (D7a2)
(D5A2) edge[dashed] (D6)
(D7a2) edge[dashed] (A7)
(D7a2) edge[dashed] (E6a1A1)
(E6) edge node[left] {$(G_2)_{12}$} (E6A1)
(A7) edge node[left] {$SU(2)_{31}$} (E8b6)
(E6a1A1)edge[dashed] (E8b6)
(E6a1A1)edge[dashed] (E7a3)
(D6) edge node[below right=-.125cm and -.125cm] {$Sp(2)_{11}$} (E7a3)
(E8b6) edge[dashed] (E6A1)
(E8b6) edge[dashed] (D7a1)
(E7a3)edge node[right] {$SU(2)_{10}$}(D7a1)
(E6A1)edge[dashed](E7a2)
(D7a1)edge[dashed](E7a2)
(D7a1)edge[dashed](E8a6)
(E7a2)edge node[left] {$SU(2)_{9}$}(E8b5)
(E8a6)edge[dashed](E8b5)
(E8a6)edge[dashed](D7)
(E8b5)edge[dashed](E7a1)
(E8b5)edge[dashed](E8a5)
(D7)edge node[right] {$SU(2)_{13}$}(E8a5)
(E7a1)edge node[below left=-.125cm and -.125cm]{$SU(2)_{8}$}(E8b4)
(E8a5)edge[dashed](E8b4)
(E8b4)edge[dashed](E7)
(E8b4)edge[dashed](E8a4)
(E7)edge node[below left=-.125cm and -.125cm]{$SU(2)_{7}$}(E8a3)
(E8a4)edge[dashed](E8a3)
(E8a3)edge[dashed](E8a2)
(E8a2)edge[dashed](E8a1)
;
\end{tikzpicture}
}
\end{equation*}

\section{Determining Unknown Current-Algebra Levels}\label{DeterminingLevels}

Fixtures are class-S theories obtained by compactifying the (2,0) theory on a sphere with three punctures. Each puncture has an associated flavour symmetry and the fixture has a flavour symmetry which is (possibly an enhancement of) the product of the flavour symmetries associated to each puncture. The latter (the ``manifest" flavour symmetry) embeds as a subalgebra of the full flavour symmetry. The flavour symmetries of fixtures obtained from the exceptional (2,0) theories we will discuss were obtained in \cite{Chacaltana:2014jba}, \cite{Chacaltana:2015bna}, and \cite{Chacaltana:2017boe}.

To each simple flavour symmetry factor one may associate a positive integer $k$ that is the flavour central charge or ``level". The flavour central charge of a simple factor of the flavour symmetry is defined by
\begin{displaymath}
J^{a}_{\mu}(x)J^{b}_{\nu}(0) \sim \frac{3k}{4 \pi ^4} \delta^{ab}\frac{x^2 g_{\mu \nu}-2x_{\mu}x_{\nu}}{x^8}+\frac{2}{\pi^2}f^{abc}\frac{x_{\mu}x_{\nu}x \cdot J^{c}(0)}{x^6}+\dots
\end{displaymath}
where the normalization is such that $k=1$ for a free half-hypermultiplet in the defining representation of $Sp(n)$. The levels of the manifest flavour symmetries are readily determined from the decomposition of the adjoint representation \cite{Chacaltana:2012zy}, as described in \S2.4.1 of \cite{Chacaltana:2014jba}.

In most cases, when the flavour symmetry is enhanced, knowing the levels of the ``manifest" subalgebra suffices to determine the levels of the full flavour symmetry. There are, however, two notable exceptions.
\begin{itemize}
\item When a manifest factor of $G_k$ (we denote the level $k$ by a subscript) is enhanced to $G_{k_1}\times G_{k_2}$, where $G$ is embedded diagonally in $G\times G$. We know that $k_1+k_2=k$, but --- without more information --- we cannot determine $k_{1,2}$ individually.
\item When a manifest $U(1)$ factor is enhanced to some nonabelian $G_k$, the freedom to change the normalization of the $U(1)$ generator prevents us, in most cases, from being able to compute the level $k$.
\end{itemize}

There are a number of fixtures with unknown levels in the exceptional case. In this paper we will determine the vast majority of these levels in the twisted and untwisted $E_6$ case as well as the $E_7$ case. There are four fixtures in the $E_7$ theory that have unknown levels that could not be determined with our methods. However our results do place constraints on some of these levels as well as relate the unknown levels of different fixtures. Finally, we note that when one of the punctures is a simple puncture ($E_n(a_1)$ in the $E_n$ theory), the levels can also be determined using the methods of \cite{Baume:2021qho}. In those cases, our results are consistent with those.

\subsection{The $E_6$ Theory}\label{the_E6_theory}

There were two fixtures in the untwisted $E_6$ theory with unknown levels

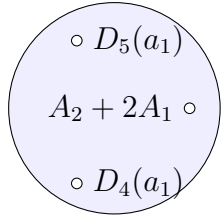
\begin{equation*}
\begin{tikzpicture}
\draw[radius=40pt,fill=lightblue] circle;
\draw[radius=2pt,fill=white]  (-.5,.9) circle node[right=2pt] {$D_5(a_1)$};
\draw[radius=2pt,fill=white]  (-.5,-1) circle node[right=2pt] {$D_4(a_1)$};
\draw[radius=2pt,fill=white]  (1,0) circle node[left=2pt] {$A_2+2A_1$};
\node at (0,-2) {$[{SU(3)}_{54-k_1-k_2}\times{SU(3)}_{k_1}\times{SU(3)}_{k_2}\times U(1)]$};
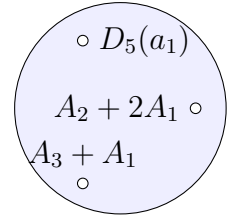
\draw[radius=40pt,fill=lightblue] (9,0) circle;
\draw[radius=2pt,fill=white]  (8.5,.9) circle node[right=2pt] {$D_5(a_1)$};
\draw[radius=2pt,fill=white]  (8.5,-1) circle node[above=2pt] {$A_3+A_1$};
\draw[radius=2pt,fill=white]  (10,0) circle node[left=2pt] {$A_2+2A_1$};
\node at (9,-2) {$[{SU(3)}_{54-k}\times{SU(3)}_k\times{SU(2)}_9\times U(1)]$};
\end{tikzpicture}
\end{equation*}
These were fixtures \#63,\#66 in section 3.4 of \cite{Chacaltana:2014jba}. In both cases, the manifest $SU(2)_{54}$ of the $A_2+2A_1$ puncture was enhanced to $SU(3)_{54}$ and thence to a product of $SU(3)$s. Hence we know the sum of the $SU(3)$ levels, but not the levels themselves. These fixtures can be obtained by Higgsing
\begin{displaymath}
A_2+A_1\xrightarrow{\quad {SU(3)}_{12}\quad} A_2+2A_1
\end{displaymath}
in fixtures \#64,\#67:

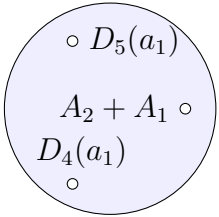
\begin{equation*}
\begin{tikzpicture}\draw[radius=40pt,fill=lightblue] circle;
\draw[radius=2pt,fill=white]  (-.5,.9) circle node[right=2pt] {$D_5(a_1)$};
\draw[radius=2pt,fill=white]  (-.5,-1) circle node[above right=2pt and -18pt] {$D_4(a_1)$};
\draw[radius=2pt,fill=white]  (1,0) circle node[left=2pt] {$A_2+A_1$};
\node at (0,-2) {$[{SU(3)}_{12}\times{SU(2)}_{18}^3\times {U(1)}^3]$};
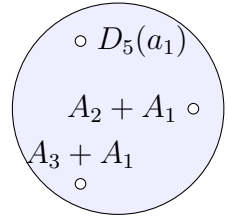
\draw[radius=40pt,fill=lightblue] (9,0) circle;
\draw[radius=2pt,fill=white]  (8.5,.9) circle node[right=2pt] {$D_5(a_1)$};
\draw[radius=2pt,fill=white]  (8.5,-1) circle node[above=2pt] {$A_3+A_1$};
\draw[radius=2pt,fill=white]  (10,0) circle node[left=2pt] {$A_2+A_1$};
\node at (9,-2) {$[{SU(3)}_{12}\times{SU(2)}_{36}\times{SU(2)}_{18}\times{SU(2)}_9\times {U(1)}^2]$};
\end{tikzpicture}
\end{equation*}
Higgsing the $SU(3)_{12}$ leaves the rest of the flavour symmetry unmolested. In the infrared, that flavour symmetry is enhanced. In the theory on the left,  the $SU(2)^3_{18}\times U(1)^3$ is enhanced ${SU(3)}_{18}^3\times U(1)$. In the theory on the right, the ${SU(2)}_{36}\times {SU(2)}_{18}\times U(1)^2$ is enhanced to ${SU(3)}_{36}\times{SU(3)}_{18}\times U(1)$ (the ${SU(2)}_9$ is unaffected). So we learn that $k_1=k_2=k=18$.

Moreover, we can Higgs

\begin{displaymath}
A_3+A_1\xrightarrow{\quad {SU(2)}_{9}\quad} D_4(a_1)
\end{displaymath}
(either before or after the first Higgsing) to obtain the fixture on the left from the fixture on the right. The effect is to enhance the ${SU(3)}_{36}$ to ${SU(3)}_{18}^2$ in the infrared, consistent with what we found.

A more complicated example (which, to our dismay, reveals a typo\footnote{Another error along these lines was pointed out to us by Martone and Zafrir \cite{MZprivate}. Mixed fixture \#5
\begin{equation*}
\begin{tikzpicture}\draw[radius=40pt,fill=lightmauve] circle;
\draw[radius=2pt,fill=white]  (-.5,.9) circle node[right=2pt] {$D_5$};
\draw[radius=2pt,fill=white]  (-.5,-1) circle node[above right=2pt and -18pt] {$A_2+2A_1$};
\draw[radius=2pt,fill=white]  (1,0) circle node[left=2pt] {$3A_1$};
\end{tikzpicture}
\end{equation*} is the rank-2 ${SU(10)}_{10}$ SCFT, first found as a fixture in the $A_4$ theory (in \cite{Chacaltana:2010ks}, where it was called ``$S_5$'') with an additional 9 free hypermultiplets.}  in one entry in the tables of \cite{Chacaltana:2014jba}) is as follows. Consider the 4-punctured sphere

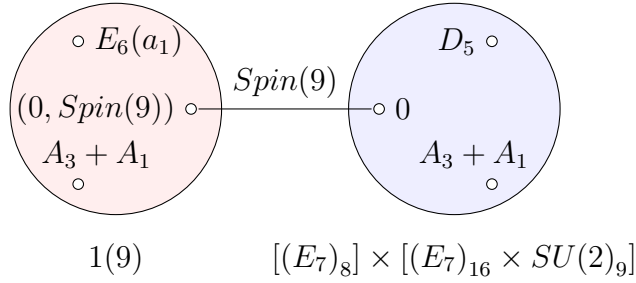
\begin{equation*}
\begin{tikzpicture}\draw[radius=40pt,fill=lightred] circle;
\draw[radius=2pt,fill=white]  (-.5,.9) circle node[right=2pt] {$E_6(a_1)$};
\draw[radius=2pt,fill=white]  (-.5,-1) circle node[above right=2pt and -18pt] {$A_3+A_1$};
\draw[radius=2pt,fill=white]  (1,0) circle node[left=2pt] {$(0,Spin(9))$};
\draw[radius=40pt,fill=lightblue] (4.5,0) circle;
\draw[radius=2pt,fill=white]  (5,.9) circle node[left=2pt] {$D_5$};
\draw[radius=2pt,fill=white]  (5,-1) circle node[above left=2pt and -18pt] {$A_3+A_1$};
\draw[radius=2pt,fill=white]  (3.5,0) circle node[right=2pt] {$0$};
\path
(1.1,0)  edge node[above] {$Spin(9)$} (3.4,0);
\node at (0,-2) {$1(9)$};
\node at (4.5,-2) {$[{(E_7)}_8]\times[{(E_7)}_{16}\times {SU(2)}_9]$};
\end{tikzpicture}
\end{equation*}
This is a $Spin(9)$ gauging of the product the rank-1 and rank-2 $E_7$ Minahan-Nemeschansky theories, with an additional hypermultiplet in the $9$. The centralizer of $Spin(9)$ in ${(E_7)}_k$ is ${SU(2)}_{2k}\times{SU(2)}_k$, so the flavour symmetry of this family of SCFTs is

\begin{displaymath}
F = {SU(2)}_{32}\times {SU(2)}_{16}^2\times {SU(2)}_9^2 \times {SU(2)}_8
\end{displaymath}
S-dualizing, we obtain

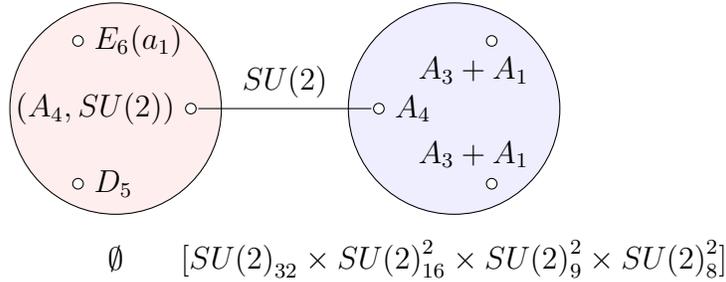
\begin{equation*}
\begin{tikzpicture}\draw[radius=40pt,fill=lightred] circle;
\draw[radius=2pt,fill=white]  (-.5,.9) circle node[right=2pt] {$E_6(a_1)$};
\draw[radius=2pt,fill=white]  (-.5,-1) circle node[right=2pt] {$D_5$};
\draw[radius=2pt,fill=white]  (1,0) circle node[left=2pt] {$(A_4, SU(2))$};
\draw[radius=40pt,fill=lightblue] (4.5,0) circle;
\draw[radius=2pt,fill=white]  (5,.9) circle node[below left=2pt and -18pt] {$A_3+A_1$};
\draw[radius=2pt,fill=white]  (5,-1) circle node[above left=2pt and -18pt] {$A_3+A_1$};
\draw[radius=2pt,fill=white]  (3.5,0) circle node[right=2pt] {$A_4$};
\path
(1.1,0)  edge node[above] {$SU(2)$} (3.4,0);
\node at (0,-2) {$\emptyset$};
\node at (4.5,-2) {$[{SU(2)}_{32}\times {SU(2)}_{16}^2\times {SU(2)}_9^2\times {SU(2)}_8^2 ]$};
\end{tikzpicture}
\end{equation*}
where the fixture on the right is \#125 from the table in section 3.4 of \cite{Chacaltana:2014jba}.

Higgsing $A_4\xrightarrow{\quad {SU(2)}_8\quad}A_4+A_1$, we obtain fixture \#89:

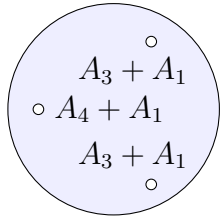
\begin{equation*}
\begin{tikzpicture}\draw[radius=40pt,fill=lightblue] (0,0) circle;
\draw[radius=2pt,fill=white]  (.5,.9) circle node[below left=2pt and -18pt] {$A_3+A_1$};
\draw[radius=2pt,fill=white]  (.5,-1) circle node[above left=2pt and -18pt] {$A_3+A_1$};
\draw[radius=2pt,fill=white]  (-1,0) circle node[right=2pt] {$A_4+A_1$};
\node at (0,-2) {$[{SU(2)}_{32}\times {SU(2)}_{16}^2\times {SU(2)}_9^2\times {SU(2)}_8]$};
\end{tikzpicture}
\end{equation*}

In the table, the flavour symmetry is listed as ``$\dots\times{SU(2)}_8^2$'', which is incorrect. One readily checks from the superconformal index that the enhancement of manifest flavour symmetry of the fixture is ${SU(2)}^2\times {U(1)}^3\to {SU(2)}^6$. Here, we've determined the levels to be as-stated.

As a further consistency check, we can Higgs $A_3+A_1\xrightarrow{\quad {SU(2)}_{9}\quad} D_4(a_1)$ as before. Before S-dualizing, this corresponds to turning on a VEV for the hypermultiplet in the $(9)$ of $Spin(9)$, which breaks $Spin(9)\to Spin(8)$:

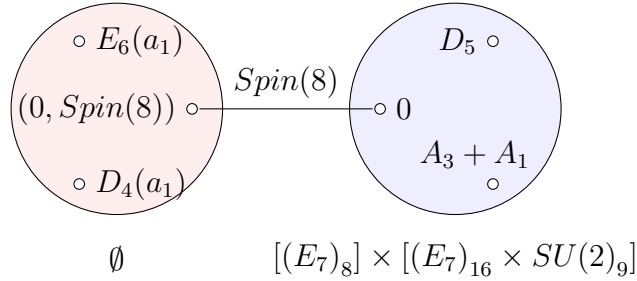
\begin{equation*}
\begin{tikzpicture}\draw[radius=40pt,fill=lightred] circle;
\draw[radius=2pt,fill=white]  (-.5,.9) circle node[right=2pt] {$E_6(a_1)$};
\draw[radius=2pt,fill=white]  (-.5,-1) circle node[right=2pt] {$D_4(a_1)$};
\draw[radius=2pt,fill=white]  (1,0) circle node[left=2pt] {$(0,Spin(8))$};
\draw[radius=40pt,fill=lightblue] (4.5,0) circle;
\draw[radius=2pt,fill=white]  (5,.9) circle node[left=2pt] {$D_5$};
\draw[radius=2pt,fill=white]  (5,-1) circle node[above left=2pt and -18pt] {$A_3+A_1$};
\draw[radius=2pt,fill=white]  (3.5,0) circle node[right=2pt] {$0$};
\path
(1.1,0)  edge node[above] {$Spin(8)$} (3.4,0);
\node at (0,-2) {$\emptyset$};
\node at (4.5,-2) {$[{(E_7)}_8]\times[{(E_7)}_{16}\times {SU(2)}_9]$};
\end{tikzpicture}
\end{equation*}
which has flavour symmetry

\begin{displaymath}
F = {SU(2)}_{16}^3\times {SU(2)}_9 \times {SU(2)}_8^3
\end{displaymath}
This theory has two other S-duality frames. In one,

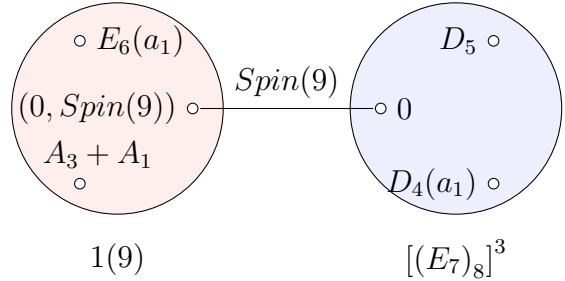
\begin{equation*}
\begin{tikzpicture}\draw[radius=40pt,fill=lightred] circle;
\draw[radius=2pt,fill=white]  (-.5,.9) circle node[right=2pt] {$E_6(a_1)$};
\draw[radius=2pt,fill=white]  (-.5,-1) circle node[above right=2pt and -18pt] {$A_3+A_1$};
\draw[radius=2pt,fill=white]  (1,0) circle node[left=2pt] {$(0,Spin(9))$};
\draw[radius=40pt,fill=lightblue] (4.5,0) circle;
\draw[radius=2pt,fill=white]  (5,.9) circle node[left=2pt] {$D_5$};
\draw[radius=2pt,fill=white]  (5,-1) circle node[left=2pt] {$D_4(a_1)$};
\draw[radius=2pt,fill=white]  (3.5,0) circle node[right=2pt] {$0$};
\path
(1.1,0)  edge node[above] {$Spin(9)$} (3.4,0);
\node at (0,-2) {$1(9)$};
\node at (4.5,-2) {${[{(E_7)}_8]}^3$};
\end{tikzpicture}
\end{equation*}
we have Higgsed the rank-2 Minahan-Nemeschansky theory to 2 copies of the rank-1 theory. In the other,

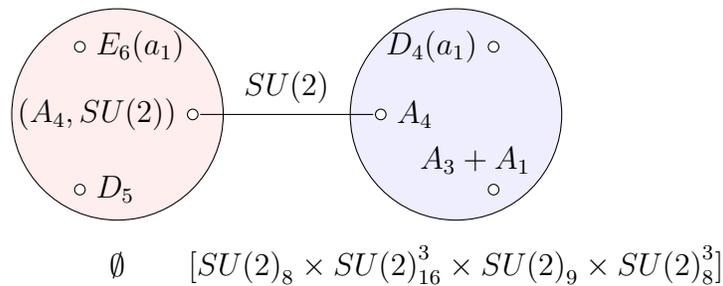
\begin{equation*}
\begin{tikzpicture}\draw[radius=40pt,fill=lightred] circle;
\draw[radius=2pt,fill=white]  (-.5,.9) circle node[right=2pt] {$E_6(a_1)$};
\draw[radius=2pt,fill=white]  (-.5,-1) circle node[right=2pt] {$D_5$};
\draw[radius=2pt,fill=white]  (1,0) circle node[left=2pt] {$(A_4, SU(2))$};
\draw[radius=40pt,fill=lightblue] (4.5,0) circle;
\draw[radius=2pt,fill=white]  (5,.9) circle node[left=2pt] {$D_4(a_1)$};
\draw[radius=2pt,fill=white]  (5,-1) circle node[above left=2pt and -18pt] {$A_3+A_1$};
\draw[radius=2pt,fill=white]  (3.5,0) circle node[right=2pt] {$A_4$};
\path
(1.1,0)  edge node[above] {$SU(2)$} (3.4,0);
\node at (0,-2) {$\emptyset$};
\node at (4.5,-2) {$[{SU(2)}_8\times {SU(2)}_{16}^3\times {SU(2)}_9\times {SU(2)}_8^3 ]$};
\end{tikzpicture}
\end{equation*}
the fixture on the right is \#121. Higgsing $A_4\xrightarrow{\quad {SU(2)}_8\quad}A_4+A_1$, we obtain fixture \#85:

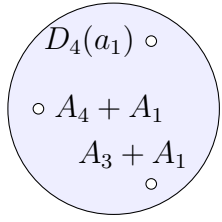
\begin{equation*}
\begin{tikzpicture}\draw[radius=40pt,fill=lightblue] (0,0) circle;
\draw[radius=2pt,fill=white]  (.5,.9) circle node[left=2pt] {$D_4(a_1)$};
\draw[radius=2pt,fill=white]  (.5,-1) circle node[above left=2pt and -18pt] {$A_3+A_1$};
\draw[radius=2pt,fill=white]  (-1,0) circle node[right=2pt] {$A_4+A_1$};
\node at (0,-2) {$[{SU(2)}_{16}^3\times {SU(2)}_9\times {SU(2)}_8^3]$};
\end{tikzpicture}
\end{equation*}

Finally, Higgsing the other $A_3+A_1\xrightarrow{\; {SU(2)}_{9}\;} D_4(a_1)$, we obtain the same relationship between fixtures \# 120 and \#84 (whose flavour symmetries are, respectively, ${SU(2)}_8^{10}$ and ${SU(2)}_8^{9}$).

\subsection{Some Product SCFTs in the $E_7$ Theory}\label{some_product_scfts_in_the__theory}

Let us start  our examination of the $E_7$ theory with the SCFTs discussed in \S3 of \cite{Distler:2018gbc}. There, we have 8 SCFTs where the ${(E_7)}_{36}$ of the full puncture, $0$, is enhanced to ${(E_7)}_{36-k}\times {(E_7)}_{k}$. They are of the form
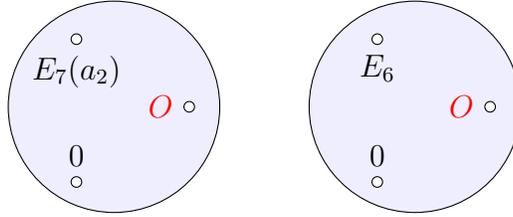
\begin{equation*}
\begin{tikzpicture}\draw[radius=40pt,fill=lightblue] circle;
\draw[radius=2pt,fill=white]  (-.5,.9) circle node[below=2pt] {$E_7(a_2)$};
\draw[radius=2pt,fill=white]  (-.5,-1) circle node[above=2pt] {$0$};
\draw[radius=2pt,fill=white]  (1,0) circle node[left=2pt] {$\color{red} O$};
%\node (sym) at (0,-2) {$\begin{gathered}SU(3)_{54-k}\times SU(3)_k\ \times SU(2)_9\times U(1)\end{gathered}$};
\draw[radius=40pt,fill=lightblue] (4,0) circle;
\draw[radius=2pt,fill=white]  (3.5,.9) circle node[below=2pt] {$E_6$};
\draw[radius=2pt,fill=white]  (3.5,-1) circle node[above=2pt] {$0$};
\draw[radius=2pt,fill=white]  (5,0) circle node[left=2pt] {$\color{red}O$};
\end{tikzpicture}
\end{equation*}
where $\color{red} O$ is one of $D_4$, $D_4+A_1$, $D_5(a_1)$ or $D_5(a_1)+A_1$. These were all determined to be product SCFTs. When ${\color{red}O}=D_5(a_1)$, one of the factors in the product was identified as the ${(E_7)}_8$ Minahan-Nemenschansky theory. Hence the level of the other $E_7$ factor in the flavour symmetry is $k=28$.

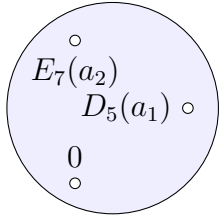
\begin{equation*}
\begin{tikzpicture}\draw[radius=40pt,fill=lightblue] circle;
\draw[radius=2pt,fill=white]  (-.5,.9) circle node[below=2pt] {$E_7(a_2)$};
\draw[radius=2pt,fill=white]  (-.5,-1) circle node[above=2pt] {$0$};
\draw[radius=2pt,fill=white]  (1,0) circle node[left=2pt] {$D_5(a_1)$};
\node at (0,-2) {$[{(E_7)}_8]\times [{(E_7)}_{28}\times {SU(2)}_{10}\times U(1)]$};
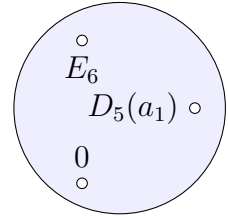
\draw[radius=40pt,fill=lightblue] (9,0) circle;
\draw[radius=2pt,fill=white]  (8.5,.9) circle node[below=2pt] {$E_6$};
\draw[radius=2pt,fill=white]  (8.5,-1) circle node[above=2pt] {$0$};
\draw[radius=2pt,fill=white]  (10,0) circle node[left=2pt] {$D_5(a_1)$};
\node at (9,-2) {$[{(E_7)}_8]\times [{(E_7)}_{28}\times {SU(2)}_{12}\times {SU(2)}_{10}\times U(1)]$};
\end{tikzpicture}
\end{equation*}

From the Higgsing diagram, we see that these are all related by the Higgsings

\begin{equation}
D_4\xrightarrow{\quad {Sp(3)}_{12}\quad}D_4+A_1
\xrightarrow{\quad {Sp(2)}_{11}\quad}D_5(a_1)
\xrightarrow{\quad {SU(2)}_{10}\quad}D_5(a_1)+A_1
\label{D4Higgsing}\end{equation}
So we see that all of these are product SCFTs with $[{(E_7)}_8]$ as one of the factors and the level of the $E_7$ in the other factor is $k=28$.

\subsubsection{S-duality}\label{sduality}

Having identified the unknown levels in these theories, related by \eqref{D4Higgsing}, we can exploit S-duality to determine others. Consider

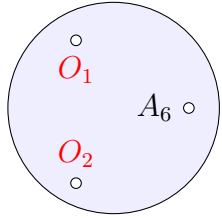
\begin{equation*}
\begin{tikzpicture}\draw[radius=40pt,fill=lightblue] circle;
\draw[radius=2pt,fill=white]  (-.5,.9) circle node[below=2pt] {$\color{red}O_1$};
\draw[radius=2pt,fill=white]  (-.5,-1) circle node[above=2pt] {$\color{red}O_2$};
\draw[radius=2pt,fill=white]  (1,0) circle node[left=2pt] {$A_6$};
\end{tikzpicture}
\end{equation*}
where $\color{red}O_{1,2}$ are again chosen from the set $\{D_4,\, D_4+A_1,\, D_5(a_1),\ D_5(a_1)+A_1\}$. There are 10 such theories in all and, in each of them, the ${SU(2)}_{36}$ symmetry of the $A_6$ puncture is enhanced to ${SU(2)}_{36-k}\times {SU(2)}_k$.

Let's pick one of these theories, say

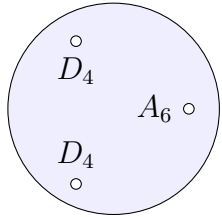
\begin{equation*}
\begin{tikzpicture}\draw[radius=40pt,fill=lightblue] circle;
\draw[radius=2pt,fill=white]  (-.5,.9) circle node[below=2pt] {$D_4$};
\draw[radius=2pt,fill=white]  (-.5,-1) circle node[above=2pt] {$D_4$};
\draw[radius=2pt,fill=white]  (1,0) circle node[left=2pt] {$A_6$};
\end{tikzpicture}
\end{equation*}
and, instead of Higgsing the $Sp(3)$ associated to the $D_4$ puncture, let's gauge (an $Sp(2)$ subgroup of) it instead. In other words, consider the 4-punctured sphere

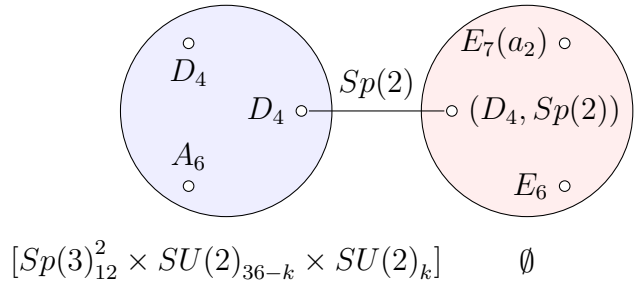
\begin{equation*}
\begin{tikzpicture}\draw[radius=40pt,fill=lightblue] circle;
\draw[radius=2pt,fill=white]  (-.5,.9) circle node[below=2pt] {$D_4$};
\draw[radius=2pt,fill=white]  (-.5,-1) circle node[above=2pt] {$A_6$};
\draw[radius=2pt,fill=white]  (1,0) circle node[left=2pt] {$D_4$};
\draw[radius=40pt,fill=lightred] (4,0) circle;
\draw[radius=2pt,fill=white]  (4.5,.9) circle node[left=2pt] {$E_7(a_2)$};
\draw[radius=2pt,fill=white]  (4.5,-1) circle node[left=2pt] {$E_6$};
\draw[radius=2pt,fill=white]  (3,0) circle node[right=2pt] {$(D_4, Sp(2))$};
\path
(1.1,0)  edge node[above] {$Sp(2)$} (2.9,0);
\node at (0,-2) {$[{Sp(3)}_{12}^2\times {SU(2)}_{36-k}\times {SU(2)}_{k}]$};
\node at (4,-2) {$\emptyset$};
\end{tikzpicture}
\end{equation*}

Now we S-dualize and obtain

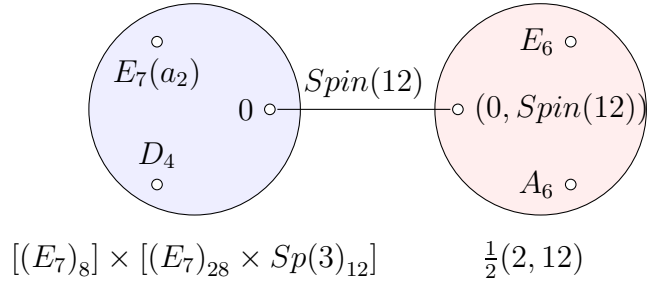
\begin{equation*}
\begin{tikzpicture}\draw[radius=40pt,fill=lightblue] circle;
\draw[radius=2pt,fill=white]  (-.5,.9) circle node[below=2pt] {$E_7(a_2)$};
\draw[radius=2pt,fill=white]  (-.5,-1) circle node[above=2pt] {$D_4$};
\draw[radius=2pt,fill=white]  (1,0) circle node[left=2pt] {$0$};
\draw[radius=40pt,fill=lightred] (4.6,0) circle;
\draw[radius=2pt,fill=white]  (5,.9) circle node[left=2pt] {$E_6$};
\draw[radius=2pt,fill=white]  (5,-1) circle node[left=2pt] {$A_6$};
\draw[radius=2pt,fill=white]  (3.5,0) circle node[right=2pt] {$(0, Spin(12))$};
\path
(1.1,0)  edge node[above] {$Spin(12)$} (3.4,0);
\node at (0,-2) {$[{(E_7)}_8]\times [{(E_7)}_{28}\times {Sp(3)}_{12}]$};
\node at (4.5,-2) {$\tfrac{1}{2}(2,12)$};
\end{tikzpicture}
\end{equation*}
and
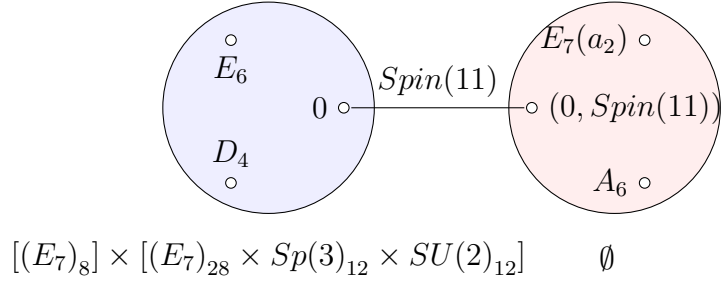
\begin{equation*}
\begin{tikzpicture}\draw[radius=40pt,fill=lightblue] circle;
\draw[radius=2pt,fill=white]  (-.5,.9) circle node[below=2pt] {$E_6$};
\draw[radius=2pt,fill=white]  (-.5,-1) circle node[above=2pt] {$D_4$};
\draw[radius=2pt,fill=white]  (1,0) circle node[left=2pt] {$0$};
\draw[radius=40pt,fill=lightred] (4.6,0) circle;
\draw[radius=2pt,fill=white]  (5,.9) circle node[left=2pt] {$E_7(a_2)$};
\draw[radius=2pt,fill=white]  (5,-1) circle node[left=2pt] {$A_6$};
\draw[radius=2pt,fill=white]  (3.5,0) circle node[right=2pt] {$(0, Spin(11))$};
\path
(1.1,0)  edge node[above] {$Spin(11)$} (3.4,0);
\node at (0,-2) {$[{(E_7)}_8]\times [{(E_7)}_{28}\times {Sp(3)}_{12}\times {SU(2)}_{12}]$};
\node at (4.5,-2) {$\emptyset$};
\end{tikzpicture}
\end{equation*}

In both cases, the SCFT on the left is one of the product SCFTs we obtained in the previous subsection. The centralizer of $Spin(12)$ (or $Spin(11)$) in ${(E_7)}_{8}\times {(E_7)}_{28}$ is ${SU(2)}_8\times  {SU(2)}_{28}$. Hence we have determined the $SU(2)$ levels in

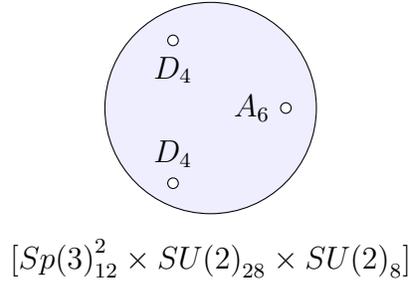
\begin{equation*}
\begin{tikzpicture}\draw[radius=40pt,fill=lightblue] circle;
\draw[radius=2pt,fill=white]  (-.5,.9) circle node[below=2pt] {$D_4$};
\draw[radius=2pt,fill=white]  (-.5,-1) circle node[above=2pt] {$D_4$};
\draw[radius=2pt,fill=white]  (1,0) circle node[left=2pt] {$A_6$};
\node at (0,-2) {$[ {Sp(3)}_{12}^2\times {SU(2)}_{28}\times {SU(2)}_{8}]$};
\end{tikzpicture}
\end{equation*}

Finally, we apply the Higgsing \eqref{D4Higgsing} to obtain the (same) $SU(2)$ levels in the other 9 theories.

\subsubsection{Matching to known SCFTs}\label{matching_to_known_scfts}

Another technique we can employ is to match the theory to known SCFTs. Consider

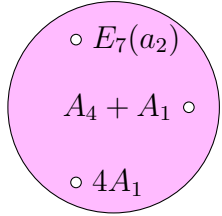
\begin{equation*}
\begin{tikzpicture}\draw[radius=40pt,fill=lightmauve] circle;
\draw[radius=2pt,fill=white]  (-.5,.9) circle node[right=2pt] {$E_7(a_2)$};
\draw[radius=2pt,fill=white]  (-.5,-1) circle node[right=2pt] {$4A_1$};
\draw[radius=2pt,fill=white]  (1,0) circle node[left=2pt] {$A_4+A_1$};
\end{tikzpicture}
\end{equation*}
This is a mixed fixture whose manifest flavour symmetry is ${Sp(3)}_{19}\times U(1)^2$. The four free hypermultiplets transform as $\tfrac{1}{2}(6)+1(1)$. In particular, they contribute $k=1$ to the level of the $Sp(3)$. Subtracting their contribution, the remaining SCFT has enhanced flavour symmetry $SU(8)_{18} \times SU(2)_{k}\times U(1)$. It is a rank-4 SCFT with $n_4=n_5=n_8=n_9=1$ and $(n_h,n_v)= (92,48)$. These data agree with an interacting fixture in the $E_6$ theory (fixture \#10 in section 3.4 of \cite{Chacaltana:2014jba})

\begin{equation*}
\begin{tikzpicture}\draw[radius=40pt,fill=lightblue] circle;
\draw[radius=2pt,fill=white]  (-.5,.9) circle node[right=2pt] {$D_5$};
\draw[radius=2pt,fill=white]  (-.5,-1) circle node[right=2pt] {$A_1$};
\draw[radius=2pt,fill=white]  (1,0) circle node[left=2pt] {$A_2+2A_1$};
\node at (0,-2) {$[{SU(8)}_{18}\times {SU(2)}_{36}\times U(1)]$};
\end{tikzpicture}
\end{equation*}

We can check this identification by computing the unrefined Schur index of both theories. After subtracting the contribution of the free hypers, we obtain

\begin{displaymath}
I_{\text{Schur}}= 1 + 67 \tau^2 + 188 \tau^3 + 2764 \tau^4 + 13496 \tau^5 + 
 102726 \tau^6 + 569632 \tau^7 + 3443569 \tau^8 +\dots
\end{displaymath}
Another example is afforded by the pair of fixtures

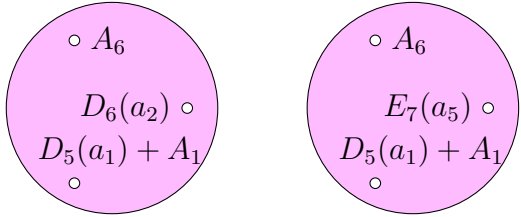
\begin{equation*}
\begin{tikzpicture}\draw[radius=40pt,fill=lightmauve] circle;
\draw[radius=2pt,fill=white]  (-.5,.9) circle node[right=2pt] {$A_6$};
\draw[radius=2pt,fill=white]  (-.5,-1) circle node[above right=2ptand -18pt] {$D_5(a_1)+A_1$};
\draw[radius=2pt,fill=white]  (1,0) circle node[left=2pt] {$D_6(a_2)$};
\draw[radius=40pt,fill=lightmauve] (4,0) circle;
\draw[radius=2pt,fill=white]  (3.5,.9) circle node[right=2pt] {$A_6$};
\draw[radius=2pt,fill=white]  (3.5,-1) circle node[above right=2pt and -18pt] {$D_5(a_1)+A_1$};
\draw[radius=2pt,fill=white]  (5,0) circle node[left=2pt] {$E_7(a_5)$};
\end{tikzpicture}
\end{equation*}
These are mixed fixtures. The fixture on the left has manifest global symmetry ${SU(2)}_{56}\times {SU(2)}_{36}\times{SU(2)}_9$ with hypermultiplets in the $\tfrac{1}{2}(3,2,1)$. Removing the contribution of the free hypers, the underlying interacting SCFT has enhanced flavour symmetry ${SU(2)}_{48-k}\times{SU(2)}_{k}\times SU(2)_9 \times{SU(2)}_{33-{k'}_1-{k'}_2}\times{SU(2)}_{{k'}_1}\times{SU(2)}_{{k'}_2}$. It is a rank-5 theory, with graded Coulomb branch dimensions $n_4=2,\, n_6=1,\, n_8=2$ and $(n_h,n_v)=(82,55)$.

The fixture on the right has manifest global symmetry ${SU(2)}_{56}\times {SU(2)}_{36}$ with hypermultiplets in the $\tfrac{1}{2}(3,2)$. Removing the contribution of the free hypers, the underlying interacting SCFT has enhanced flavour symmetry ${SU(2)}_{48-k_1-k_2}\times{SU(2)}_{k_1}\times{SU(2)}_{k_2}\times{SU(2)}_{33-k_3-k_4-k_5}\times{SU(2)}_{k_3}\times{SU(2)}_{k_4}\times{SU(2)}_{k_5}$. This theory is also rank-5, with $n_4=3,\, n_6=n_8=1$ and $(n_h,n_v)=(73,47)$.

We can pass from the former to the latter by Higgsing $D_6(a_2)\xrightarrow{\;{SU(2)}_9\;}E_7(a_5)$ (which preserves the rank). Comparing the invariants (and Schur indices), we recognize these as fixtures \#89 and \#85 of the $E_6$ theory, discussed in \S\ref{the_E6_theory}. So $k=k'_1=k_1=k_2=16$, $k'_2=k_3=k_4=k_5=8$.

As a more complicated example, consider the quartet of fixtures

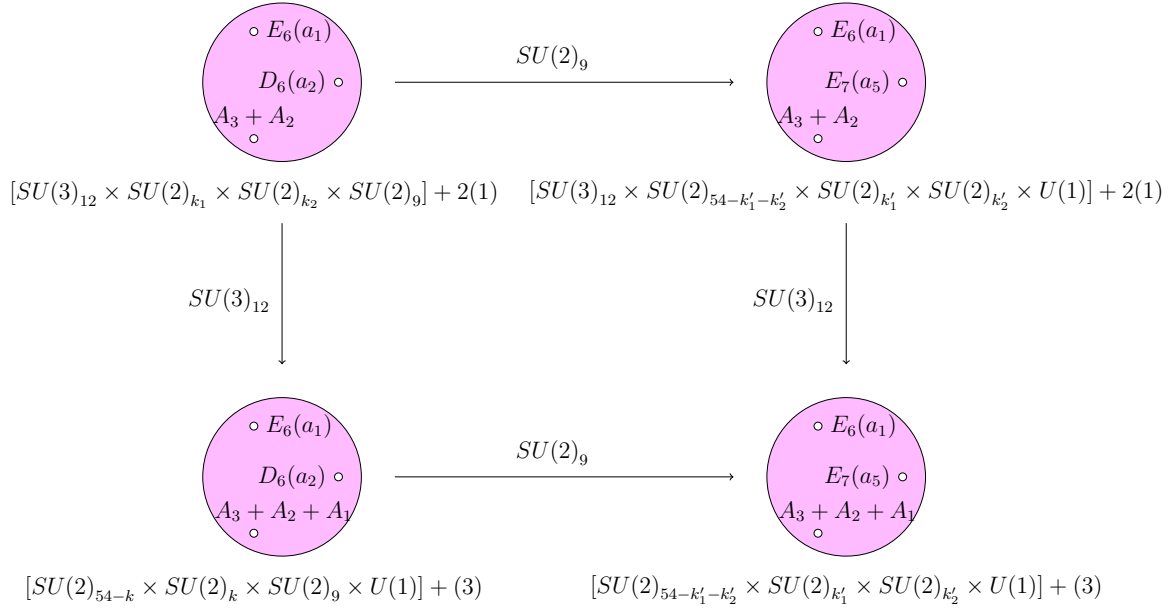
\begin{equation*}
\scalebox{.75}{
\begin{tikzpicture}\draw[radius=40pt,fill=lightmauve] circle;
\draw[radius=2pt,fill=white]  (-.5,.9) circle node[right=2pt] {$E_6(a_1)$};
\draw[radius=2pt,fill=white]  (-.5,-1) circle node[above=2pt] {$A_3+A_2$};
\draw[radius=2pt,fill=white]  (1,0) circle node[left=2pt] {$D_6(a_2)$};
\node at (-.5,-2) {$[{SU(3)}_{12}\times{SU(2)}_{k_1}\times{SU(2)}_{k_2}\times{SU(2)}_9]+2(1)$};
\draw[radius=40pt,fill=lightmauve] (10,0) circle;
\draw[radius=2pt,fill=white]  (9.5,.9) circle node[right=2pt] {$E_6(a_1)$};
\draw[radius=2pt,fill=white]  (9.5,-1) circle node[above=2pt] {$A_3+A_2$};
\draw[radius=2pt,fill=white]  (11,0) circle node[left=2pt] {$E_7(a_5)$};
\node at (10,-2) {$[{SU(3)}_{12}\times {SU(2)}_{54-k'_1-k'_2}\times{SU(2)}_{k'_1}\times{SU(2)}_{k'_2}\times U(1)]+2(1)$};
\draw[radius=40pt,fill=lightmauve] (0,-7) circle;
\draw[radius=2pt,fill=white]  (-.5,-6.1) circle node[right=2pt] {$E_6(a_1)$};
\draw[radius=2pt,fill=white]  (-.5,-8) circle node[above right=2pt and -24pt] {$A_3+A_2+A_1$};
\draw[radius=2pt,fill=white]  (1,-7) circle node[left=2pt] {$D_6(a_2)$};
\node at (-.5,-9) {$[{SU(2)}_{54-k}\times{SU(2)}_{k}\times{SU(2)}_9\times U(1)]+(3)$};
\draw[radius=40pt,fill=lightmauve] (10,-7) circle;
\draw[radius=2pt,fill=white]  (9.5,-6.1) circle node[right=2pt] {$E_6(a_1)$};
\draw[radius=2pt,fill=white]  (9.5,-8) circle node[above right=2pt and -24pt] {$A_3+A_2+A_1$};
\draw[radius=2pt,fill=white]  (11,-7) circle node[left=2pt] {$E_7(a_5)$};
\node at (10,-9) {$[{SU(2)}_{54-k'_1-k'_2}\times{SU(2)}_{k'_1}\times{SU(2)}_{k'_2}\times U(1)]+(3)$};
\path[->]
(2,0) edge (8,0) node[above right=2pt and 2cm] {${SU(2)}_9$}
(0,-2.5) edge (0,-5) node[below left=1cm and 2pt] {${SU(3)}_{12}$}
(2,-7) edge (8,-7) node[above right=2pt and 2cm] {${SU(2)}_9$}
(10,-2.5) edge (10,-5) node[below left=1cm and 2pt] {${SU(3)}_{12}$};
\end{tikzpicture}
}
\end{equation*}
which are related by the nilpotent Higgsings
$D_6(a_2)\xrightarrow{\;{SU(2)}_9\;}E_7(a_5)$ and/or by %\linebreak
 $A_3+A_2\xrightarrow{\;{SU(3)}_{12}\;}A_3+A_2+A_1$.
In the top row, the two hypermultiplets are singlets under the nonabelian part of the manifest flavour symmetry; in the bottom row, they transform as a triplet of the manifest ${SU(2)}_{224}$. After subtracting the contributions of the free hypermultiplets, we recognize the theories on the bottom row as fixtures \#66 and \#63 from the $E_6$ theory that we discussed in  \S\ref{the_E6_theory}. Since we determined the current algebra levels for those theories, we can fill in the other two SCFTs: $k_1=36$ and $k_2=k=k'_1=k'_2=18$.

\subsubsection{Examples where we fail}\label{examples_where_we_fail}

Unfortunately, there remain a handful of fixtures in the $E_7$ theory, which are not amenable to the above techniques. Consider
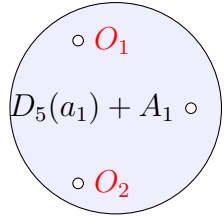
\begin{equation*}
\begin{tikzpicture}\draw[radius=40pt,fill=lightblue] circle;
\draw[radius=2pt,fill=white]  (-.5,.9) circle node[right=2pt] {$\color{red}O_1$};
\draw[radius=2pt,fill=white]  (-.5,-1) circle node[right=2pt] {$\color{red}O_2$};
\draw[radius=2pt,fill=white]  (1,0) circle node[left=2pt] {$D_5(a_1)+A_1$};
\end{tikzpicture}
\end{equation*}
where $\color{red}O_{1,2}$ are both chosen from Higgsing diagram

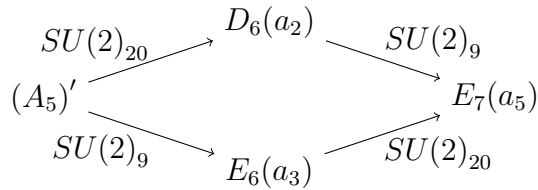
\begin{equation*}
\begin{tikzpicture}
\node (A5p) at (0,0) {${(A_5)}'$};
\node (D6a2) at (3,1) {$D_6(a_2)$};
\node (E6a3) at (3,-1) {$E_6(a_3)$};
\node (E7a5) at (6,0) {$E_7(a_5)$};
\path[->] 
(A5p) edge node[above left=-.125cm and -.125cm] {${SU(2)}_{20}$} (D6a2)
(A5p) edge node[below left=-.125cm and -.125cm] {${SU(2)}_{9}$} (E6a3)
(D6a2) edge node[above right=-.125cm and -.125cm] {${SU(2)}_{9}$} (E7a5)
(E6a3) edge node[below right=-.125cm and -.125cm] {${SU(2)}_{20}$} (E7a5);
\end{tikzpicture}
\end{equation*}

There are 10 fixtures of this form. If \emph{both} $\color{red}O_{1,2}$ are chosen from the subset $\{D_6(a_2),E_7(a_5)\}$, then the ${SU(2)}_{56}$ of the $D_5(a_1)+A_1$ puncture is enhanced to ${SU(2)}_{56-k}\times {SU(2)}_{k}$. Otherwise, it is unenhanced. Unfortunately, none of these 10 fixtures can be gauged, so we cannot use S-duality to determine $k$ in the 3 fixtures with enhanced flavour symmetry.

Another case, where we are able to constrain but not completely fix the levels, consists of the fixtures

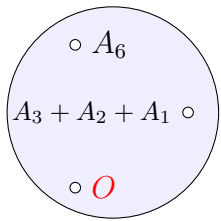
\begin{equation*}
\begin{tikzpicture}\draw[radius=40pt,fill=lightblue] circle;
\draw[radius=2pt,fill=white]  (-.5,.9) circle node[right=2pt] {$A_6$};
\draw[radius=2pt,fill=white]  (-.5,-1) circle node[right=2pt] {$\color{red}O$};
\draw[radius=2pt,fill=white]  (1,0) circle node[left=2pt] {\footnotesize{$A_3+A_2+A_1$}};
\end{tikzpicture}
\end{equation*}
where $\color{red}O$ is taken from the set $\{D_{5}+A_1,D_{6}(a_1),E_7(a_4),A_6 \}$. The Higgsing Diagram for the this set is

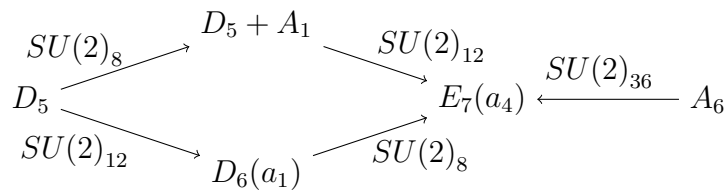
\begin{equation*}
\begin{tikzpicture}
\node (D5) at (0,0) {$D_5$};
\node (D5A1) at (3,1) {$D_5+A_1$};
\node (D6a1) at (3,-1) {$D_6(a_1)$};
\node (E7a4) at (6,0) {$E_7(a_4)$};
\node (A6) at (9,0) {$A_6$};
\path[->] 
(D5) edge node[above left=-.125cm and -.125cm] {${SU(2)}_{8}$} (D5A1)
(D5) edge node[below left=-.125cm and -.125cm] {${SU(2)}_{12}$} (D6a1)
(D5A1) edge node[above right=-.125cm and -.125cm] {${SU(2)}_{12}$} (E7a4)
(D6a1) edge node[below right=-.125cm and -.125cm] {${SU(2)}_{8}$} (E7a4)
(A6) edge node[above left=-.0cm and -.75cm] {${SU(2)}_{36}$} (E7a4);
\end{tikzpicture}
\end{equation*}

The case of ${\color{red}O} = E_7(a_4)$ has manifest symmetry $SU(2)_{224} \times SU(2)_{36}$ and free hypermutiplets in the representation $\frac{1}{2}(3,2)$ The remaining SCFT then has a $SU(2)_{216} \times SU(2)_{33}$ symmetry which is enhanced to $SU(2)_{128} \times SU(2)_{88} \times SU(2)_{33}$ which is then further enhanced to $SU(2)_{128-k} \times SU(2)_{k} \times Sp(3)_{11}$. 

One can set ${\color{red}O}= A_6$, this fixture has the enhanced symmetry $SU(2)_{152} \times Spin(7)_{36}$. This tells us the manifest $SU(2)_{224}$ was enhanced to a $SU(2)_{152} \times SU(2)_{72}$. Higgsing an $SU(2)_{36}$ gives the original fixture with unknown levels. After subtracting the contribution from the hypermultiplets we have flavour symmetry $SU(2)_{152-(8-l)} \times SU(2)_{72-l} \times SU(2)_{33}$ where $0 \leq l \leq 8$. This is then embedded into $SU(2)_{128-k} \times SU(2)_{k} \times Sp(3)_{11}$. For this to happen we need a further enhancement of $SU(2)_{152-(8-l)} \times SU(2)_{72-l}$ to something containing a factor of $SU(2)_{88}$. This must come from the $SU(2)_{152-(8-l)}$ and so we are left with a flavour symmetry of the form $SU(2)_{64-(8-l)} \times SU(2)_{72-l} \times Sp(3)_{11}$. No other Higgsings tell us any information so this is the best we can do with these methods.

\subsection{Results}\label{results}
Here we  present our results for the previously-unknown levels in the $E_6$ and $E_7$ theories. The enhanced flavour symmetry factors, whose levels were undetermined, are highlighted in {\color{red}red}, as are the manifest flavour symmetry factors that underwent the enhancement. The \href{https://golem.ph.utexas.edu/class-S/E7/}{online application} for the $E_7$ theory has been updated accordingly.

{
\footnotesize
\renewcommand{\arraystretch}{1.2}

\begin{longtable}{|c|c|c|}
\caption{Untwisted $E_6$}\\
\hline
Fixture&Manifest Flavour Symmetry&Enhanced Flavour Symmetry\\
\hline
\endfirsthead
\hline
Fixture&Manifest Flavour Symmetry&Enhanced Flavour Symmetry\\
\hline 
\endhead
\endfoot
$\begin{matrix}D_5(a_1)\\ D_4(a_1)\end{matrix}\quad A_2+2A_1$&${\color{red}{SU(2)}_{54}\times {U(1)}^3}\times U(1)$&${\color{red}{SU(3)}_{18}^3 }\times U(1)$\\
\hline
$\begin{matrix}D_5(a_1)\\ A_3+A_1\end{matrix}\quad A_2+2A_1$&${\color{red}{SU(2)}_{54}\times {U(1)}^2 }\times {SU(2)}_9\times U(1)$&${\color{red}{SU(3)}_{36}\times {SU(3)}_{18} }\times {SU(2)}_9\times U(1)$\\
\hline
\end{longtable}
}

{
\footnotesize
\setlength\LTleft{-.125in}
\renewcommand{\arraystretch}{1.5}

\begin{longtable}{|c|c|c|}
\caption{Twisted $E_6$}\\
\hline
Fixture&Manifest Flavour Symmetry&Enhanced Flavour Symmetry\\
\endfirsthead
\hline
Fixture&Manifest Flavour Symmetry&Enhanced Flavour Symmetry\\
\endhead
\endfoot
\hline 
$\begin{matrix} \underline{F_4}\\ A_1 \end{matrix}\quad \underline{A_1+\tilde{A}_1}$&${SU(6)}_{18}\times{\color{red}{SU(2)}_{64}}\times SU(2)_{10}$&${SU(6)}_{18}\times{\color{red}{SU(2)}_{48} \times SU(2)_{16}} \times SU(2)_{10}$\\
\hline
$\begin{matrix}\underline{F_4(a_2)}\\ \underline{F_4(a_2)}\end{matrix}\quad A_2+2A_1$&${\color{red}{SU(2)}_{54}}\times {U(1)}$&${\color{red}{SU(2)}_{36}\times{SU(2)}_{18}}\times U(1)$\\
\hline
$\begin{matrix}\underline{F_4(a_2)}\\ E_{6}(a_3)\end{matrix}\quad \underline{A_1+\tilde{A}_1}$&${\color{red}{SU(2)}_{64}}\times {SU(2)}_{10}$&${\color{red}{SU(2)}_{32}\times{SU(2)}_{16}\times{SU(2)}_{16}}\times SU(2)_{10}$\\
\hline
$\begin{matrix}\underline{F_4(a_2)}\\ D_5(a_1)\end{matrix}\quad \underline{A_1+\tilde{A}_1}$&${\color{red} {SU(2)}_{64}}\times {SU(2)}_{10} \times U(1)$&${\color{red}{SU(2)}_{48}\times{SU(2)}_{16}}\times SU(2)_{10}\times U(1)$\\
\hline
$\begin{matrix}\underline{F_4(a_2)}\\ D_4\end{matrix}\quad \underline{A_1+\tilde{A}_1}$&$SU(3)_{12} \times{\color{red}{SU(2)}_{64}}\times {SU(2)}_{10}$&${SU(3)}_{12}\times{\color{red}{SU(2)}_{48}\times{SU(2)}_{16}}\times SU(2)_{10}$\\
\hline
$\begin{matrix}\underline{C_3}\\ D_5(a_1)\end{matrix}\quad \underline{A_1+\tilde{A}_1}$&$
\begin{gathered}
{\color{red}{SU(2)}_{64}}\times {SU(2)}_{10}\\ \times SU(2)_6 \times U(1)\end{gathered}$&$\begin{gathered}{\color{red}{SU(2)}_{48}\times{SU(2)}_{16}}\times SU(2)_{10}\\ \times SU(2)_6 \times U(1)\end{gathered}$\\
\hline
$\begin{matrix}\underline{C_3}\\ D_4\end{matrix}\quad \underline{A_1+\tilde{A}_1}$&$\begin{gathered}SU(3)_{12} \times{\color{red}{SU(2)}_{64}}\\ \times {SU(2)}_{10} \times SU(2)_6\end{gathered}$&$\begin{gathered}{SU(3)}_{12}\times{\color{red}{SU(2)}_{48}\times{SU(2)}_{16}}\\ \times SU(2)_{10} \times SU(2)_6\end{gathered}$\\
\hline
$\begin{matrix}\underline{B_3}\\ E_6(a_3)\end{matrix}\quad \underline{A_2+\tilde{A}_1}$&${\color{red}SU(2)_{39}} \times SU(2)_{24}$&${\color{red}SU(2)_{26} \times SU(2)_{13}} \times SU(2)_{24}$\\
\hline
$\begin{matrix}\underline{\tilde{A_2}}\\ D_5\end{matrix}\quad \underline{A_1+\tilde{A}_1}$&$(G_{2})_{10} \times{\color{red}{SU(2)}_{64}}\times {SU(2)}_{10} \times U(1)$&$(G_{2})_{10}\times{\color{red}{SU(2)}_{48}\times{SU(2)}_{16}}\times SU(2)_{10} \times U(1)$\\
\hline
$\begin{matrix}\underline{F_4(a_2)}\\ \underline{F_4(a_2)}\end{matrix}\quad 2A_2+A_1$&${\color{red}{SU(2)}_{26}}$&${\color{red}SU(2)_{16} \times SU(2)_{9} \times U(1)} +{\color{red}\tfrac{1}{2}{(2)}}$\\
\hline
$\begin{matrix}\underline{F_4(a_2)}\\ D_5(a_1)\end{matrix}\quad \underline{A_2+\tilde{A}_1}$&${\color{red}{SU(2)}_{39}} \times U(1)$&${\color{red}{SU(2)}_{20} \times SU(2)_{18}} \times U(1)+{\color{red}\tfrac{1}{2}{(2)}_0}$\\
\hline
\end{longtable}
}

{
\scriptsize
\renewcommand{\arraystretch}{1.75}

\begin{longtable}{|c|c|c|}
\caption{$E_7$}\\
\hline
Fixture&Manifest Flavour Symmetry&Enhanced Flavour Symmetry\\
\endfirsthead
\hline
Fixture&Manifest Flavour Symmetry&Enhanced Flavour Symmetry\\
\endhead
\hline 
$\begin{matrix} A_5+A_1\\ D_6(a_2) \end{matrix}\quad D_{5}(a_1)+A_1$&${SU(2)}_{26}\times {\color{red}{{SU(2)}_{56}}} \times {SU(2)}_{9}$&${SU(2)}_{26}\times {\color{red}{{SU(2)}_{28}\times {SU(2)}_{28}}}\times {SU(2)}_{9}$\\
\hline
$\begin{matrix} A_5+A_1\\ E_7(a_5) \end{matrix}\quad D_{5}(a_1)+A_1$&${SU(2)}_{26}\times {\color{red}{{SU(2)}_{56}}}$&${SU(2)}_{26}\times \color{red}{{SU(2)}_{28}\times {SU(2)}_{28}}$\\
\hline
$\begin{matrix} D_4\\ D_4  \end{matrix}\quad A_6$&${Sp(3)}_{12}^2\times \color{red}{{SU(2)}_{36}}$&${Sp(3)}_{12}^2\times \color{red}{{SU(2)}_{28}\times {SU(2)}_{8}}$\\
\hline
$\begin{matrix} D_4\\ D_4+A_1  \end{matrix}\quad A_6$&${Sp(3)}_{12} \times Sp(2)_{11} \times  \color{red}{{SU(2)}_{36}}$&${Sp(3)}_{12} \times Sp(2)_{11} \times  \color{red}{{SU(2)}_{28}\times {SU(2)}_{8}}$\\
\hline
$\begin{matrix} D_4+A_1\\ D_4+A_1  \end{matrix}\quad A_6$&$Sp(2)_{11}^2 \times  \color{red}{{SU(2)}_{36}}$&$Sp(2)_{11}^2 \times  \color{red}{{SU(2)}_{28}\times {SU(2)}_{8}}$\\
\hline
$\begin{matrix} D_4\\ D_5(a_1)  \end{matrix}\quad A_6$&$Sp(3)_{12} \times SU(2)_{10} \times U(1) \times \color{red}{{SU(2)}_{36}}$&$Sp(3)_{12} \times SU(2)_{10} \times U(1) \times \color{red}{{SU(2)}_{28}\times {SU(2)}_{8}}$\\
\hline
$\begin{matrix} D_4+A_1\\ D_{5}(a_1)  \end{matrix}\quad A_6$&$Sp(2)_{11} \times SU(2)_{10} \times U(1) \times  \color{red}{{SU(2)}_{36}}$&$Sp(2)_{11} \times SU(2)_{10} \times U(1) \times  \color{red}{{SU(2)}_{28}\times {SU(2)}_{8}}$\\
\hline
$\begin{matrix} D_5(a_1)\\ D_{5}(a_1)  \end{matrix}\quad A_6$&$SU(2)_{10}^2 \times U(1)^2 \times  \color{red}{{SU(2)}_{36}}$&$SU(2)_{10}^2 \times {U(1)}^2 \times  \color{red}{{SU(2)}_{28}\times {SU(2)}_{8}}$\\
\hline
$\begin{matrix} D_4\\ D_5(a_1)+A_1  \end{matrix}\quad A_6$&$Sp(3)_{12} \times SU(2)_{56} \times  \color{red}{{SU(2)}_{36}}$&$Sp(3)_{12} \times SU(2)_{56} \times  \color{red}{{SU(2)}_{28}\times {SU(2)}_{8}}$\\
\hline
$\begin{matrix} D_4+A_1\\ D_5(a_1)+A_1  \end{matrix}\quad A_6$&$Sp(2)_{11} \times SU(2)_{56} \times  \color{red}{{SU(2)}_{36}}$&$Sp(2)_{11} \times SU(2)_{56} \times  \color{red}{{SU(2)}_{28}\times {SU(2)}_{8}}$\\
\hline
$\begin{matrix} D_5(a_1)\\ D_{5}(a_1)+A_1  \end{matrix}\quad A_6$&$SU(2)_{10} \times U(1)\times SU(2)_{56} \times  \color{red}{{SU(2)}_{36}}$&$SU(2)_{10} \times {U(1)} \times SU(2)_{56} \times  \color{red}{{SU(2)}_{28}\times {SU(2)}_{8}}$\\
\hline
$\begin{matrix} D_5(a_1)+A_1\\ D_{5}(a_1)+A_1  \end{matrix}\quad A_6$&$SU(2)_{56}^2 \times  \color{red}{{SU(2)}_{36}}$&$SU(2)_{56}^2 \times  \color{red}{{SU(2)}_{28}\times {SU(2)}_{8}}$\\
\hline
$\begin{matrix} D_5(a_1)\\ E_6(a_3)  \end{matrix}\quad A_6$&$SU(2)_{10} \times U(1) \times  {\color{red}{{SU(2)}_{36}} \times SU(2)_{20}}$&${SU(2)}_{10}\times {\color{red}{SU(4)}_{20}\times {SU(2)}_{8}^2} \times U(1)$\\
\hline
$\begin{matrix} D_5(a_1)+A_1\\ E_6(a_3)  \end{matrix}\quad A_6$&${\color{red}{SU(2)_{56} \times {SU(2)}_{36}} \times SU(2)_{20}}$&${\color{red}{Spin(7)}_{20} \times SU(2)_{16} \times {SU(2)}_{8}^2}$\\
\hline
$\begin{matrix} D_5(a_1)\\ A_4+A_2  \end{matrix}\quad D_5+A_1$&$SU(2)_{10} \times U(1) \times  {\color{red}{{SU(2)}_{108}}} \times SU(2)_{12}$&$SU(2)_{10} \times U(1) \times  {\color{red}{{SU(2)}_{96}}\times SU(2)_{12}} \times SU(2)_{12}$\\
\hline
$\begin{matrix} A_4+A_2\\ (A_5)'  \end{matrix}\quad D_5+A_1$&$SU(2)_{20} \times SU(2)_9 \times  {\color{red}{{SU(2)}_{108}}} \times SU(2)_{12}$&$SU(2)_{20} \times SU(2)_9 \times  {\color{red}{{SU(2)}_{96}}\times SU(2)_{12}} \times SU(2)_{12}$\\
\hline
$\begin{matrix} D_5(a_1)+A_1\\ A_4+A_2  \end{matrix}\quad D_5+A_1$&$SU(2)_{56} \times  {\color{red}{{SU(2)}_{108}}} \times SU(2)_{12}$&$SU(2)_{56} \times  {\color{red}{{SU(2)}_{96}}\times SU(2)_{12}} \times SU(2)_{12}$\\
\hline
$\begin{matrix} A_4+A_2\\ D_6(a_2)  \end{matrix}\quad D_5+A_1$&$SU(2)_{9} \times  {\color{red}{{SU(2)}_{108}} \times SU(2)_{12}}$&$SU(2)_{9} \times  {\color{red}{{SU(2)}_{48}}\times SU(2)_{24} \times (G_2)_{12}}$\\
\hline
$\begin{matrix} A_4+A_2\\ E_6(a_3)  \end{matrix}\quad D_5+A_1$&$SU(2)_{20} \times SU(2)_{12} \times  {\color{red}{{SU(2)}_{108}}}$&$SU(2)_{20} \times SU(2)_{12}\times  {\color{red}{SU(2)}_{24}^2 \times SU(2)_{60}}$\\
\hline
$\begin{matrix} A_4+A_2\\ E_7(a_5)  \end{matrix}\quad D_5+A_1$&${\color{red}{{SU(2)}_{108}}\times SU(2)_{12}}$&${\color{red}(G_{2})_{12} \times {SU(2)}_{24}^3}$\\
\hline
$\begin{matrix} (A_3+A_1)' \\ A_6  \end{matrix}\quad D_5+A_1$&$SU(2)_{13} \times SU(2)_{24} \times SU(2)_{12}^2 \times  {\color{red}{{SU(2)}_{36}}}$&$SU(2)_{13} \times SU(2)_{12}^2 \times SU(2)_{24} \times  {\color{red}{SU(2)}_{24} \times SU(2)_{12}}$\\
\hline
$\begin{matrix} D_4(a_1) \\ A_6  \end{matrix}\quad D_5+A_1$&$SU(2)_{12}^4 \times  {\color{red}{{SU(2)}_{36}}}$&$SU(2)_{12}^4 \times  {\color{red}{SU(2)_{12}^3}}$\\
\hline
$\begin{matrix} D_4(a_1)+A_1 \\ A_6  \end{matrix}\quad D_5+A_1$&$SU(2)_{12}^2 \times  {\color{red}{{SU(2)}_{36}} \times SU(2)_{12}}$&$SU(2)_{12}^2 \times  {\color{red}{SU(2)_{12}^2 \times Sp(2)_{12}}}$\\
\hline
$\begin{matrix} A_2+2A_1 \\ D_5+A_1 \end{matrix}\quad D_5+A_1$&$SU(2)_{16} \times SU(2)_{28} \times SU(2)_{12}^2 \times  {\color{red}{{SU(2)}_{84}}}$&$SU(2)_{16} \times SU(2)_{12}^2 \times SU(2)_{28} \times  {\color{red}{SU(2)}_{28} \times SU(2)_{56}}$\\
\hline
$\begin{matrix} D_5(a_1) \\ A_4+A_2 \end{matrix}\quad E_7(a_4)$&$SU(2)_{10} \times U(1) \times  {\color{red}{{SU(2)}_{108}}}$&$SU(2)_{10} \times U(1) \times  {\color{red}{SU(2)}_{96} \times SU(2)_{12}}$\\
\hline
$\begin{matrix} A_3+A_2+A_1 \\ A_5+A_1 \end{matrix}\quad E_7(a_4)$&$SU(2)_{26} \times  {\color{red}{{SU(2)}_{224}}}$&$SU(2)_{26} \times  {\color{red}{SU(2)}_{112} \times SU(2)_{112}}$\\
\hline
$\begin{matrix} D_5(a_1)+A_1 \\ A_4+A_2 \end{matrix}\quad E_7(a_4)$&$SU(2)_{56} \times  {\color{red}{{SU(2)}_{108}}}$&$SU(2)_{56} \times  {\color{red}{SU(2)}_{96} \times SU(2)_{12}}$\\
\hline
$\begin{matrix} A_3+A_2+A_1 \\ D_6(a_2) \end{matrix}\quad E_7(a_4)$&$SU(2)_{9} \times  {\color{red}{{SU(2)}_{224}}}$&$SU(2)_{9} \times  {\color{red}{SU(2)}_{112} \times SU(2)_{112}}$\\
\hline
$\begin{matrix} A_4+A_2 \\ E_6(a_3) \end{matrix}\quad E_7(a_4)$&${\color{red}{SU(2)_{20} \times{SU(2)}_{108}}}$&${\color{red}{SU(2)}_{24}^2 \times (G_2)_{20}}$\\
\hline
$\begin{matrix} A_3+A_2+A_1 \\ E_7(a_5) \end{matrix}\quad E_7(a_4)$&${\color{red}{{SU(2)}_{224}}}$&${\color{red}{SU(2)}_{112} \times SU(2)_{112}}$\\
\hline
$\begin{matrix} A_2+2A_1 \\ D_5+A_1 \end{matrix}\quad E_7(a_4)$&$SU(2)_{16} \times SU(2)_{28} \times SU(2)_{12} \times  {\color{red}{{SU(2)}_{84}}}$&$SU(2)_{16} \times SU(2)_{12} \times SU(2)_{28} \times  {\color{red}{SU(2)}_{28} \times SU(2)_{56}}$\\
\hline
$\begin{matrix} A_2+A_1 \\ E_7(a_4) \end{matrix}\quad E_7(a_4)$&$SU(4)_{18} \times {\color{red}{{U(1)}}}$&$SU(4)_{18} \times   {\color{red} U(1) \times SU(2)_{56}}$\\
\hline
$\begin{matrix} D_4 \\ A_3+A_2 \end{matrix}\quad E_6(a_1)$&$Sp(3)_{12} \times SU(2)_{12} \times U(1) \times {\color{red}{{U(1)}}}$&$Sp(3)_{12} \times SU(2)_{12} \times U(1) \times {\color{red}{U(1) \times SU(2)_{56}}}$\\
\hline
$\begin{matrix} D_4+A_1 \\ A_3+A_2 \end{matrix}\quad E_6(a_1)$&$Sp(2)_{11} \times SU(2)_{12} \times U(1) \times {\color{red}{{U(1)}}}$&$Sp(2)_{11} \times SU(2)_{12} \times U(1) \times {\color{red}{U(1) \times SU(2)_{56}}}$\\
\hline
$\begin{matrix} D_5(a_1) \\ A_3+A_2 \end{matrix}\quad E_6(a_1)$&$SU(2)_{10} \times SU(2)_{12} \times U(1)^2 \times {\color{red}{{U(1)}}}$&$SU(2)_{10} \times SU(2)_{12} \times U(1)^2 \times {\color{red}{U(1) \times SU(2)_{56}}}$\\
\hline
$\begin{matrix} A_4 \\ A_4+A_2 \end{matrix}\quad E_6(a_1)$&$SU(3)_{12} \times U(1)^2 \times {\color{red}{{SU(2)}_{108}}}$&$SU(3)_{12} \times U(1)^2 \times  {\color{red}{SU(2)}_{54} \times SU(2)_{54}}$\\
\hline
$\begin{matrix} A_4+A_1 \\ A_4+A_2 \end{matrix}\quad E_6(a_1)$&$U(1)^3 \times {\color{red}{{SU(2)}_{108}}}$&$U(1)^3 \times  {\color{red}{SU(2)}_{54} \times SU(2)_{54}}$\\
\hline
$\begin{matrix} D_5(a_1)+A_1 \\ A_3+A_2 \end{matrix}\quad E_6(a_1)$&$SU(2)_{56} \times SU(2)_{12} \times U(1) \times {\color{red}{{U(1)}}}$&$SU(2)_{56} \times SU(2)_{12} \times U(1) \times {\color{red}{U(1) \times SU(2)_{56}}}$\\
\hline
$\begin{matrix} A_3+A_2 \\ E_6(a_3) \end{matrix}\quad E_6(a_1)$&$SU(2)_{12} \times {\color{red}{{U(1)^2}}} \times {\color{red}{{SU(2)}_{20}}}$&$SU(2)_{12} \times   {\color{red}{U(1)^2 \times SU(4)}_{20} \times SU(2)_{36}}$\\
\hline
$\begin{matrix} A_2 \\ A_6 \end{matrix}\quad E_6(a_1)$&$SU(6)_{20} \times U(1) \times {\color{red}{{SU(2)}_{36}}}$&$SU(6)_{20} \times U(1) \times  {\color{red}{SU(2)}_{18} \times SU(2)_{18}}$\\
\hline
$\begin{matrix} A_2+A_1 \\ A_6 \end{matrix}\quad E_6(a_1)$&$SU(4)_{18} \times U(1) \times {\color{red}{{SU(2)}_{36}}\times U(1)}$&$SU(4)_{18} \times U(1) \times  {\color{red}{SU(3)}_{18} \times SU(3)_{18}}$\\
\hline
$\begin{matrix} A_2+2A_1 \\ (A_3+A_1)' \end{matrix}\quad E_6$&$\begin{gathered}SU(2)_{16} \times SU(2)_{28} \times SU(2)_{13}\\ \times SU(2)_{24} \times SU(2)_{12}^2 \times {\color{red}{{SU(2)}_{84}}}\end{gathered}$&$\begin{gathered}SU(2)_{16} \times SU(2)_{28} \times SU(2)_{13}\\ \times SU(2)_{24} \times SU(2)_{12}^2 \times  {\color{red}{SU(2)}_{28} \times SU(2)_{56}}\end{gathered}$\\
\hline
$\begin{matrix} A_2+2A_1 \\ D_4(a_1) \end{matrix}\quad E_6$&$SU(2)_{16} \times SU(2)_{28} \times SU(2)_{12}^4 \times {\color{red}{{SU(2)}_{84}}}$&$SU(2)_{16} \times SU(2)_{28} \times SU(2)_{12}^4 \times  {\color{red}{SU(2)}_{28}^3}$\\
\hline
$\begin{matrix} A_2+2A_1 \\ A_3+2A_1 \end{matrix}\quad E_6$&$\begin{gathered}SU(2)_{16} \times SU(2)_{28} \times SU(2)_{13}\\ \times SU(2)_{24} \times SU(2)_{12} \times {\color{red}{{SU(2)}_{84}}}\end{gathered}$&$\begin{gathered}SU(2)_{16} \times SU(2)_{28} \times SU(2)_{13}\\ \times SU(2)_{24} \times SU(2)_{12} \times  {\color{red}{SU(2)}_{28} \times SU(2)_{56}}\end{gathered}$\\
\hline
$\begin{matrix} 0 \\ D_4 \end{matrix}\quad E_6$&$Sp(3)_{12} \times SU(2)_{12} \times {\color{red}{{(E_7)_{36}}}}$&$Sp(3)_{12} \times SU(2)_{12} \times {\color{red}{{(E_7)_{28}\times (E_7)_{8}}}}$\\
\hline
$\begin{matrix} A_2+2A_1 \\ D_4(a_1)+A_1 \end{matrix}\quad E_6$&$SU(2)_{16} \times SU(2)_{28} \times SU(2)_{12}^3 \times {\color{red}{{SU(2)}_{84}}}$&$SU(2)_{16} \times SU(2)_{28} \times SU(2)_{12}^3 \times  {\color{red}{SU(2)}_{28}^3}$\\
\hline
$\begin{matrix} A_2 \\ A_3+A_2 \end{matrix}\quad E_6$&$SU(6)_{20} \times SU(2)_{12}^2 \times {\color{red} U(1)}$&$SU(6)_{20} \times SU(2)_{12}^2 \times   {\color{red}{U(1) \times SU(2)}_{56}}$\\
\hline
$\begin{matrix} A_2+A_1 \\ A_3+A_2 \end{matrix}\quad E_6$&$SU(4)_{18} \times SU(2)_{12}^2 \times U(1) \times {\color{red} U(1)}$&$SU(4)_{18} \times SU(2)_{12}^2 \times U(1) \times  {\color{red}{U(1) \times SU(2)}_{56}}$\\
\hline
$\begin{matrix} 0 \\ D_4+A_1 \end{matrix}\quad E_6$&$Sp(2)_{11} \times SU(2)_{12} \times {\color{red}{{(E_7)_{36}}}}$&$Sp(2)_{11} \times SU(2)_{12} \times {\color{red}{{(E_7)_{28}\times (E_7)_{8}}}}$\\
\hline
$\begin{matrix} 0 \\ D_5(a_1)+A_1 \end{matrix}\quad E_6$&$SU(2)_{56} \times SU(2)_{12} \times {\color{red}{{(E_7)_{36}}}}$&$SU(2)_{56} \times SU(2)_{12} \times {\color{red}{{(E_7)_{28}\times (E_7)_{8}}}}$\\
\hline
$\begin{matrix} A_{3}+2A_1 \\ A_3+A_2+A_1 \end{matrix}\quad E_7(a_3)$&$SU(2)_{13}\times SU(2)_{24} \times {\color{red}{{SU(2)}_{224}}}$&${SU(2)}_{13}\times {SU(2)}_{24}\times {\color{red}{{SU(2)}_{112}^2}}$\\
\hline
$\begin{matrix} D_4(a_1)+A_1 \\ A_3+A_2+A_1 \end{matrix}\quad E_7(a_3)$&$SU(2)_{12}^2 \times {\color{red}{{SU(2)}_{224}}}$&${SU(2)}_{12}^2 \times {\color{red}{{SU(2)}_{112}^2}}$\\
\hline
$\begin{matrix} A_2+A_1 \\ D_5(a_1) \end{matrix}\quad E_7(a_3)$&$SU(4)_{18} \times SU(2)_{10} \times U(1) \times {\color{red}{{U(1)}}}$&$SU(4)_{18} \times SU(2)_{10} \times U(1) \times {\color{red}{{SU(2)_{28}^2}}}$\\
\hline
$\begin{matrix} 4A_1 \\ D_5(a_1)+A_1 \end{matrix}\quad E_7(a_3)$&$Sp(3)_{19} \times {\color{red}{{SU(2)_{56}}}}$&$Sp(3)_{19} \times {\color{red}{{SU(2)_{28}^2}}}$\\
\hline
$\begin{matrix} A_2+2A_1 \\ (A_3+A_1)' \end{matrix}\quad E_7(a_2)$&$\begin{gathered}  {SU(2)}_{16}\times {SU(2)}_{28}\times {SU(2)}_{13}\\\times {SU(2)}_{24}\times {SU(2)}_{12} \times {\color{red}{{SU(2)_{84}}}}\end{gathered}$&$\begin{gathered}{SU(2)}_{16}\times {SU(2)}_{28}\times {SU(2)}_{13}\\ \times {SU(2)}_{24}\times {SU(2)}_{12} \times {\color{red}{{SU(2)_{56} \times SU(2)_{28}}}}\end{gathered}$\\
\hline
$\begin{matrix} A_2+2A_1 \\ D_4(a_1) \end{matrix}\quad E_7(a_2)$&${SU(2)}_{16}\times {SU(2)}_{28}\times {SU(2)}_{12}^3 \times {\color{red}{{SU(2)_{84}}}}$&${SU(2)}_{16}\times {SU(2)}_{28}\times  {SU(2)}_{12}^3 \times {\color{red}{{ SU(2)_{28}^3}}}$\\
\hline
$\begin{matrix} A_2+A_1 \\ A_3+2A_1 \end{matrix}\quad E_7(a_2)$&$SU(4)_{18} \times SU(2)_{13} \times SU(2)_{24} \times {\color{red}{U(1) }}$&$SU(4)_{18} \times SU(2)_{13} \times SU(2)_{24} \times   {\color{red}{U(1) \times SU(2)}_{56}}$\\
\hline
$\begin{matrix} 0 \\ D_4 \end{matrix}\quad E_7(a_2)$&$Sp(3)_{12} \times {\color{red}{{(E_7)_{36}}}}$&$Sp(3)_{12} \times {\color{red}{{(E_7)_{28}\times (E_7)_{8}}}}$\\
\hline
$\begin{matrix} A_2+A_1 \\ D_4(a_1)+A_1 \end{matrix}\quad E_7(a_2)$&$SU(4)_{18} \times SU(2)_{12}^2 \times {\color{red}{U(1) }}$&$SU(4)_{18} \times SU(2)_{12}^2 \times   {\color{red}{U(1) \times SU(2)}_{56}}$\\
\hline
$\begin{matrix} A_2+2A_1 \\ D_4(a_1)+A_1 \end{matrix}\quad E_7(a_2)$&$SU(2)_{16} \times SU(2)_{12}^2 \times   {\color{red}{SU(2)_{28} \times SU(2)}_{84}}$&$SU(2)_{16} \times SU(2)_{12}^2  \times  {\color{red}{Sp(2)}_{28} \times SU(2)_{28}^2}$\\
\hline
$\begin{matrix} A_2\\ A_3+A_2 \end{matrix}\quad E_7(a_2)$&$SU(6)_{20} \times SU(2)_{12} \times {\color{red}{U(1) }}$&$SU(6)_{20} \times SU(2)_{12}  \times  {\color{red}{U(1) \times SU(2)}_{56}}$\\
\hline
$\begin{matrix} A_2+A_1 \\ A_3+A_2 \end{matrix}\quad E_7(a_2)$&$SU(4)_{18} \times SU(2)_{12} \times U(1) \times {\color{red}{U(1)}}$&$SU(4)_{18} \times SU(2)_{12} \times U(1) \times  {\color{red}{SU(3)}_{56}}$\\
\hline
$\begin{matrix} 4A_1 \\ A_3+A_2+A_1 \end{matrix}\quad E_7(a_2)$&$Sp(3)_{19} \times {\color{red}{{SU(2)_{224}}}}$&$Sp(3)_{19} \times {\color{red}{{SU(2)_{112}^2}}}$\\
\hline
$\begin{matrix} 0 \\ D_4+A_1 \end{matrix}\quad E_7(a_2)$&$Sp(2)_{11} \times {\color{red}{{(E_7)_{36}}}}$&$Sp(2)_{11} \times {\color{red}{{(E_7)_{28}\times (E_7)_{8}}}}$\\
\hline
$\begin{matrix} 0 \\ D_5(a_1)+A_1 \end{matrix}\quad E_7(a_2)$&$SU(2)_{56} \times {\color{red}{{(E_7)_{36}}}}$&$SU(2)_{56} \times {\color{red}{{(E_7)_{28}\times (E_7)_{8}}}}$\\
\hline
$\begin{matrix} A_5+A_1 \\ D_5(a_1)+A_1 \end{matrix}\quad A_6$&$SU(2)_{26} \times SU(2)_{56} \times {\color{red}{{SU(2)_{36}}}}$&$SU(2)_{26} \times SU(2)_{48} \times {\color{red}{{SU(2)_{24}\times SU(2)_{9}}}+\frac{1}{2}(1,3,2)}$\\
\hline
$\begin{matrix} D_6(a_2) \\ D_5(a_1) \end{matrix}\quad A_6$&$SU(2)_{9} \times U(1) \times {\color{red}{{SU(2)_{10} \times SU(2)_{36}}}}$&$\begin{gathered}SU(2)_{9} \times U(1) \times {\color{red}{SU(2)}_{16}\times}\\ {\color{red}{SU(2)}_{8}\times {Sp(2)}_{10}+\frac{1}{2}(1,1,2)}\end{gathered}$\\
\hline
$\begin{matrix} D_6(a_2) \\ D_5(a_1)+A_1 \end{matrix}\quad A_6$&$SU(2)_{9} \times {\color{red}{{SU(2)_{56} \times SU(2)_{36}}}}$&$\begin{gathered}{SU(2)}_{9} \times {\color{red}{SU(2)}_{16}^2\times SU(2)_{32}\times}\\ {\color{red}{SU(2)}_{9} \times {SU(2)}_{8}+\frac{1}{2}(1,3,2)}\end{gathered}$\\
\hline
$\begin{matrix} E_7(a_5) \\ D_5(a_1) \end{matrix}\quad A_6$&$U(1) \times {\color{red}{{SU(2)_{10} \times SU(2)_{36}}}}$&$U(1) \times {\color{red}{{U(1)^2 \times  SU(2)_{8}^3 \times Sp(2)_{10}}}+\frac{1}{2}(1,2)}$\\
\hline
$\begin{matrix} E_7(a_5) \\ D_5(a_1)+A_1 \end{matrix}\quad A_6$&${\color{red}{{SU(2)_{56} \times SU(2)_{36}}}}$&${\color{red}{{SU(2)_{16}^3 \times SU(2)_{9} \times SU(2)_{8}^3}}+\frac{1}{2}(3,2)}$\\
\hline
$\begin{matrix} 2A_2+A_1 \\ D_5+A_1 \end{matrix}\quad D_5+A_1$&$SU(2)_{12}^2 \times {\color{red}{{SU(2)_{38} \times SU(2)_{36}}}}$&$SU(2)_{12}^2 \times {\color{red}{{SU(2)_{24}^2 \times SU(2)_{16} \times SU(2)_{13}}}+\frac{1}{2}(1,1,2,1)}$\\
\hline
$\begin{matrix} A_4+A_1 \\ A_5+A_1 \end{matrix}\quad E_7(a_4)$&$SU(2)_{26} \times U(1) \times {\color{red}{U(1)}}$&$SU(2)_{26} \times U(1)  \times {\color{red}{U(1)  \times SU(2)_{54}}+(1)}$\\
\hline
$\begin{matrix} A_4+A_2 \\ A_5+A_1 \end{matrix}\quad E_7(a_4)$&$SU(2)_{26} \times  {\color{red}{SU(2)_{108}}}$&$SU(2)_{26} \times {\color{red}{SU(2)_{76}\times SU(2)_{26}}+\frac{1}{2}(1,4)}$\\
\hline
$\begin{matrix} A_4+A_1 \\ D_6(a_2) \end{matrix}\quad E_7(a_4)$&$SU(2)_{9} \times U(1) \times {\color{red}{U(1)}}$&$SU(2)_{9} \times U(1)  \times {\color{red}{SU(2)_{54} \times U(1)^2}+(1)}$\\
\hline
$\begin{matrix} A_4+A_2 \\ D_6(a_2) \end{matrix}\quad E_7(a_4)$&$SU(2)_{9} \times  {\color{red}{SU(2)_{108}}}$&$SU(2)_{9} \times {\color{red}{SU(2)_{48}\times SU(2)_{26} \times SU(2)_{24}}+\frac{1}{2}(1,4)}$\\
\hline
$\begin{matrix} A_4+A_1 \\ E_7(a_5) \end{matrix}\quad E_7(a_4)$&$U(1) \times{\color{red}{U(1)}}$&$U(1)  \times {\color{red}{SU(2)_{54} \times U(1)^3}+(1)}$\\
\hline
$\begin{matrix} A_4+A_2 \\ E_7(a_5) \end{matrix}\quad E_7(a_4)$&${\color{red}{SU(2)_{108}}}$&${\color{red}{ SU(2)_{26} \times SU(2)_{24}^3}+\frac{1}{2}(4)}$\\
\hline
$\begin{matrix} (A_3+A_1)'' \\ A_6 \end{matrix}\quad E_7(a_4)$&$Spin(7)_{16} \times {\color{red}{SU(2)_{36}}}$&$Spin(7)_{16} \times {\color{red}{ SU(2)_{24} \times SU(2)_{11}}+\frac{1}{2}(1,2)}$\\
\hline
$\begin{matrix} 2A_2+A_1 \\ D_5+A_1 \end{matrix}\quad E_7(a_4)$&${\color{red}{SU(2)_{36} \times SU(2)_{38} \times SU(2)_{12}}}$&${\color{red}{ Sp(2)_{12} \times SU(2)_{24}^2 \times SU(2)_{13}}+\frac{1}{2}(1,2,1)}$\\
\hline
$\begin{matrix} 2A_2 \\ E_7(a_4) \end{matrix}\quad E_7(a_4)$&${\color{red}{(G_2)_{16} \times SU(2)_{36}}}$&${\color{red}{Spin(7)_{16} \times SU(2)_{24} \times SU(2)_{11}}+\frac{1}{2}(1,2)}$\\
\hline
$\begin{matrix} 2A_2+A_1 \\ E_7(a_4) \end{matrix}\quad E_7(a_4)$&${\color{red}{SU(2)_{36} \times SU(2)_{38}}}$&${\color{red}{Sp(2)_{24} \times SU(2)_{13} \times SU(2)_{11}}+\frac{1}{2}(1,2)+\frac{1}{2}(2,1)}$\\
\hline
$\begin{matrix} A_3+A_2 \\ D_6(a_2) \end{matrix}\quad E_6(a_1)$&$SU(2)_{9} \times {\color{red}{SU(2)_{12} \times U(1)^2}}$&$SU(2)_9 \times  {\color{red}{SU(3)_{12} \times SU(2)_{36} \times SU(2)_{18}}+2(1,1)}$\\
\hline
$\begin{matrix} A_3+A_2+A_1 \\ D_6(a_2) \end{matrix}\quad E_6(a_1)$&$SU(2)_{9} \times U(1) \times  {\color{red}{SU(2)_{224}}}$&$SU(2)_9 \times U(1) \times {\color{red}{SU(3)_{36} \times SU(3)_{18} }+(1,3)}$\\
\hline
$\begin{matrix} A_3+A_2 \\ E_7(a_5) \end{matrix}\quad E_6(a_1)$&$U(1) \times {\color{red}{SU(2)_{12} \times U(1)}}$&$U(1) \times  {\color{red}{SU(3)_{12} \times SU(2)_{18}^3}+2(1)}$\\
\hline
$\begin{matrix} A_3+A_2+A_1 \\ E_7(a_5) \end{matrix}\quad E_6(a_1)$&$U(1) \times  {\color{red}{SU(2)_{224}}}$&$U(1) \times {\color{red}{ SU(3)_{18}^3 }+(3)}$\\
\hline
$\begin{matrix} 2A_2+A_1 \\ (A_3+A_1)' \end{matrix}\quad E_6$&$\begin{gathered}SU(2)_{24} \times SU(2)_{12}^2 \times\\ {\color{red}{SU(2)_{36} \times SU(2)_{38} \times SU(2)_{13}}}\end{gathered}$&$\begin{gathered}{SU(2)}_{24} \times {SU(2)}_{12}^2 \times{\color{red}{Sp(2)}_{13} \times}\\ {\color{red} {SU(2)}_{24}^2 \times {SU(2)}_{12}+\frac{1}{2}(1,1,1,1,2,1)}\end{gathered}$\\
\hline
$\begin{matrix} 2A_2+A_1 \\ D_4(a_1) \end{matrix}\quad E_6$&$SU(2)_{12}^4 \times {\color{red}{SU(2)_{36} \times SU(2)_{38} }}$&$SU(2)_{12}^4 \times  {\color{red}{SU(2)_{12}^6}+\frac{1}{2}(1,1,1,1,1,2)}$\\
\hline
$\begin{matrix} 2A_2+A_1 \\ D_4(a_1)+A_1 \end{matrix}\quad E_6$&$SU(2)_{12} \times {\color{red}{SU(2)_{12}^2 \times SU(2)_{36} \times SU(2)_{38} }}$&$SU(2)_{12} \times  {\color{red}{SU(2)_{12}^2 \times Sp(2)_{12}^3}+\frac{1}{2}(1,1,1,1,2)}$\\
\hline
$\begin{matrix} A_3+2A_1 \\ A_4+A_1 \end{matrix}\quad E_7(a_3)$&$SU(2)_{13} \times U(1) \times {\color{red}{SU(2)_{24} \times U(1) }}$&$SU(2)_{13} \times U(1) \times{\color{red}{SU(3)_{24} \times SU(2)_{54}+(1,1)}}$\\
\hline
$\begin{matrix} D_4(a_1)+A_1 \\ A_4+A_1 \end{matrix}\quad E_7(a_3)$&$U(1) \times {\color{red}{SU(2)_{12}^2 \times U(1) }}$&$U(1) \times{\color{red}{SU(3)_{12}^2 \times SU(2)_{54}+(1,1)}}$\\
\hline
$\begin{matrix} A_3+A_2 \\ A_4+A_1 \end{matrix}\quad E_7(a_3)$&$U(1) \times {\color{red}{SU(2)_{12} \times U(1)^2 }}$&$U(1) \times{\color{red}{SU(3)_{12} \times SU(3)_{54}+(1)}}$\\
\hline
$\begin{matrix} A_2+2A_1\\ E_6(a_3) \end{matrix}\quad E_7(a_3)$&${\color{red}{SU(2)_{16} \times SU(2)_{28} \times SU(2)_{84} \times SU(2)_{20} }}$&${\color{red}{Spin(7)_{16} \times SU(2)_{24}^3+\frac{1}{2}(1,2,2,2)}}$\\
\hline
$\begin{matrix} 2A_2+A_1 \\ D_4(a_1) \end{matrix}\quad E_7(a_2)$&${\color{red}{SU(2)_{36} \times SU(2)_{38} \times SU(2)_{12}^3 }}$&${\color{red}{SU(2)_{12}^3 \times Sp(2)_{12}^3}+\frac{1}{2}(1,2,1,1,1)}$\\
\hline
$\begin{matrix} 4A_1 \\ A_4+A_1 \end{matrix}\quad E_7(a_2)$&$U(1) \times {\color{red}{Sp(3)_{19} \times U(1)}}$&$U(1) \times  {\color{red}{SU(8)_{18} \times SU(2)_{36}}+\frac{1}{2}(6)+(1)}$\\
\hline
\end{longtable}
}

{
\footnotesize
\renewcommand{\arraystretch}{1.25}

\begin{longtable}{|c|c|c|}
\caption{Undetermined levels in the $E_7$ theory}\\
\hline
Fixture&Manifest Flavour Symmetry&Enhanced Flavour Symmetry\\
\endfirsthead
\hline
Fixture&Manifest Flavour Symmetry&Enhanced Flavour Symmetry\\
\endhead
\hline 
$\begin{matrix} D_6(a_2)\\ D_6(a_2) \end{matrix}\quad D_5(a_1)+A_1$&${\color{red}{SU(2)}_{56}} \times {SU(2)}_{9}^2$&${\color{red}{SU(2)}_{56-k}\times {SU(2)}_{k}}\times {SU(2)}_{9}^2$\\
\hline 
$\begin{matrix} D_6(a_2)\\ E_7(a_5)  \end{matrix}\quad D_5(a_1)+A_1$&${\color{red}{SU(2)}_{56}} \times {SU(2)}_{9}$&${\color{red}{SU(2)}_{56-k}\times {SU(2)}_{k}}\times {SU(2)}_{9}$\\
\hline 
$\begin{matrix} E_7(a_5)\\ E_7(a_5)  \end{matrix}\quad D_5(a_1)+A_1$&$\color{red}{SU(2)}_{56}$&$\color{red} {SU(2)}_{56-k}\times {SU(2)}_{k}$\\
\hline 
$\begin{matrix} A_3+A_2+A_1 \\ A_6 \end{matrix}\quad E_7(a_4)$&${\color{red}{SU(2)_{224} \times SU(2)_{36}}}$&${\color{red}{ SU(2)_{128-k} \times SU(2)_{k} \times Sp(3)_{11}}+\frac{1}{2}(3,2)}$\\
\hline 
\end{longtable}
}

\section{Theories with the Same ``Conventional Invariants"}\label{Distinguishing}

\subsection{Global Form of the Flavour Symmetry Group}\label{global_form_of_the_flavour_symmetry_group_for_classs_theories}

Each puncture $p_i$, in a theory of class-S, contributes a factor of $\mathfrak{f}_{p_{i}}$ to the flavour symmetry algebra of the theory. The ``manifest" flavour symmetry algebra, $\mathfrak{f}_{\text{manifest}}= \oplus_i \mathfrak{f}_{p_{i}}$. In general, the full symmetry algebra is some enhancement $\mathfrak{f}\supset \mathfrak{f}_{\text{manifest}}$. Let $F$ be the corresponding compact Lie group, whose semi-simple part is simply-connected. In addition, every $\mathcal{N}=2$ SCFT has a $\mathbb{Z}_2$ global symmetry\footnote{$S$ is the center of the $\mathcal{N}=2$ superconformal supergroup,  $SU(2,2|2)$. So the symmetries that act effectively are $\bigl(SU(2,2|2)\times F\bigr)/\Gamma$. For brevity, we'll omit the $SU(2,2|2)$ factor from our formul\ae\ below.}, $S$, generated by $\gamma = e^{2\pi i(R+j_1+j_2)}$. In general, $F\times S$ does not act effectively. Some subgroup $\Gamma\subset Z(F)\times S$ acts trivially on all of the local operators in the theory. We call $\hat{F}=(F\times S)/\Gamma$ the global form of the flavour symmetry group. Determining $\Gamma$, and hence $\hat{F}$, is the subject of this subsection. Without $S$, a method for determining $\Gamma$ was proposed in  \cite{Bhardwaj:2021ojs}. Here we extend that prescription to include $S$.

To motivate the proposal of  \cite{Bhardwaj:2021ojs}, consider the Schur Index of an untwisted class S theory, of type $\mathfrak{j}\in\text{ADE}$, with $n$ punctures $p_{i}$ determined in \cite{Gadde:2011uv}
\begin{equation}\label{schur}
I_{\text{Schur}}(\tau) = \sum_{\Lambda} (\psi^{\rho}_{\Lambda}(\tau))^{2-2g-n}\prod_{i=1}^{n}\psi^{p_{i}}_{\Lambda}( \mathbf{a}_{i}, \tau)
\end{equation}
The sum is over highest weight representations $\Lambda$ of $\mathfrak{j}$. Here $\rho$ corresponds to the trivial puncture (the regular embedding $\mathfrak{su}(2)\hookrightarrow \mathfrak{j}$) and the wave functions $\psi^{p_{i}}_{\Lambda}( \mathbf{a}_{i}, \tau)$ are given by

\begin{displaymath}
\psi^{p_{i}}_{\Lambda}( \mathbf{a}_{i}, \tau) = K^{p_{i}}_{\Lambda}( \mathbf{a}_{i}, \tau)\chi^{p_{i}}_{\Lambda}( \mathbf{a}_{i}, \tau)
\end{displaymath}
The characters, $\chi^{p_{i}}_{\Lambda}( \mathbf{a}_{i}, \tau)$, are those corresponding to the decomposition of $\Lambda$ under the embedding of $\mathfrak{su}(2) \times \mathfrak{f}_{p_{i}}\hookrightarrow \mathfrak{j}$. The $K$-factors are given by the plethystic exponential

\begin{displaymath}
\text{P.E.}\left[\frac{1}{1-\tau^2}\chi_{\text{adj}}^{p_{i}}( \mathbf{a}_{i}, \tau)\right]
\end{displaymath}

For $\mathfrak{j} \neq D_{2n}$, there's a highest weight irreducible representation $R$ of $\mathfrak{j}$ such that all of the other highest weight representations are contained in the tensor powers of $R$. Let $R_{p_{i}}$ denotes the decomposition of $R$ under $\mathfrak{su}(2) \times \mathfrak{f}_{p_{i}}$. Then $\Gamma$ is the subgroup that leaves $R_{\rho}^n \otimes \Big(\bigotimes_{i} R_{p_{i}} \Big)$ invariant (when we take $\gamma$ to act as the center of $SU(2)$, i.e. as $\tau\to -\tau$ on the level of the superconformal index). This can be seen from the expression  \eqref{schur} for the superconformal index. If the above condition is satisfied then, assuming the $K$-factors are invariant as will be shown below, $\Gamma$ will leave the contribution of $\Lambda =R$ to the sum in \eqref{schur} invariant. Since all other reps can be generated by tensor powers of $R$, it will leave those terms in the sum invariant as well.

To see that the $K$-factors are invariant, note an element of the center of the flavour symmetry must act as a multiple of the identity on each irreducible representation in the decomposition of $R$, hence it will act as a diagonal matrix on the decomposition $R_{p_{i}}$. If the tensor product of diagonal matrices is plus or minus the identity, then each diagonal matrix must be a multiple of the identity. Therefore if an element of the center of $F \times S$ leaves the $\bigotimes_{i} R_{p_{i}}$ invariant or anti-invariant, then the element of the center must act as a multiple of the identity on each $R_{p_{i}}$. Since the adjoint can be generated by tensor powers of R, it must act as a multiple of the identity on the decomposition of the adjoint of $\mathfrak{j}$. Since this is an embedding, the decomposition of the adjoint will contain $(3;1)$ which is invariant under the center of $S \times F$ and hence the entire decomposition is invariant. Thus the $K$-factors are invariant.

Note that in the $E_8$ case the above implies that $\Gamma$ should leave each $R_{p_{i}}$ invariant. Thus one does not need to worry about any tensor products. In general, the action of elements of $\Gamma$ on $R_{p_{i}}$ will be multiplication by an $N$th root of unity where $N$ is the smallest power such that $R^N$ contains the adjoint representation.

To generalize to $\mathfrak{j} = D_{2n}$, one must repeat the procedure for $R=S,S'$, the two chiral spinor representations. For twisted theories, the Schur index is a sum over highest weight representations of the Langlands dual, $\mathfrak{g}$, of the subalgebra $\mathfrak{g}^\vee\subset \mathfrak{j}$ invariant under the outer automorphism \cite{Mekareeya:2012tn}. In this case there is a map $\sigma$ from the set of representations of $\mathfrak{g}$ to the representations of $\mathfrak{j}$. One modifies the above procedure by finding a representation $R$ of  $\mathfrak{g}$ that generates all other representations via tensor products with itself, decomposing $R$ for the twisted punctures, and decomposing the represetation $\sigma(R)$ for untwisted punctures.

Note that  this method only works for the manifest flavour symmetry $F_{\text{manifest}}$ and those cases where the global form of $\hat{F}_{\text{manifest}}$ determines the global form of $\hat{F}$. It also assumes that if all the Schur operators are invariant under $\Gamma$, then same is true of all local operators in the SCFT.

\subsection{Theories with the Same Conventional Invariants}\label{isomorphic_theories}

In \cite{Distler:2020tub} it was proposed that the global form of the flavour symmetry might distinguish theories with the same ``conventional invariants": flavour symmetry algebras (and the associated current-algebra levels), $a,c$ central charges, and graded Coulomb branch dimensions.

Using the nilpotent Higgsings introduced above, we find a way to generate many theories with the same conventional invariants. We find many families of pairs where each fixture in each pair is related to a fixture in another pair via nilpotent Higgsings. Among these families we find examples of pairs whose conventional invariants coincide but with different global form of their flavour symmetry groups, pairs whose invariants coincide and are actually the same theory, as well as pairs with the same invariants and flavour symmetry groups but which are nonetheless distinct theories, disproving the conjecture of \cite{Distler:2020tub}.

%The first family has compatible global forms of their flavour symmetry and their Schur indices are found to agree. The second family has a more complicated story consisting of pairs that are distinct and some which are the same theory. In the last family we find theories whose global form of the flavour symmetry group is the same but whose Schur indices disagree at order $\tau^4$.

\subsubsection{$D_{2n}$}

Using very even punctures in the $D_{2n}$ series many theories were found with the same invariants in \cite{Distler:2020tub}. For example the pairs of $D_{2n}$ theories

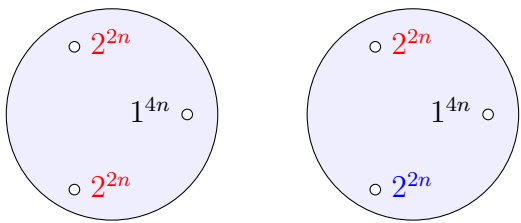
\begin{equation*}
\begin{tikzpicture}
\draw[radius=40pt,fill=lightblue] circle;
\draw[radius=2pt,fill=white]  (-.5,.9) circle node[above right=-7pt and 2pt] {$\color{red}{2^{2n}}$};
\draw[radius=2pt,fill=white]  (-.5,-1) circle node[above right=-7pt and 2pt] {$\color{red}{2^{2n}}$};
\draw[radius=2pt,fill=white]  (1,0) circle node[above left=-7pt and 2pt] {$1^{4n}$};
\draw[radius=40pt,fill=lightblue] (4,0) circle;
\draw[radius=2pt,fill=white]  (3.5,.9) circle node[above right=-7pt and 2pt] {$\color{red}{2^{2n}}$};
\draw[radius=2pt,fill=white]  (3.5,-1) circle node[above right=-7pt and 2pt] {$\color{blue}{2^{2n}}$};
\draw[radius=2pt,fill=white]  (5,0) circle node[above left=-7pt and 2pt] {$1^{4n}$};
\end{tikzpicture}
\end{equation*}
have the same conventional invariants, but are nonetheless distinct SCFTs.
The flavour symmetry is $\mathfrak{sp}(n)_{4n} \oplus \mathfrak{sp}(n)_{4n} \oplus \mathfrak{so}(4n)_{4(2n-1)}$. Let $\gamma_{1}$ generate the center of the $Sp(n)$ of the top puncture and $\gamma_{2}$ generate the center of the $Sp(n)$ of the bottom puncture. Let $\gamma_{3},\gamma_{4}$ generate the center of $Spin(4n)$ with $\gamma_{3}$ acting as $-1$ on the left handed spinor and vector representation, while $\gamma_{4}$ acts as $-1$ on the right handed spinor and vector representation. For $n$ even, $\Gamma$ for the theories on the left is given by $\langle \gamma, \gamma_{1} \gamma_{2}, \gamma_1 \gamma_3 \rangle$ while for the theories on the right it is $\langle \gamma \gamma_1 \gamma_2, \gamma_1\gamma_3, \gamma_2\gamma_4 \rangle$. For $n$ odd, $\Gamma$ for the theories on the left is $\langle \gamma \gamma_{3}\gamma_4, \gamma_1 \gamma_2, \gamma_1 \gamma_4 \rangle$, and for the theories on the right it is $\langle \gamma, \gamma_1 \gamma_3,  \gamma_2 \gamma_4 \rangle$. We see that they do indeed have different flavour symmetry groups.

\subsubsection{Generating examples from nilpotent Higgsing}

Using the technology of nilpotent Higgsings it is rather straightforward to generate additional examples of pairs of theories in class-S, with the same conventional invariants. Consider a fixture of the form
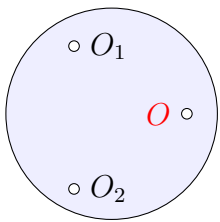
\begin{equation*}
\begin{tikzpicture}\draw[radius=40pt,fill=lightblue] circle;
\draw[radius=2pt,fill=white]  (-.5,.9) circle node[right=2pt] {$O_1$};
\draw[radius=2pt,fill=white]  (-.5,-1) circle node[right=2pt] {$O_2$};
\draw[radius=2pt,fill=white]  (1,0) circle node[left=2pt] {${\color{red}{O}}$};
\end{tikzpicture}
\end{equation*} 
in some theory of class-S, where there are nilpotent Higgsings
\begin{equation*}
O_1\xrightarrow{\; \mathfrak{f}_k\;} O_3,\qquad 
O_2\xrightarrow{\; \mathfrak{f}_k\;} O_4
\end{equation*}
such that the manifest flavour symmetries of the punctures satisfy
\begin{equation}\label{fcompat}
\mathfrak{f}_{O_1} = \mathfrak{f}_k \oplus \mathfrak{f}_{O_3},\qquad
\mathfrak{f}_{O_2} = \mathfrak{f}_k \oplus \mathfrak{f}_{O_4}.
\end{equation}
for the \emph{same} $\mathfrak{f}_k$. Then (provided that we choose the third puncture ${\color{red}{O}}$ sufficiently high up on the Hasse diagram), the theories
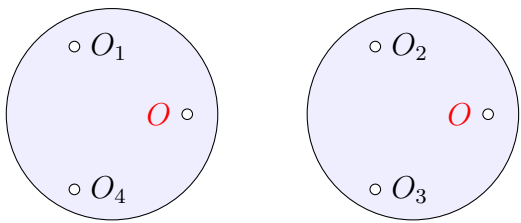
\begin{equation*}
\begin{tikzpicture}\draw[radius=40pt,fill=lightblue] circle;
\draw[radius=2pt,fill=white]  (-.5,.9) circle node[right=2pt] {$O_1$};
\draw[radius=2pt,fill=white]  (-.5,-1) circle node[right=2pt] {$O_4$};
\draw[radius=2pt,fill=white]  (1,0) circle node[left=2pt] {${\color{red}{O}}$};
\draw[radius=40pt,fill=lightblue] (4,0) circle;
\draw[radius=2pt,fill=white]  (3.5,.9) circle node[right=2pt] {$O_2$};
\draw[radius=2pt,fill=white]  (3.5,-1) circle node[right=2pt] {$O_3$};
\draw[radius=2pt,fill=white]  (5,0) circle node[left=2pt] {${\color{red}{O}}$};
\end{tikzpicture}
\end{equation*}
have exactly the same conventional invariants. (If we choose ${\color{red}{O}}$ too low on the Hasse diagram, then the Higgsing might yield a bad pair of theories, or one where the IR flavour symmetries are enhanced in different ways.)

%The requirement of the IR enhancements being the same may seem restrictive, however one can often choose Higgsings such that the manifest symmetries will be the same. Given enough punctures, it can be assured that the manifest flavour symmetry is the full flavour symmetry, and so these pairs can be guaranteed to have the same invariants. However if enhancements due occur to the manifest flavour symmetry, it is conceivably possible that the flavour symmetries can be different, though this can only happen with a low amount of punctures. In these cases one must check by hand if the invariants are the same. Additionally, even if the manifest flavour symmetries are different, the full flavour symmetries can be the same. We shall see some examples of this at the end of the section.

%The method involving the very even punctures above in some cases coincides with the method using nilpotent Higgsings, however not all very even punctures can be reached via different nilpotent Higgsings, so in general they are different.

As an example consider the fixture 

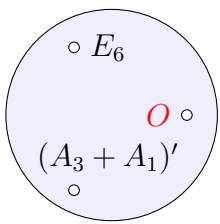
\begin{equation*}
\begin{tikzpicture}\draw[radius=40pt,fill=lightblue] circle;
\draw[radius=2pt,fill=white]  (-.5,.9) circle node[right=2pt] {$E_6$};
\draw[radius=2pt,fill=white]  (-.5,-1) circle node[above right=2pt and -18pt] {$(A_3+A_1)'$};
\draw[radius=2pt,fill=white]  (1,0) circle node[left=2pt] {${\color{red}{O}}$};
\node at (0,-2) {$F_{\text{manifest}}=SU(2)_{13} \times SU(2)_{24} \times SU(2)_{12}\times SU(2)_{12} \times F_{\color{red}{O}}$};
\end{tikzpicture}
\end{equation*} 
in the $E_7$ theory. Higgsing one or the other of the $SU(2)_{12}$s yields the pair of theories
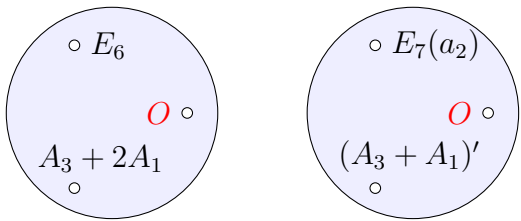
\begin{equation*}
\begin{tikzpicture}\draw[radius=40pt,fill=lightblue] circle;
\draw[radius=2pt,fill=white]  (-.5,.9) circle node[right=2pt] {$E_6$};
\draw[radius=2pt,fill=white]  (-.5,-1) circle node[above right=2ptand -18pt] {$A_3+2A_1$};
\draw[radius=2pt,fill=white]  (1,0) circle node[left=2pt] {${\color{red}{O}}$};
\draw[radius=40pt,fill=lightblue] (4,0) circle;
\draw[radius=2pt,fill=white]  (3.5,.9) circle node[right=2pt] {$E_7(a_2)$};
\draw[radius=2pt,fill=white]  (3.5,-1) circle node[above right=2pt and -18pt] {$(A_3+A_1)'$};
\draw[radius=2pt,fill=white]  (5,0) circle node[left=2pt] {${\color{red}{O}}$};
\end{tikzpicture}
\end{equation*}
With ${\color{red}{O}}$ chosen from the $E_7$ Higgsing diagram
\begin{equation}\label{familyHasse}
\scalebox{.75}{
\begin{tikzpicture}
\node (0) at (0,0) {$0$};
\node[right=1cm of 0] (A1) {$A_1$};
\node[right=2cm of A1] (A1A1) {$2A_1$};
\node[above right=1cm and 1cm of A1A1] (3A1pp) {$\color{midgreen}(3A_1)''$};
\node[below right=1cm and 1cm of A1A1] (3A1p) {$(3A_1)'$};
\node[right=1cm of 3A1pp] (A1A1A1A1) {$\color{midgreen}4A_1$};
\node[right=1cm of 3A1p] (A2) {$A_2$};
\node[right=5cm of A1A1] (A2A1) {$\color{midgreen}A_2+A_1$};
\node[right=1.5cm of A2A1] (A2A1A1) {$\color{midgreen}{A_2+2A_1}$};
\node[right=6.5cm of A1A1A1A1] (A2A2) {$\color{purple}2A_2$};
\node[right=6.5cm of A2] (A2A1A1A1) {$\color{midgreen}A_2+3A_1$};
\node[right=2.5cm of A2A1A1] (A2A2A1) {${\color{midgreen}2A_2+A_1}$};
\path[->] (0) edge node[above] {$(E_7)_{36}$} (A1)
(A1) edge node[above] {$Spin(12)_{28}$} (A1A1)
(A1A1) edge node[above left=-.125cm and -.125cm] {$SU(2)_{20}$} (3A1pp)
(A1A1) edge node[below left=-.125cm and -.125cm] {$Spin(9)_{24}$} (3A1p)
(3A1pp) edge node[above] {$(F_4)_{24}$} (A1A1A1A1)
(3A1p) edge node[above left=-.125cm and -.125cm] {$Sp(3)_{20}$} (A1A1A1A1)
(3A1p) edge node[below] {$SU(2)_{19}$} (A2)
(A1A1A1A1) edge node[above right=-.125cm and -.125cm] {$Sp(3)_{19}$} (A2A1)
(A2) edge node[below right=-.125cm and -.125cm] {$SU(6)_{20}$} (A2A1)
(A2A1) edge node[above] {$SU(4)_{18}$} (A2A1A1)
(A2A1A1) edge node[above left=-.125cm and -.125cm] {$SU(2)_{28}$} (A2A2)
(A2A1A1) edge node[below left=-.125cm and -.125cm] {$SU(2)_{16}$} (A2A1A1A1)
(A2A2) edge node[above right=-.125cm and -.125cm] {$(G_2)_{16}$} (A2A2A1)
(A2A1A1A1) edge node[below right=-.125cm and -.125cm] {$(G_2)_{28}$} (A2A2A1)
;
\end{tikzpicture}
}
\end{equation}
the manifest flavour symmetry algebras are the same for each pair in this family (except for $2A_2$, as we will explain presently). The other conventional invariants also coincide. But the theories in black have different $\Gamma$s and are non-isomorphic SCFTs, as verified by computing the respective Schur indices. 

This pattern continues until we Higgs $2A_1\xrightarrow{SU(2)_{20}}{\color{midgreen}(3A_1)''}$. For that pair, the $\Gamma$s are the same and a computation of the Schur index to $O(\tau^{12})$ 
\begin{equation*}
\begin{split}
I_{\text{Schur}}({\color{red}O}=(3A_1)'')=1&+61 \tau^2+10 \tau^3+2017 \tau^4+798 \tau^5+47969 \tau^6+32690\tau^7+920783 \tau^8\\&+927216 \tau^9+15202440 \tau^{10}+20597848 \tau^{11}+224805960 \tau^{12}+O(\tau^{13})
\end{split}
\end{equation*}
leads us to believe that the SCFTs are also isomorphic. The same holds for subsequent Higgsings, and all of the pairs marked in {\color{midgreen}green} appear to be isomorphic SCFTs.  

A new phenomenon, however, occurs one we reach $\color{midgreen}A_2+A_1$. For the remaining 5 punctures, the flavour symmetry is enhanced over the manifest one, and our methods do not determine the full $\Gamma$, but only the subgroup $\Gamma_{\text{manifest}}$.

Computing the Schur indices, however, leads us to believe that the pairs of theories marked in {\color{midgreen}green}  are isomorphic, which leads to a \emph{prediction} for $\Gamma$ for 3 of those 4 cases.

\begin{itemize}
\item For ${\color{red}{O}} = {\color{midgreen}A_2+A_1}$, a manifest $U(1)$ is enhanced to $U(1)^2$. Computing the manifest flavour symmetries one finds for the theory on the left that the manifest $\Gamma$ is generated by $\langle \gamma_{1} \delta_{L} \gamma_4 \rangle$ while on the right it is generated by $\langle \gamma_{1} \delta_{R} \gamma_2 \gamma_3 \rangle$. Here $\delta_{L}$ is the generator of the $\mathbb{Z}_{4}$ subgroup of the manifest $U(1)$ for the theory on the left and similarly for $\delta_{R}$. Assuming the two theories are the same due to their matching Schur indices this implies the manifest $U(1)$ symmetries are different. Thus we may write the $U(1)^2$ of both theories as $U(1)_{L} \times U(1)_{R}$ where $U(1)_{L}$ is the manifest $U(1)$ of the theory on the left and $U(1)_{R}$ is the manifest $U(1)$ on the theory on the right. Requiring the global forms of the flavour symmetry groups to be the same determines $\Gamma$ to be $\langle \gamma_{1} \delta_{L} \gamma_4 ,\gamma_{1} \delta_{R} \gamma_2 \gamma_3 \rangle$.
\item For ${\color{red}{O}} = {\color{midgreen}A_2+2A_1}$, the manifest flavour symmetry is $SU(2)_{16} \times SU(2)_{28} \times SU(2)_{84} \times SU(2)_{13} \times SU(2)_{24} \times SU(2)_{12}$.  Let $\Gamma_{i}$ be the generator of the center of the $i$th factor in the manifest flavour symmetry group above. Then $\Gamma_{L,\text{manifest}} = \langle \gamma_1 \gamma_2 \gamma_{6}, \gamma \gamma_{3} \gamma_4 \gamma_5  \rangle$ and $\Gamma_{R,\text{manifest}} = \langle \gamma_1 \gamma_2 \gamma_5 \gamma_6, \gamma \gamma_3 \gamma_4  \rangle$. While the manifest flavour symmetry groups have different global form, the Schur indices nonetheless agree up to $O(\tau^{10})$, which leads us to believe that these are isomorphic theories. In fact, the $SU(2)_{84}$ is enhanced to an $SU(2)_{56} \times SU(2)_{28}$. So $F=SU(2)_{16} \times SU(2)_{28}^2 \times SU(2)_{56} \times SU(2)_{13} \times SU(2)_{24} \times SU(2)_{12}$ and its center has an additional generator. Demanding that $\Gamma_L=\Gamma_R\equiv\Gamma$ then determines $\Gamma=\langle \gamma_1 \gamma_2 \gamma_7, \gamma \gamma_3 \gamma_4 \gamma_5 \gamma_6, \gamma_1 \gamma_3 \gamma_6 \gamma_7 \rangle$, where $\gamma_{i}$ is the generator of the center of the $i^{\text{th}}$ factor in $F$.
\item Since there are two $SU(2)_{28}$s, the Higgsing to $\color{purple}2A_2$ involves a \emph{different} $SU(2)_{28}$ in the theory on the left versus the theory on the right. And, indeed, the two theories with $\color{purple}2A_2$ have \emph{different} flavour symmetry algebras: $F_L=Spin(7)_{16} \times SU(2)_{24}^2\times SU(2)_{13}\times Sp(2)_{12}$, while $F_R= Spin(7)_{16}\times Sp(2)_{24}\times SU(2)_{13} \times  Sp(2)_{12}$.
\item For ${\color{red}{O}} = {\color{midgreen}A_2+3A_1}$, the flavour symmetry enhancement is from $(G_2)_{28}$ to $Spin(7)_{28}$ adding element to the center. However we can determine the possible additional generator of $\Gamma$ by Higgsing from the ${\color{red}{O}} = {\color{midgreen}A_2+2A_1}$ theory. The element of $\Gamma$ of the UV theory $\gamma_{2}\gamma_{3}\gamma_{6}$ embeds into the IR theory as $\gamma_{1}\gamma_{3}$. Thus we are able to completely determine $\Gamma$. 
\item For ${\color{red}{O}} = {\color{midgreen}2A_2+A_1}$, there could be additional elements of $\Gamma$ due to the enhancements, however we cannot determine them with our methods.
\end{itemize}

We assemble our results for this family of pairs of SCFTs in the table below. They seem to support the conjecture of \cite{Distler:2020tub} that the distinct SCFTs with the same ``conventional" invariants are distinguished by the global form of the flavour symmetry group and, conversely, when the conventional invariants and the global forms of the flavour symmetry group coincide, the theories are isomorphic.

{
\footnotesize
\setlength\LTleft{-.175in}
\renewcommand{\arraystretch}{3}

\begin{longtable}{|c|c|c|c|c|}
\hline
$\color{red}O$&Manifest Flavour Symmetry&Flavour Symmetry Group&$\Gamma_{L}$&$\Gamma_{R}$\\
\endhead
\hline 
$0$&$\begin{gathered}(E_7)_{36} \times SU(2)_{13} \times\\ SU(2)_{24}\times SU(2)_{12}\end{gathered}$&$\begin{gathered}(E_7)_{36} \times SU(2)_{13} \times\\ SU(2)_{24}\times SU(2)_{12}\end{gathered}$&$\begin{aligned}\langle \gamma_{1} &\gamma_{4},\\& \gamma \gamma_{2} \gamma_{3}  \rangle \end{aligned}$&$\begin{aligned} \langle \gamma_{1}& \gamma_{2} \gamma_{3},\\& \gamma  \gamma_{4} \rangle \end{aligned}$\\
\hline
$A_1$&$\begin{gathered}Spin(12)_{28} \times SU(2)_{13}\\ \times SU(2)_{24}\times SU(2)_{12} \end{gathered}$&$\begin{gathered}Spin(12)_{28} \times SU(2)_{13}\\ \times SU(2)_{24}\times SU(2)_{12}\end{gathered}$&$\begin{aligned}\langle \gamma_{1}& \gamma_{5},\\ &\gamma \gamma_2 \gamma_3 \gamma_4 \rangle\end{aligned}$&$\begin{aligned}\langle \gamma_{1}& \gamma_{3} \gamma_{4},\\ &\gamma \gamma_2 \gamma_5 \rangle\end{aligned}$\\
\hline
$2A_1$&$\begin{gathered}Spin(9)_{24} \times SU(2)_{20} \times\\ SU(2)_{13} \times SU(2)_{24}\times SU(2)_{12}\end{gathered}$&$\begin{gathered}Spin(9)_{24} \times SU(2)_{20} \times\\ SU(2)_{13} \times SU(2)_{24}\times SU(2)_{12}\end{gathered}$&$\begin{aligned}\langle \gamma_{1}& \gamma_{2}\gamma_{5}, \\ & \gamma \gamma_{3} \gamma_{4}  \rangle\end{aligned}$&$\begin{aligned}\langle \gamma_{1}&\gamma_{2} \gamma_{3} \gamma_{4}, \\ & \gamma \gamma_5\rangle\end{aligned}$\\
\hline
$(3A_1)'$&$\begin{gathered}Sp(3)_{20} \times SU(2)_{19} \times\\ SU(2)_{13} \times SU(2)_{24}\times SU(2)_{12}\end{gathered}$&$\begin{gathered}Sp(3)_{20} \times SU(2)_{19} \times\\ SU(2)_{13} \times SU(2)_{24}\times SU(2)_{12}\end{gathered}$&$\begin{aligned}
\langle \gamma_{1}& \gamma_{2}\gamma_{5},\\ &\gamma \gamma_2 \gamma _3 \gamma_4 \rangle\end{aligned}$&$\begin{aligned}\langle \gamma_{1}&\gamma_{2} \gamma_{3} \gamma_{4},\\ &\gamma \gamma_2 \gamma_5 \rangle\end{aligned}$\\
\hline
$A_2$&$\begin{gathered}SU(6)_{20} \times SU(2)_{13} \times\\ SU(2)_{24}\times SU(2)_{12} \times U(1)\end{gathered}$&$\begin{gathered}SU(6)_{20} \times SU(2)_{13} \times\\ SU(2)_{24}\times SU(2)_{12} \times U(1)\end{gathered}$&$\begin{aligned}\langle \gamma_{1}^3& \gamma_{4},\\ &\gamma \gamma_2 \gamma_3  \rangle\end{aligned}$&$\begin{aligned}\langle \gamma_{1}^3& \gamma_{2} \gamma_{3},\\ &\gamma  \gamma_4 \rangle\end{aligned}$\\
\hline
$(3A_1)''$&$\begin{gathered}(F_4)_{24} \times SU(2)_{13} \times\\ SU(2)_{24}\times SU(2)_{12}\end{gathered}$&$\begin{gathered}(F_4)_{24} \times SU(2)_{13}\\ \times SU(2)_{24}\times SU(2)_{12}\end{gathered}$&$\langle \gamma\gamma_{2}\gamma_{3} \gamma_{4} \rangle $&$\langle \gamma \gamma_{2} \gamma_{3} \gamma_{4} \rangle $\\
\hline
$4A_1$&$\begin{gathered}Sp(3)_{19} \times SU(2)_{13} \times\\ SU(2)_{24}\times SU(2)_{12}\end{gathered}$&$\begin{gathered}Sp(3)_{19} \times SU(2)_{13} \times\\ SU(2)_{24}\times SU(2)_{12}\end{gathered}$&$\langle \gamma \gamma _1 \gamma_{2} \gamma_{3} \gamma_{4}\rangle$&$ \langle \gamma \gamma_1 \gamma_{2} \gamma_{3} \gamma_{4} \rangle$\\
\hline
$A_2+A_1$&$\begin{gathered}SU(4)_{18} \times SU(2)_{13} \times\\ SU(2)_{24}\times SU(2)_{12} \times U(1)\end{gathered}$&$\begin{gathered}SU(4)_{18} \times SU(2)_{13} \times SU(2)_{24}\\ \times SU(2)_{12} \times U(1)_{L} \times U(1)_{R} \end{gathered}$&$\begin{aligned}
\langle \gamma_{1}& \delta_{L} \gamma_4,\\ &\gamma_{1} \delta_{R} \gamma_2 \gamma_3, \\& \gamma \gamma_{1}^2 \gamma_2 \gamma_3 \gamma_4   \rangle
\end{aligned}$&$\begin{aligned}\langle \gamma_{1}& \delta_{L} \gamma_4,\\ &\gamma_{1} \delta_{R} \gamma_2 \gamma_3,\\& \gamma \gamma_{1}^2 \gamma_2 \gamma_3 \gamma_4 \rangle\end{aligned}$\\
\hline
$A_2+2A_1$&$\begin{gathered}SU(2)_{16} \times SU(2)_{28} \times SU(2)_{84}\\ \times SU(2)_{13} \times SU(2)_{24}\times SU(2)_{12}\end{gathered}$&$\begin{gathered}SU(2)_{16} \times SU(2)_{28}^2 \times SU(2)_{56}\\ \times SU(2)_{13} \times SU(2)_{24}\times SU(2)_{12}\end{gathered}$&$\begin{aligned}\langle \gamma_1 &\gamma_2 \gamma_7,\\ & \gamma \gamma_3 \gamma_4 \gamma_5 \gamma_6,\\ & \gamma_1 \gamma_3 \gamma_6 \gamma_7 \rangle\end{aligned}$&$\begin{aligned}\langle \gamma_1 &\gamma_2 \gamma_7,\\& \gamma \gamma_3 \gamma_4 \gamma_5 \gamma_6,\\& \gamma_1 \gamma_3 \gamma_6 \gamma_7 \rangle\end{aligned}$\\
\hline
$A_2+3A_1$&$\begin{gathered}(G_2)_{28} \times SU(2)_{13}\\ \times SU(2)_{24}\times SU(2)_{12} \end{gathered}$&$\begin{gathered}Spin(7)_{28} \times SU(2)_{13}\\ \times SU(2)_{24}\times SU(2)_{12} \end{gathered}$&$\begin{aligned}\langle \gamma \gamma_{2}& \gamma_{3} \gamma_{4}, \\& \gamma_{1} \gamma_{3} \rangle \end{aligned}$&$\begin{aligned}\langle \gamma \gamma_{2}& \gamma_{3} \gamma_{4}, \\& \gamma_1 \gamma_3 \rangle \end{aligned}$\\
\hline
$2A_2+A_1$&$\begin{gathered} SU(2)_{36} \times SU(2)_{38} \times\\ SU(2)_{13} \times SU(2)_{24}\times SU(2)_{12}\end{gathered}$&$\begin{gathered}Sp(2)_{24} \times SU(2)_{24}\\ \times Sp(2)_{12} \times Sp(2)_{13}\end{gathered}$&$\langle \gamma \gamma_1 \gamma _2 \gamma_3 \gamma_4 \rangle$&$\langle \gamma \gamma_1 \gamma _2 \gamma_3 \gamma_4 \rangle$\\
\hline
\end{longtable}
}

\subsubsection{Another family}

The previous family of examples seemed in perfect accord with the conjecture of \cite{Distler:2020tub}. When the global form of the flavour symmetry groups differ, the SCFTs were distinct; when the global forms were the same, the theories were isomorphic. The next family of examples will not be so obliging.

Again, take the $E_7$ theory and the fixture 
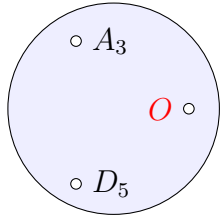
\begin{equation*}
\begin{tikzpicture}\draw[radius=40pt,fill=lightblue] circle;
\draw[radius=2pt,fill=white]  (-.5,.9) circle node[right=2pt] {$A_3$};
\draw[radius=2pt,fill=white]  (-.5,-1) circle node[right=2pt] {$D_5$};
\draw[radius=2pt,fill=white]  (1,0) circle node[left=2pt] {${\color{red}{O}}$};
\node at (0,-2) {$F_{\text{manifest}}=Spin(7)_{16} \times SU(2)_{12} \times SU(2)_{12}\times SU(2)_{8} \times F_{\color{red}{O}}$};
\end{tikzpicture}
\end{equation*}
We can Higgs either $SU(2)_{12}$ to obtain a pair of theories
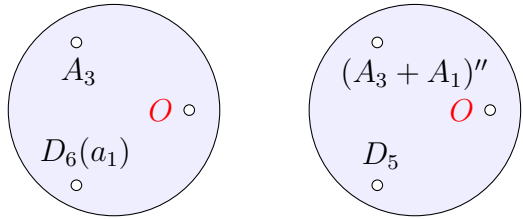
\begin{equation*}
\begin{tikzpicture}
\draw[radius=40pt,fill=lightblue] circle;
\draw[radius=2pt,fill=white]  (-.5,.9) circle node[below right=2pt and -10pt] {$A_3$};
\draw[radius=2pt,fill=white]  (-.5,-1) circle node[above right=2pt and -18pt] {$D_6(a_1)$};
\draw[radius=2pt,fill=white]  (1,0) circle node[left=2pt] {${\color{red}{O}}$};
\draw[radius=40pt,fill=lightblue] (4,0) circle;
\draw[radius=2pt,fill=white]  (3.5,.9) circle node[below right=2pt and -18pt] {$(A_3+A_1)''$};
\draw[radius=2pt,fill=white]  (3.5,-1) circle node[above right=2pt and -10pt] {$D_5$};
\draw[radius=2pt,fill=white]  (5,0) circle node[left=2pt] {${\color{red}{O}}$};
\end{tikzpicture}
\end{equation*} 
Every pair has the same conventional invariants. Moveover, for 
$$
\begin{aligned}
{\color{red}O} \in \{
(3A_1)'',\;&
4A_1,\;
A_2+3A_1,\;
(A_3+A_1)'',\;
2A_2+A_1,\;
A_3+2A_1,\;
D_4(a_1)+A_1,\;
A_4,\;\\
&A_3+A_2+A_1,\;
(A_5)'',\;
D_4+A_1,\;
A_5+A_1,\;
D_5(a_1)+A_1,\;
D_6(a_2),\;
E_7(a_5)\}
\end{aligned}
$$
they even have the same global form of the flavour symmetry group. For instance, for ${\color{red}{O}}=4A_1$ the flavour symmetry for both theories is 
\begin{equation*}
\hat{F}=\left(Spin(7)_{16} \times Sp(3)_{19}\times SU(2)_{8}\times SU(2)_{12}\times S\right)/\Gamma
\end{equation*}
 where $\Gamma = \langle \gamma \gamma_{2}\gamma_{3} \gamma_{4} \rangle$. 
 
Nevertheless, for 11 of those 15 punctures, one quickly discovers that the theories are not isomorphic. The Schur indices differ at $O(\tau^4)$ because the theory on the right has an additional $\hat{B}_2$ operator, in the $8$ of the $Spin(7)_{16}$ associated to the $(A_3+A_1)''$ puncture, that is absent in the theory on the left.

At least in some case, perhaps we should have \emph{expected} this to be the case.  While it's true that the two theories with ${\color{red}{O}}=4A_1$ have the same global form of the flavour symmetry group,  we can Higgs $4A_1\xrightarrow{\; Sp(3)_{19}\;} A_2+A_1$ . The theories with ${\color{red}{O}}=A_2+A_1$ have \emph{distinct} global forms of the flavour symmetry group, which \emph{ought} to tell us that the parent theories with ${\color{red}{O}}=4A_1$ must also be distinct\footnote{Note that this is unlike the family of examples in \eqref{familyHasse}, where there were two distinct $A_2+2A_1\xrightarrow{\; SU(2)_{28}\;}2A_2$ Higgsings, which led to distinct child SCFTs, even though the parents were isomorphic.}. And once we've determined that the two theories with ${\color{red}{O}}=4A_1$ are distinct, then the Higgsing $(3A_1)''\xrightarrow{\;{(F_4)}_{24}\;}4A_1$ implies that the theories with ${\color{red}{O}}=(3A_1)''$ must also be distinct.

In similar fashion, this ``explains" why the pairs associated to ${\color{red}{O}}=(A_3+A_1)''$, $A_4$ and $D_4+A_1$ are distinct SCFTs. Alas, there's no similar explanation for remaining six. More surprising, the remaining four choices for $ {\color{red}{O}}$, which are related by the nilpotent Higgsings
\begin{equation*}
{\color{midgreen}(A_5)''}\xrightarrow{\;(G_2)_{12}\;}{\color{midgreen}(A_5+A_1)},\qquad {\color{midgreen}D_6(a_2)}\xrightarrow{\;SU(2)_9\;}{\color{midgreen}E_7(a_5)},
\end{equation*}
\emph{do} appear to lead to isomorphic pairs of SCFTs. We have checked their Schur indices up to $O(\tau^{12})$, and they agree.

\subsubsection{Sporadic examples}
In sporadic cases, the condition \eqref{fcompat} can be relaxed, with an infrared enhancement of the flavour symmetry compensating for the lack of agreement of the manifest symmetries.
As an example, start with the $E_7$ fixture 

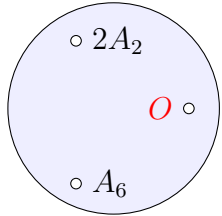
\begin{equation*}
\begin{tikzpicture}\draw[radius=40pt,fill=lightblue] circle;
\draw[radius=2pt,fill=white]  (-.5,.9) circle node[right=2pt] {$2A_2$};
\draw[radius=2pt,fill=white]  (-.5,-1) circle node[right=2pt] {$A_6$};
\draw[radius=2pt,fill=white]  (1,0) circle node[left=2pt] {${\color{red}{O}}$};
\node at (0,-2) {$F_{\text{manifest}}=(G_2)_{16} \times SU(2)_{36} \times SU(2)_{36}\times F_{\color{red}{O}}$};
\end{tikzpicture}
\end{equation*} 
For general ${\color{red}{O}}$, Higgsing one or the other of the $SU(2)_{36}$ yields IR theories with \emph{different} flavour symmetries as Higgsing $2A_2\xrightarrow{\;SU(2)_{36}\;}(A_3+A_1)''$ enhances the $(G_2)_{16}$ symmetry of the $2A_2$ puncture to the $Spin(7)_{16}$ symmetry of the $(A_3+A_1)''$ puncture.

However, if we choose ${\color{red}{O}}$ from the Higgsing diagram
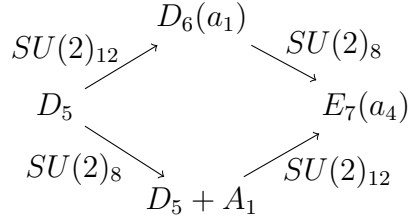
\begin{equation*}
\begin{tikzpicture}
\node (D5) at (0,0) {$D_5$};
\node[above right=.5cm and .8cm of D5] (D6a1) {$D_6(a_1)$};
\node[below=1.75cm of D6a1] (D5A1) {$D_5+A_1$};
\node[right=3cm of D5] (E7a4) {$E_7(a_4)$};
\path[->] (D5) edge node[above left=-.125cm and -.125cm] {$SU(2)_{12}$} (D6a1)
(D5) edge node[below left=-.125cm and -.125cm] {$SU(2)_{8}$} (D5A1)
(D6a1) edge node[above right=-.125cm and -.125cm] {$SU(2)_{8}$} (E7a4)
(D5A1) edge node[below right=-.125cm and -.125cm] {$SU(2)_{12}$} (E7a4)
;
\end{tikzpicture}
\end{equation*}
then the $(G_2)_{16}$ is enhanced to $Spin(7)_{16}$ for both fixtures in the pair
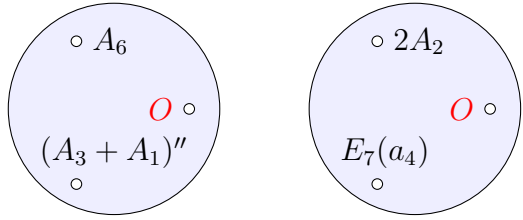
\begin{equation*}
\begin{tikzpicture}\draw[radius=40pt,fill=lightblue] circle;
\draw[radius=2pt,fill=white]  (-.5,.9) circle node[right=2pt] {$A_6$};
\draw[radius=2pt,fill=white]  (-.5,-1) circle node[above right=2ptand -18pt] {$(A_3+A_1)''$};
\draw[radius=2pt,fill=white]  (1,0) circle node[left=2pt] {${\color{red}{O}} $};
\draw[radius=40pt,fill=lightblue] (4,0) circle;
\draw[radius=2pt,fill=white]  (3.5,.9) circle node[right=2pt] {$2A_2$};
\draw[radius=2pt,fill=white]  (3.5,-1) circle node[above right=2pt and -18pt] {$E_7(a_4)$};
\draw[radius=2pt,fill=white]  (5,0) circle node[left=2pt] {${\color{red}{O}} $};
\end{tikzpicture}
\end{equation*}
While the manifest symmetries are different (the theory on the right has a $(G_2)_{16}$ where the theory on the left has a $Spin(7)_{16}$ factor), the enhanced flavour symmetry algebras are the same. The Schur indices agree up to $O(\tau^9)$. Due to enhancements we are unable to compute the global form of the full flavour symmetry, however the elements we are able to compute are the same. Thus we believe that each pair represents isomorphic SCFTs.

\section*{Acknowledgements}
We would like to thank Monica Kang and Craig Lawrie for useful discussions, and Mario Martone for the collaboration which inspired us to find applications for this class of nilpotent Higgsings. Some of this work was pursued at the 2022 Aspen Winter Workshop ``Geometrization of (S)QFTs in $D\leq6$". The first author gratefully acknowledges the Aspen Center for Physics, which is supported by the National Science Foundation Grant No.~PHY--1607611. This work was supported in part by the National Science Foundation under Grant No.~PHY--1914679. 
\bibliographystyle{utphys}
%\small\baselineskip=.93\baselineskip
%\let\bbb\bibitem\def\bibitem{\itemsep1pt\bbb}
\bibliography{ref}

\providecommand{\href}[2]{#2}\begingroup\raggedright\begin{thebibliography}{10}

\bibitem{Distler:2020tub}
J.~Distler, B.~Ergun, and A.~Shehper, ``Distinguishing {$d = 4$ $\mathcal{N} =
  2$ SCFTs},'' \href{http://arxiv.org/abs/2012.15249}{{\ttfamily
  arXiv:2012.15249 [hep-th]}}.

\bibitem{Chacaltana:2014jba}
O.~Chacaltana, J.~Distler, and A.~Trimm, ``Tinkertoys for the {$E_6$} theory,''
  \href{http://dx.doi.org/10.1007/JHEP09(2015)007}{{\em JHEP} {\bfseries 09}
  (2015) 007},
\href{http://arxiv.org/abs/1403.4604}{{\ttfamily arXiv:1403.4604 [hep-th]}}.
%%CITATION = ARXIV:1403.4604;%%.

\bibitem{Chacaltana:2015bna}
O.~Chacaltana, J.~Distler, and A.~Trimm, ``Tinkertoys for the twisted {$E_6$}
  theory,'' \href{http://dx.doi.org/10.1007/JHEP04(2015)173}{{\em JHEP}
  {\bfseries 04} (2015) 173},
\href{http://arxiv.org/abs/1501.00357}{{\ttfamily arXiv:1501.00357 [hep-th]}}.
%%CITATION = ARXIV:1501.00357;%%.

\bibitem{Chacaltana:2017boe}
O.~Chacaltana, J.~Distler, A.~Trimm, and Y.~Zhu, ``Tinkertoys for the {$E_7$}
  theory,''
\href{http://arxiv.org/abs/1704.07890}{{\ttfamily arXiv:1704.07890 [hep-th]}}.
%%CITATION = ARXIV:1704.07890;%%.

\bibitem{Chacaltana:2018vhp}
O.~Chacaltana, J.~Distler, A.~Trimm, and Y.~Zhu, ``Tinkertoys for the {$E_8$}
  theory,'' \href{http://arxiv.org/abs/1802.09626}{{\ttfamily arXiv:1802.09626
  [hep-th]}}.

\bibitem{Tachikawa:2015bga}
Y.~Tachikawa, ``A review of the {$T_N$} theory and its cousins,''
  \href{http://dx.doi.org/10.1093/ptep/ptv098}{{\em PTEP} {\bfseries 2015}
  no.~11, (2015) 11B102},
\href{http://arxiv.org/abs/1504.01481}{{\ttfamily arXiv:1504.01481 [hep-th]}}.
%%CITATION = ARXIV:1504.01481;%%.

\bibitem{Beem:2019snk}
C.~Beem, C.~Meneghelli, W.~Peelaers, and L.~Rastelli, ``{VOAs} and rank-two
  instanton {SCFTs},'' \href{http://dx.doi.org/10.1007/s00220-020-03746-9}{{\em
  Commun. Math. Phys.} {\bfseries 377} no.~3, (2020) 2553--2578},
  \href{http://arxiv.org/abs/1907.08629}{{\ttfamily arXiv:1907.08629
  [hep-th]}}.

\bibitem{Distler-Martone}
J.~Distler and M.~Martone. In preparation.

\bibitem{Beem:2014rza}
C.~Beem, W.~Peelaers, L.~Rastelli, and B.~C. van Rees, ``Chiral algebras of
  class {S},'' \href{http://dx.doi.org/10.1007/JHEP05(2015)020}{{\em JHEP}
  {\bfseries 05} (2015) 020},
\href{http://arxiv.org/abs/1408.6522}{{\ttfamily arXiv:1408.6522 [hep-th]}}.
%%CITATION = ARXIV:1408.6522;%%.

\bibitem{Bhardwaj:2021ojs}
L.~Bhardwaj, ``Global form of flavor symmetry groups in 4d {$\mathcal{N}=2$}
  theories of class {S},'' \href{http://arxiv.org/abs/2105.08730}{{\ttfamily
  arXiv:2105.08730 [hep-th]}}.

\bibitem{MR3570131}
B.~Fu, D.~Juteau, P.~Levy, and E.~Sommers, ``Generic singularities of nilpotent
  orbit closures,'' \href{http://dx.doi.org/10.1016/j.aim.2016.09.010}{{\em
  Adv. Math.} {\bfseries 305} (2017) 1--77},
  \href{http://arxiv.org/abs/1502.05770}{{\ttfamily arXiv:1502.05770
  [math.RT]}}.

\bibitem{MR604841}
H.~Kraft and C.~Procesi, ``Minimal singularities in {${\rm GL}_{n}$},'' {\em
  Invent. Math.} {\bfseries 62} no.~3, (1981) 503--515.

\bibitem{MR694606}
H.~Kraft and C.~Procesi, ``On the geometry of conjugacy classes in classical
  groups,'' \href{http://dx.doi.org/10.1007/BF02565876}{{\em Comment. Math.
  Helv.} {\bfseries 57} no.~4, (1982) 539--602}.

\bibitem{Chacaltana:2012zy}
O.~Chacaltana, J.~Distler, and Y.~Tachikawa, ``Nilpotent orbits and
  codimension-two defects of 6d {$N=(2,0)$} theories,''
  \href{http://dx.doi.org/10.1142/S0217751X1340006X}{{\em Int. J. Mod. Phys.}
  {\bfseries A28} (2013) 1340006},
\href{http://arxiv.org/abs/1203.2930}{{\ttfamily arXiv:1203.2930 [hep-th]}}.
%%CITATION = ARXIV:1203.2930;%%.

\bibitem{Baume:2021qho}
F.~Baume, M.~J. Kang, and C.~Lawrie, ``Two 6d origins of 4d {SCFTs}: class
  {$\mathcal{S}$} and 6d (1,0) on a torus,''
  \href{http://arxiv.org/abs/2106.11990}{{\ttfamily arXiv:2106.11990
  [hep-th]}}.

\bibitem{MZprivate}
M.~Martone and G.~Zafrir. Private communication.

\bibitem{Chacaltana:2010ks}
O.~Chacaltana and J.~Distler, ``Tinkertoys for {G}aiotto duality,''
  \href{http://dx.doi.org/10.1007/JHEP11(2010)099}{{\em JHEP} {\bfseries 1011}
  (2010) 099},
\href{http://arxiv.org/abs/1008.5203}{{\ttfamily arXiv:1008.5203 [hep-th]}}.
%%CITATION = ARXIV:1008.5203;%%.

\bibitem{Distler:2018gbc}
J.~Distler and B.~Ergun, ``Product {SCFTs} for the {$E_7$} theory,''
  \href{http://arxiv.org/abs/1803.02425}{{\ttfamily arXiv:1803.02425
  [hep-th]}}.

\bibitem{Gadde:2011uv}
A.~Gadde, L.~Rastelli, S.~S. Razamat, and W.~Yan, ``Gauge theories and
  {M}acdonald polynomials,''
  \href{http://dx.doi.org/10.1007/s00220-012-1607-8}{{\em Commun. Math. Phys.}
  {\bfseries 319} (2013) 147--193},
\href{http://arxiv.org/abs/1110.3740}{{\ttfamily arXiv:1110.3740 [hep-th]}}.
%%CITATION = ARXIV:1110.3740;%%.

\bibitem{Mekareeya:2012tn}
N.~Mekareeya, J.~Song, and Y.~Tachikawa, ``2d {TQFT} structure of the
  superconformal indices with outer-automorphism twists,''
  \href{http://dx.doi.org/10.1007/JHEP03(2013)171}{{\em JHEP} {\bfseries 1303}
  (2013) 171},
\href{http://arxiv.org/abs/1212.0545}{{\ttfamily arXiv:1212.0545 [hep-th]}}.
%%CITATION = ARXIV:1212.0545;%%.

\end{thebibliography}\endgroup

\end{document}